\documentclass[submission, Phys]{SciPost}
\pdfoutput=1 
\usepackage{slashed}
\usepackage{amssymb}
\usepackage{arydshln}
\usepackage{adjustbox}
\usepackage{easybmat}
\usepackage{bbm}
\usepackage{bm}
\usepackage{adjustbox}
\usepackage{booktabs}
\usepackage{multirow} 
\usepackage{lscape}
\usepackage{booktabs}

\newcommand{\ii}{\ensuremath{\mathrm{i}}}
\newcommand{\mr}[1]{\textcolor{black}{#1}}
\newcommand{\jg}[1]{\textcolor{black}{#1}}
\newcommand{\gtick}{\textcolor{green}{\mathbf{\checkmark}}}
\newcommand{\otick}{\textcolor{orange}{\mathbf{\checkmark}}}
\newcommand{\rtick}{\textcolor{red}{\text{\sffamily X}}}
\allowdisplaybreaks

\begin{document}
\begin{center}
{\Large \textbf{Towards the renormalisation of the Standard Model
  effective field theory to dimension eight: Bosonic interactions I
  }
}\end{center}
\begin{center}
M. Chala\textsuperscript{1*},
G. Guedes\textsuperscript{1,2},
M. Ramos\textsuperscript{1,2},
J. Santiago\textsuperscript{1},
\end{center}

\begin{center}
{\bf 1} CAFPE and Departamento de F\'isica Te\'orica y del Cosmos,
Universidad de Granada, Campus de Fuentenueva, E--18071 Granada, Spain
\\
{\bf 2} Laborat\'orio de Instrumenta\c cao e F\'isica Experimental de Part\'iculas, Departamento de
F\'isica da Universidade do Minho, Campus de Gualtar, 4710-057 Braga, Portugal
\\[0.5cm]
* mikael.chala@ugr.es
\end{center}

\begin{center}
\today
\end{center}


\section*{Abstract}
{\bf
We compute the one-loop renormalisation group running of the bosonic Standard Model effective operators to order $v^4/\Lambda^4$, with $v\sim 246$ GeV being the electroweak scale and  $\Lambda$ the unknown new physics threshold. We concentrate on the effects triggered by pairs of the leading dimension-six interactions, namely those that can arise at tree level in weakly-coupled ultraviolet completions of the Standard Model.
We highlight some interesting consequences, including the interplay
between positivity bounds and the form of the anomalous dimensions;
the non renormalisation of the $S$ and $U$ parameters; or the
importance of radiative corrections to the Higgs potential for the
electroweak phase transition. As a byproduct of this work, we provide
a complete Green basis of operators involving only the Higgs and derivatives at
dimension-eight, comprising 13 redundant interactions.
}

\vspace{10pt}
\noindent\rule{\textwidth}{1pt}
\tableofcontents\thispagestyle{fancy}
\noindent\rule{\textwidth}{1pt}
\vspace{10pt}


\section{Introduction}
The Standard Model (SM) extended with effective interactions, also
known as SM effective field theory (SMEFT)~\cite{Brivio:2017vri}, is
increasingly becoming one of the favourite    
options for describing particle physics at currently explored
energies. The main reasons are the apparent absence of new resonances
below the  
TeV scale~\cite{Yuan:2020fyf}
and the fact that in general the SMEFT explains the
experimental data better than the SM alone~\cite{Ellis:2020unq}. 

Relatively to the SM, the impact of effective operators of dimension
$d>4$ on observables computed at energy $\sim E$ is of order
$(E/\Lambda)^{d-4}$, with 
$\Lambda\gg E$ being the (unknown) new physics threshold. Thus, the
most relevant interactions are those of lowest energy dimension, which,
ignoring lepton 
number violation (LNV), are the ones of dimension six. These operators
have been experimentally tested from very different angles at all kind
of particle physics 
facilities. In particular, the knowledge of the corresponding
renormalisation group
running~\cite{Grojean:2013kd,Elias-Miro:2013gya,Elias-Miro:2013mua,Jenkins:2013zja,Jenkins:2013wua,Alonso:2013hga,Buchalla:2019wsc} has
allowed the 
high-energy physics community to probe the SMEFT to order
$E^2/\Lambda^2$ combining experimental information gathered across
very different energies; see for example Refs.~\cite{Descotes-Genon:2018foz,Bissmann:2019gfc,Terol-Calvo:2019vck,Degrande:2020evl}. 

However, there is by now convincing evidence that dimension-six
operators do not suffice for making predictions within the SMEFT in a
number of 
situations. For example, dimension-six interactions do not provide the
dominant contribution to some observables~\cite{Azatov:2016sqh} or
even they 
do not arise at tree level in concrete ultraviolet (UV) completions of
the SM~\cite{Chala:2018ari,Murphy:2020rsh}. It can be also that
relatively low values of $\Lambda$ are favoured 
by data in some interactions, and therefore corrections involving higher powers of $E/\Lambda$
are not negligible~\cite{Chala:2018ari,Corbett:2021eux}; or simply that some
observables 
are so well measured (or constrained) that they are sensitive to
higher-dimensional
operators~\cite{Panico:2018hal,Corbett:2021eux,Ardu:2021koz}.

In either case, dimension-eight operators must be retained when using
the SMEFT. (Dimension-seven
interactions~\cite{Lehman:2014jma,Liao:2016hru} are also LNV.) This
has been in fact the approach adopted in a number of recent
theoretical 
works~\cite{Chala:2018ari,Hays:2018zze,Panico:2018hal,Ellis:2019zex,Alioli:2020kez,Ellis:2020ljj,Hays:2020scx,Gu:2020ldn,Corbett:2021eux,Ardu:2021koz}, but so far mostly at tree level. Our goal is to make a
first step forward towards the renormalisation of the SMEFT to order 
$E^4/\Lambda^4$.
We think that, beyond opening the door to using the SMEFT precisely
and consistently across energy scales, there are several motivations
to address this 
challenge. For example:

 \textbf{1.} Several classes of dimension-eight operators
 (including purely bosonic)  
that arise only at one loop in weakly-coupled UV completions of the SM
can be renormalised by dimension-eight terms that can be generated at tree level~\cite{Craig:2019wmo}. (While at dimension six this occurs solely in one case.)  This implies that the
running of some operators can provide the leading SMEFT corrections 
to SM predictions in observables in which only loop-induced
interactions contribute at tree level. 

 \textbf{2.} %
 Eight is the lowest dimension at which
 there exist two 
 co-leading contributions to renormalisation within the SMEFT: one
 involving single insertions of dimension-eight operators, and another
 one consisting of pairs of dimension-six
 interactions. (Pairs of dimension-five operators renormalise dimension-six ones~\cite{Davidson:2018zuo}, but they are LNV and therefore sub-leading with respect to single dimension-six terms.) Non-renormalisation theorems have been established only
 in relation to the first
 contribution~\cite{Craig:2019wmo,Murphy:2020rsh}. Thus, whether
 tree-level dimension-six operators renormalise loop-induced
 dimension-eight interactions is, to the best 
of our knowledge, still unknown.

 \textbf{3.} Dimension-eight operators are subject to positivity bounds~\cite{Zhang:2018shp,Bi:2019phv,Remmen:2019cyz,Remmen:2020vts,Remmen:2020uze,Bonnefoy:2020yee,Bellazzini:2020cot}. Thus, precisely because dimension-six interactions mix
into dimension-eight ones, it is \textit{a priori} conceivable that theoretical constraints on combinations of dimension-six
Wilson coefficients can be established if the corresponding renormalisation group equations (RGEs) are precisely known.

Inspired by these observations, and in particular by \textbf{2}, in this paper we will focus on renormalisation triggered by dimension-six operators. (We will consider the effects of higher-dimensional operators in loops in subsequent works.) Also, we will concentrate on the running of the bosonic sector of the
SMEFT.

This article is organised as follows. In section~\ref{sec:theory} we introduce the relevant Lagrangian and clarify the notation used thorough the rest of the paper.
In section~\ref{sec:computation} we describe the technical details of the renormalisation programme. In section~\ref{sec:rgeform} we 
unravel the global structure of the renormalisation group equations (RGEs). We finalise with a discussion of the results in section~\ref{sec:discussion}. We dedicate Appendix~\ref{sec:redundancies} to relations that hold on-shell between different operators. In Appendix~\ref{sec:fullrges} we write explicitly all RGEs, while in Appendix~\ref{sec:UVmodel} we describe briefly a UV model that accounts for generic tree-level generated dimension-six bosonic operators.

\section{Theory and conventions}
\label{sec:theory}
\begin{table}[h!]
 \begin{center}
  \resizebox{\textwidth}{!}{\begin{tabular}{cclcl}
   \toprule\\[-0.3cm]
   & \textbf{Operator} & \textbf{Notation} & \textbf{Operator} & \textbf{Notation}\\[0.5cm]
   \rotatebox[origin=c]{90}{\boldmath{$\phi^8$}} & $( \phi^\dagger\phi)^4$ & $\mathcal{O}_{\phi^8}$ & &\\[0.1cm]
   \hline\\[-0.3cm]
   \rotatebox[origin=c]{90}{\boldmath{$\phi^6 D^2$}} &  $(\phi^{\dag} \phi)^2 (D_{\mu} \phi^{\dag} D^{\mu} \phi)$ & $\mathcal{O}_{\phi^6}^{(1)}$  &  $(\phi^{\dag} \phi) (\phi^{\dag} \sigma^I \phi) (D_{\mu} \phi^{\dag} \sigma^I D^{\mu} \phi)$ &  
$\mathcal{O}_{\phi^6}^{(2)}$ \\[0.4cm]
   \hline\\[-0.3cm]
   \multirow{2}{*}{\rotatebox[origin=c]{90}{\boldmath{$\phi^4 D^4$}}}  &  $(D_{\mu} \phi^{\dag} D_{\nu} \phi) (D^{\nu} \phi^{\dag} D^{\mu} \phi)$ &  $\mathcal{O}_{\phi^4}^{(1)}$ &
 $(D_{\mu} \phi^{\dag} D_{\nu} \phi) (D^{\mu} \phi^{\dag} D^{\nu} \phi)$ & $\mathcal{O}_{\phi^4}^{(2)}$\\[0.2cm]
& $(D^{\mu} \phi^{\dag} D_{\mu} \phi) (D^{\nu} \phi^{\dag} D_{\nu} \phi)$ & $\mathcal{O}_{\phi^4}^{(3)}$ & & 
\\[0.3cm]
   \hline\\[-0.3cm]
   \multirow{2}{*}{\rotatebox[origin=c]{90}{\boldmath{\textcolor{gray}{$X^3 \phi^2$}}}}  &  \textcolor{gray}{$\epsilon^{IJK} (\phi^\dag \phi) W_{\mu}^{I\nu} W_{\nu}^{J\rho} W_{\rho}^{K\mu}$} & \textcolor{gray}{$\mathcal{O}_{W^3\phi^2}^{(1)}$} & \textcolor{gray}{$\epsilon^{IJK} (\phi^\dag \phi) W_{\mu}^{I\nu} W_{\nu}^{J\rho} \widetilde{W}_{\rho}^{K\mu}$} &  
\textcolor{gray}{$\mathcal{O}_{W^3\phi^2}^{(2)}$} \\[0.2cm]
& \textcolor{gray}{$\epsilon^{IJK} (\phi^\dag \sigma^I \phi) B_{\mu}^{\,\nu} W_{\nu}^{J\rho} W_{\rho}^{K\mu}$} & \textcolor{gray}{$\mathcal{O}_{W^2B\phi^2}^{(1)}$} & \textcolor{gray}{$\epsilon^{IJK} (\phi^\dag \sigma^I \phi) (\widetilde{B}^{\mu\nu} W_{\nu\rho}^J W_{\mu}^{K\rho} + B^{\mu\nu} W_{\nu\rho}^J \widetilde{W}_{\mu}^{K\rho})$} & \textcolor{gray}{$\mathcal{O}_{W^2B\phi^2}^{(2)}$}\\[0.4cm]
   \hline\\[-0.3cm]
   \multirow{6}{*}{\rotatebox[origin=c]{90}{\boldmath{$X^2 \phi^4$}}} & $(\phi^\dag \phi)^2 G_{\mu\nu}^A G^{A\mu\nu}$ & $O_{G^2\phi^4}^{(1)}$ & $(\phi^\dag \phi)^2 \widetilde{G}_{\mu\nu}^A G^{A\mu\nu}$ & $O_{G^2\phi^4}^{(2)}$\\[0.2cm]
   &$(\phi^\dag \phi)^2 W^I_{\mu\nu} W^{I\mu\nu}$ & $\mathcal{O}_{W^2\phi^4}^{(1)}$ &   $(\phi^\dag \phi)^2 \widetilde W^I_{\mu\nu} W^{I\mu\nu}$ & 
$\mathcal{O}_{W^2\phi^4}^{(2)}$\\[0.2cm]
& $(\phi^\dag \sigma^I \phi) (\phi^\dag \sigma^J \phi) W^I_{\mu\nu} W^{J\mu\nu}$ & $\mathcal{O}_{W^2\phi^4}^{(3)}$  &
$(\phi^\dag \sigma^I \phi) (\phi^\dag \sigma^J \phi) \widetilde W^I_{\mu\nu} W^{J\mu\nu}$ & $\mathcal{O}_{W^2\phi^4}^{(4)}$  \\[0.2cm]
& $ (\phi^\dag \phi) (\phi^\dag \sigma^I \phi) W^I_{\mu\nu} B^{\mu\nu}$ & $\mathcal{O}_{WB\phi^4}^{(1)}$ & $(\phi^\dag \phi) (\phi^\dag \sigma^I \phi) \widetilde W^I_{\mu\nu} B^{\mu\nu}$ & $\mathcal{O}_{WB\phi^4}^{(2)}$ \\[0.2cm]
& $ (\phi^\dag \phi)^2 B_{\mu\nu} B^{\mu\nu}$ & $\mathcal{O}_{B^2\phi^4}^{(1)}$ & $(\phi^\dag \phi)^2 \widetilde B_{\mu\nu} B^{\mu\nu}$ & $\mathcal{O}_{B^2\phi^4}^{(2)}$\\[0.4cm]
   \hline\\[-0.3cm]
   \multirow{12}{*}{\rotatebox[origin=c]{90}{\textcolor{gray}{\boldmath{$X^2 \phi^2 D^2$}}}} &   \textcolor{gray}{$(D^{\mu} \phi^{\dag} D^{\nu} \phi) W_{\mu\rho}^I W_{\nu}^{I \rho}$} & \textcolor{gray}{$\mathcal{O}_{W^2\phi^2D^2}^{(1)}$}  &
\textcolor{gray}{$(D^{\mu} \phi^{\dag} D_{\mu} \phi) W_{\nu\rho}^I W^{I \nu\rho}$} & \textcolor{gray}{$\mathcal{O}_{W^2\phi^2D^2}^{(2)}$} \\[0.2cm]
&  \textcolor{gray}{$(D^{\mu} \phi^{\dag} D_{\mu} \phi) W_{\nu\rho}^I \widetilde{W}^{I \nu\rho}$} & \textcolor{gray}{$\mathcal{O}_{W^2\phi^2D^2}^{(3)}$} &
\textcolor{gray}{$i \epsilon^{IJK} (D^{\mu} \phi^{\dag} \sigma^I D^{\nu} \phi) W_{\mu\rho}^J W_{\nu}^{K \rho}$} & \textcolor{gray}{$\mathcal{O}_{W^2\phi^2D^2}^{(4)}$} \\[0.2cm]
&  \textcolor{gray}{$\epsilon^{IJK} (D^{\mu} \phi^{\dag} \sigma^I D^{\nu} \phi) (W_{\mu\rho}^J \widetilde{W}_{\nu}^{K \rho} - \widetilde{W}_{\mu\rho}^J W_{\nu}^{K \rho})$} & \textcolor{gray}{$\mathcal{O}_{W^2\phi^2D^2}^{(5)}$} &  \textcolor{gray}{$i \epsilon^{IJK} (D^{\mu} \phi^{\dag} \sigma^I D^{\nu} \phi) (W_{\mu\rho}^J \widetilde{W}_{\nu}^{K \rho} + \widetilde{W}_{\mu\rho}^J W_{\nu}^{K \rho})$} & \textcolor{gray}{$\mathcal{O}_{W^2\phi^2D^2}^{(6)}$} \\[0.2cm]
&  \textcolor{gray}{$(D^{\mu} \phi^{\dag} \sigma^I D_{\mu} \phi) B_{\nu\rho} W^{I \nu\rho}$} & \textcolor{gray}{$\mathcal{O}_{WB\phi^2D^2}^{(1)}$} &  \textcolor{gray}{$(D^{\mu} \phi^{\dag} \sigma^I D_{\mu} \phi) B_{\nu\rho} \widetilde{W}^{I \nu\rho}$} & \textcolor{gray}{$\mathcal{O}_{WB\phi^2D^2}^{(2)}$} \\[0.2cm]
&  \textcolor{gray}{$i (D^{\mu} \phi^{\dag} \sigma^I D^{\nu} \phi) (B_{\mu\rho} W_{\nu}^{I \rho} - B_{\nu\rho} W_{\mu}^{I\rho})$} & \textcolor{gray}{$\mathcal{O}_{WB\phi^2D^2}^{(3)}$}  &  \textcolor{gray}{$(D^{\mu} \phi^{\dag} \sigma^I D^{\nu} \phi) (B_{\mu\rho} W_{\nu}^{I \rho} + B_{\nu\rho} W_{\mu}^{I\rho})$} & 
\textcolor{gray}{$\mathcal{O}_{WB\phi^2D^2}^{(4)}$} \\[0.2cm]
& \textcolor{gray}{$i (D^{\mu} \phi^{\dag} \sigma^I D^{\nu} \phi) (B_{\mu\rho} \widetilde{W}_\nu^{^I \rho} - B_{\nu\rho} \widetilde{W}_\mu^{^I \rho})$} & \textcolor{gray}{$\mathcal{O}_{WB\phi^2D^2}^{(5)}$}   & \textcolor{gray}{$(D^{\mu} \phi^{\dag} \sigma^I D^{\nu} \phi) (B_{\mu\rho} \widetilde{W}_\nu^{^I \rho} + B_{\nu\rho} \widetilde{W}_\mu^{^I \rho})$} & \textcolor{gray}{$\mathcal{O}_{WB\phi^2D^2}^{(6)}$} \\[0.2cm]
& \textcolor{gray}{$(D^{\mu} \phi^{\dag} D^{\nu} \phi) B_{\mu\rho} B_{\nu}^{\,\,\,\rho}$} & \textcolor{gray}{$\mathcal{O}_{B^2\phi^2D^2}^{(1)}$} &
 \textcolor{gray}{$(D^{\mu} \phi^{\dag} D_{\mu} \phi) B_{\nu\rho} B^{\nu\rho}$} & \textcolor{gray}{$\mathcal{O}_{B^2\phi^2D^2}^{(2)}$}   \\[0.2cm]
& \textcolor{gray}{$(D^{\mu} \phi^{\dag} D_{\mu} \phi) B_{\nu\rho} \widetilde{B}^{\nu\rho}$} & \textcolor{gray}{$\mathcal{O}_{B^2\phi^2D^2}^{(3)}$} \\[0.4cm] 
   \hline\\[-0.3cm]
   \multirow{4}{*}{\rotatebox[origin=c]{90}{\boldmath{$X \phi^4 D^2$}}} & $\text{i}(\phi^{\dag} \phi) (D^{\mu} \phi^{\dag} \sigma^I D^{\nu} \phi) W_{\mu\nu}^I$ & $\mathcal{O}_{W\phi^4D^2}^{(1)}$  & $\text{i}(\phi^{\dag} \phi) (D^{\mu} \phi^{\dag} \sigma^I D^{\nu} \phi) \widetilde{W}_{\mu\nu}^I$ & $\mathcal{O}_{W\phi^4D^2}^{(2)}$   \\[0.2cm]
& $\text{i}\epsilon^{IJK} (\phi^{\dag} \sigma^I \phi) (D^{\mu} \phi^{\dag} \sigma^J D^{\nu} \phi) W_{\mu\nu}^K$  &
$\mathcal{O}_{W\phi^4D^2}^{(3)}$  & $\text{i}\epsilon^{IJK} (\phi^{\dag} \sigma^I \phi) (D^{\mu} \phi^{\dag} \sigma^J D^{\nu} \phi) \widetilde{W}_{\mu\nu}^K$ &  $\mathcal{O}_{W\phi^4D^2}^{(4)}$ \\[0.2cm]
& $\text{i}(\phi^{\dag} \phi) (D^{\mu} \phi^{\dag} D^{\nu} \phi) B_{\mu\nu}$ & $\mathcal{O}_{B\phi^4D^2}^{(1)}$ & $\text{i}(\phi^{\dag} \phi) (D^{\mu} \phi^{\dag} D^{\nu} \phi) \widetilde{B}_{\mu\nu}$ & $\mathcal{O}_{B\phi^4D^2}^{(2)}$\\[0.4cm]
   \bottomrule
  \end{tabular}}
 \end{center}
 \caption{\it Basis of bosonic dimension-eight operators involving the Higgs. We follow the notation from Ref.~\cite{Murphy:2020rsh}. All interactions are hermitian. The operators in grey arise only at one loop in weakly-coupled renormalisable UV completions of the SM~\cite{Craig:2019wmo}.}\label{tab:dim8ops}
\end{table}
We denote by $e$, $u$ and $d$ the right-handed (RH) leptons and
quarks; while $l$ and $q$ refer to the left-handed (LH)
counterparts. The electroweak (EW) gauge bosons and the gluon are named by $W, B$ and by $G$, respectively. We represent the Higgs doublet by $\phi =  (\phi^+, \phi^0)^T$, and $\tilde{\phi} = \ii\sigma_2\phi^*$ with $\sigma_I$ ($I=1,2,3$) being the Pauli matrices. Thus, the renormalisable SM Lagrangian reads: 
\begin{align}\nonumber
 \mathcal{L}_\text{SM} = & -\frac{1}{4}G_{\mu\nu}^{A}G^{A\,\mu\nu} -\frac{1}{4}W_{\mu\nu}^{a}W^{a\,\mu\nu} -\frac{1}{4}B_{\mu\nu}B^{\mu\nu}\\ \nonumber
 &
+\overline{q_{L}^{\alpha}}\ii\slashed{D}q_{L}^{\alpha}
+\overline{l_{L}^{\alpha}}\ii\slashed{D}l_{L}^{\alpha}
+\overline{u_{R}^{\alpha}}\ii\slashed{D}u_{R}^{\alpha}
+\overline{d_{R}^{\alpha}}\ii\slashed{D}d_{R}^{\alpha}
+\overline{e_{R}^{\alpha}}\ii\slashed{D}e_{R}^{\alpha}
\\
& +\left(D_{\mu}\phi\right)^{\dagger}\left(D^{\mu}\phi\right)
\jg{+}\mu^{2}|\phi|^{2}-\lambda|\phi|^{4}
-\left(
y_{\alpha\beta}^{u}\overline{q_{L}^{\alpha}}\widetilde{\phi}u_{R}^{\beta}
+y_{\alpha\beta}^{d}\overline{q_{L}^{\alpha}}\phi d_{R}^{\beta}
+y_{\alpha\beta}^{e}\overline{l_{L}^{\alpha}}\phi e_{R}^{\beta}
+\text{h.c.}\right)~.
\end{align}
We adopt the minus-sign convention for the covariant derivative:
\begin{equation}
 D_\mu = \partial_\mu - \ii g_1 Y B_\mu -ig_2\frac{\sigma^I}{2} W_\mu^I -\ii g_3\frac{\lambda^A}{2} G_\mu^A\,,
\end{equation}
where $g_1, g_2$ and $g_3$ represent, respectively, the $U(1)_Y$, $SU(2)_L$ and $SU(3)_c$ gauge couplings, $Y$ stands for the hypercharge and $\lambda^A$ are the Gell-Mann matrices.

We use the Warsaw basis~\cite{Grzadkowski:2010es} for the dimension-six SMEFT Lagrangian $\mathcal{L}^{(6)}$, and the basis of Ref.~\cite{Murphy:2020rsh} for the dimension-eight part $\mathcal{L}^{(8)}$.
(An equivalent basis can be found in Ref.~\cite{Li:2020gnx}.) While the renormalisation of $\mathcal{L}^{(6)}$ has been studied at length~\cite{Jenkins:2013zja,Jenkins:2013wua,Alonso:2013hga}, the running of $\mathcal{L}^{(8)}$ is largely unknown. 
Assuming lepton-number conservation, 
the running of dimension-eight Wilson coefficients  receives
contributions from loops involving single insertions of
dimension-eight couplings as well as from pairs of dimension-six
operators. Schematically: %
\begin{equation}\label{eq:gprime}
16\pi^2\mu \frac{d c_{i}^{(8)}}{d\mu} = \gamma_{ij} c_j^{(8)} + \gamma_{ijk}' c_j^{(6)}c_k^{(6)}\,. 
\end{equation}
Although $c^{(6)}$ (and $c^{(8)}$) are in general unknown, fits of the
SMEFT to the data favour relatively large values of some of these
coefficients~\cite{Ellis:2020unq}. This implies that the $\gamma'$ term,
which is quadratic in the dimension-six couplings, can dominate the
running of dimension-eight Wilson coefficients even if the latter are
equally large. As such, the computation of this piece of the running
is especially appealing. 

Moreover, non-renormalisation
theorems~\cite{Elias-Miro:2014eia,Cheung:2015aba,Bern:2019wie,Craig:2019wmo} have
not been yet established for the mixing triggered by pairs of
dimension-six operators. Consequently, for now the zeros in $\gamma'$ can
be only obtained upon explicit calculation. 

We therefore focus on this part of the dimension-eight running in what
follows. Likewise, and as a first attack to the problem, we will
concentrate on the bosonic sector of the theory.  The advantage of
this is that bosonic operators are not renormalised by
field-redefining away redundant operators involving fermions (the
opposite is not true).  
Besides, we consider loops involving only dimension-six operators that
can arise at tree level in weakly-coupled UV completions of the
SM. These can be found in
Refs.~\cite{delAguila:2000rc,delAguila:2008pw,delAguila:2010mx,deBlas:2014mba,deBlas:2017xtg}. 
Thus, 
our starting Lagrangian is: 
\begin{align}\label{eq:luv}\nonumber
\mathcal{L}_{\text{UV}} &= \mathcal{L}_\text{SM} + \frac{1}{\Lambda^2}\bigg\lbrace c_{\phi} (\phi^\dagger\phi)^3 + c_{\phi \square} (\phi^\dagger\phi)\square(\phi^\dagger\phi) + c_{\phi D} (\phi^\dagger D^\mu\phi)^* (\phi^\dagger D_\mu\phi)\\\nonumber
&+c_{\phi \psi_L}^{(1)} (\phi^\dagger i \overleftrightarrow{D}_\mu\phi)(\overline{\psi_L}\gamma^\mu \psi_L) + c_{\phi \psi_L}^{(3)} (\phi^\dagger i \overleftrightarrow{D}^I_\mu\phi)(\overline{\psi_L}\gamma^\mu\sigma^I \psi_L) + c_{\phi \psi_R} (\phi^\dagger i \overleftrightarrow{D}_\mu\phi)(\overline{\psi_R}\gamma^\mu \psi_R)\\
& + \left[c_{\phi ud} (\widetilde{\phi}iD_\mu\phi)(\overline{u_R}\gamma^\mu d_R) + c_{\psi_R\phi} (\phi^\dagger\phi)\overline{\psi_L}\widetilde{\phi}\psi_R +\text{h.c.} \right]
\bigg\rbrace\,,
\end{align}
with $\psi_R=u_R,d_R,e_R$ and $\psi_L=q_L,l_L$.
Restricting to the bosonic sector of the SMEFT, only dimension-eight operators involving Higgses can be renormalised at one loop from the Lagrangian above. For clarity, we reproduce them in Table~\ref{tab:dim8ops} following the notation of Ref.~\cite{Murphy:2020rsh}. 

\section{Computation} 
\label{sec:computation}
\begin{figure}[t]
  \includegraphics[width=0.19\columnwidth]{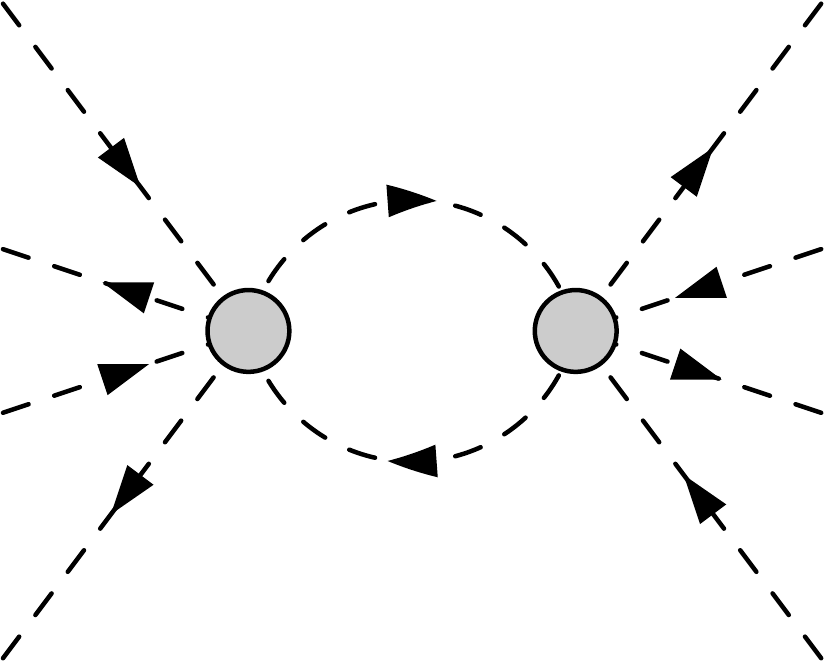}
  \includegraphics[width=0.19\columnwidth]{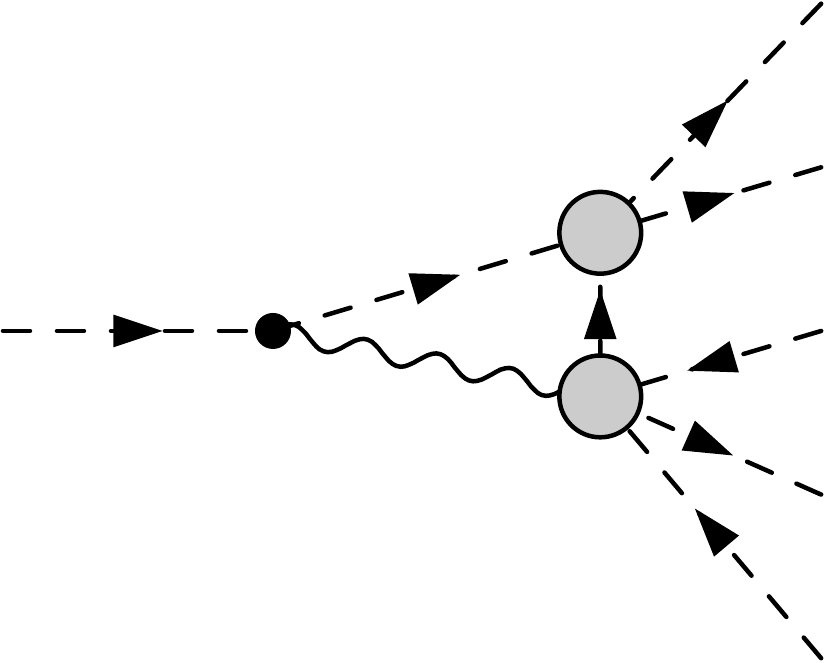}
 \includegraphics[width=0.19\columnwidth]{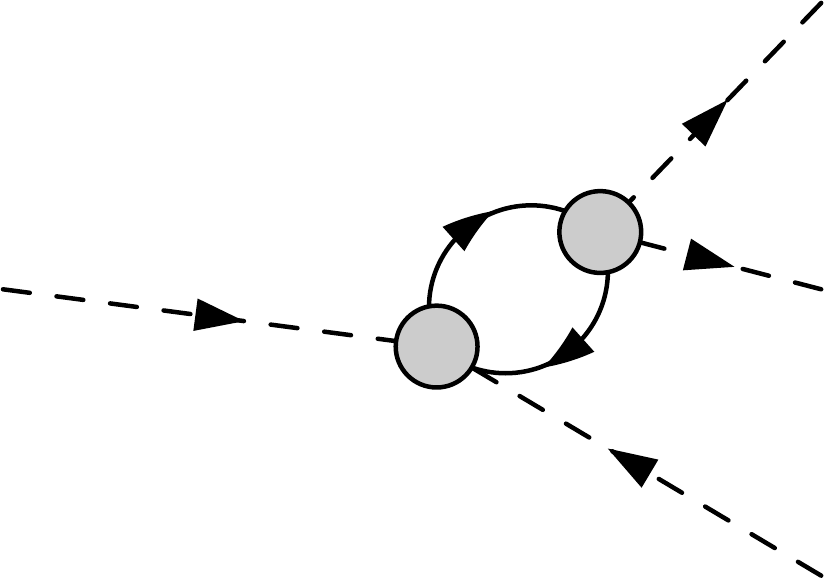}
 \includegraphics[width=0.19\columnwidth]{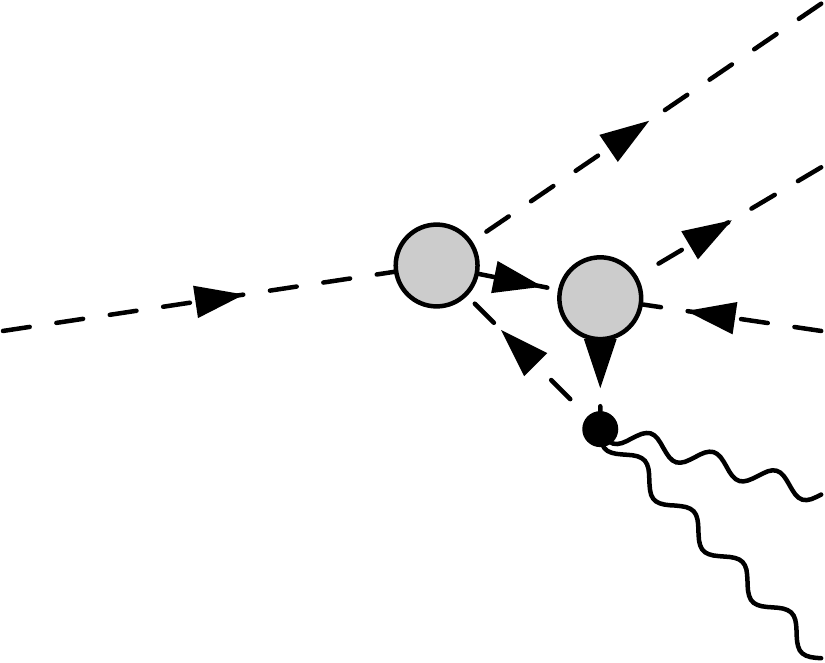}
 \includegraphics[width=0.19\columnwidth]{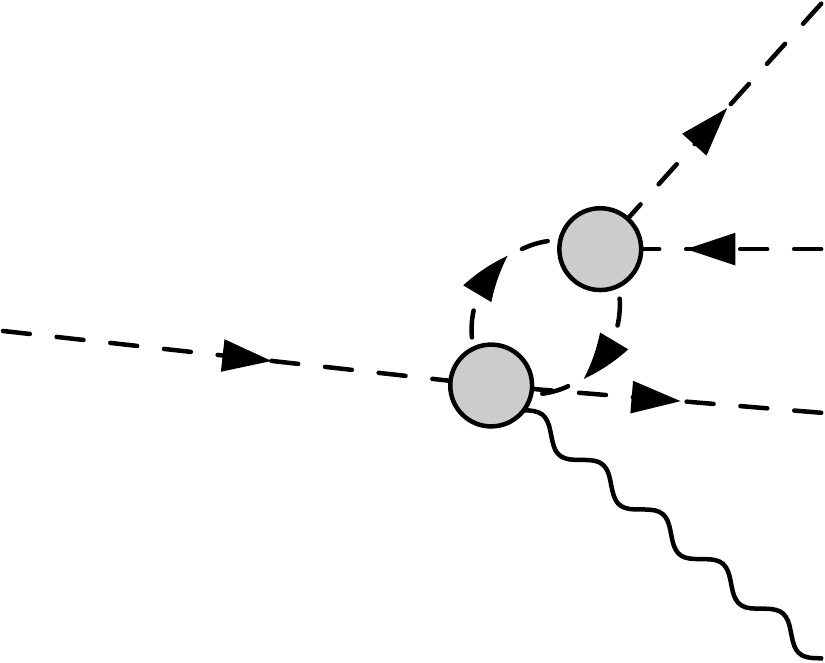}
 \caption{\it Example diagrams for the renormalisation of operators in classes $\phi^8$ (first), $\phi^6 D^2$ (second), $\phi^4 D^2$ (third), $X^2\phi^4$ (fourth) and $X\phi^4 D^2$ (fifth). The gray blobs represent dimension-six interactions.}\label{fig:Diagrams}
\end{figure}
We use the background field method and work in the Feynman gauge in
dimensional regularisation with space-time dimension $\text{d} =
4-2\epsilon$. We compute the  one-loop divergences generated by
$\mathcal{L}_{\text{UV}}$ using dedicated routines that rely on
\texttt{FeynRules}~\cite{Alloul:2013bka},
\texttt{FeynArts}~\cite{Hahn:2000kx} and
\texttt{FormCalc}~\cite{Hahn:1998yk}.
Most of the calculations have been cross-checked using
\texttt{Match-Maker}~\cite{matchmaker}. All amplitudes
of the kind $X^3\phi^2$ and $X^2\phi^2 D^2$ are  finite, hence
operators in these classes do not renormalise within our theory. 

The bosonic one-loop divergent Lagrangian, involving Higgses, can be written as: 
\begin{align}\label{eq:divergences}
16\pi^2\epsilon\,\mathcal{L}_\text{DIV} &= \tilde{K}_\phi (D_\mu\phi)^\dagger (D^\mu\phi) \jg{+} \tilde{\mu}^2|\phi|^2-\tilde{\lambda}|\phi|^4+
\tilde{c}_i^{(6)}\frac{\mathcal{O}_i^{(6)}}{\Lambda^2} +
\tilde{c}_j^{(8)}\frac{\mathcal{O}_j^{(8)}}{\Lambda^4}\,, 
\end{align}
where $i$ and $j$ run over  elements in the Green bases of operators of dimension-six and dimension-eight, respectively.
The former extends the Warsaw basis with the 
interactions given in Table~\ref{tab:red-six}.
The bosonic Higgs operators expanding the dimension-eight Green basis and which are redundant in the basis of Table~\ref{tab:dim8ops} are shown in Table~\ref{tab:red-eight}. To the best of our knowledge, this last result is completely new. (\jg{We do not include other bosonic redundant operators not involving the Higgs as those are not renormalised by the interactions in Eq. \ref{eq:luv}.})
%
%
\begin{table}[t]
 \begin{center}
  \resizebox{0.8\textwidth}{!}{\begin{tabular}{cclcl}
   \toprule\\[-0.3cm]
   & \textbf{Operator} & \textbf{Notation} & \textbf{Operator} & \textbf{Notation}\\[0.5cm]
   \boldmath{$\phi^2 D^4$} & $(D_\mu D^\mu\phi^\dagger)(D_\nu D^\nu\phi)$  & $\mathcal{O}_{D\phi}$ & &\\[0.2cm] 
   \hline\\[-0.3cm]
   \boldmath{$\phi^4 D^2$} & $(\phi^\dagger\phi) (D_\mu\phi)^\dagger (D^\mu\phi)$ & $\mathcal{O}_{\phi D}^\prime$ & $(\phi^\dagger\phi)D^\mu(\phi^\dagger\ii\overleftrightarrow{D}_\mu\phi)$ & $\mathcal{O}_{\phi D}^{\prime\prime}$ \\[0.2cm] 
   \hline\\[-0.3cm]
   \boldmath{$X\phi^2 D^2$} & $D_\nu W^{I\mu\nu} (\phi^\dagger\ii\overleftrightarrow{D}_\mu^I\phi)$ & $\mathcal{O}_{WD\phi}$ & $\partial_\nu B^{\mu\nu} (\phi^\dagger\ii\overleftrightarrow{D}_\mu\phi)$ & $\mathcal{O}_{BD\phi}$\\[0.2cm]
   \bottomrule
  \end{tabular}}
  \caption{\it Independent dimension-six bosonic operators involving the Higgs which are redundant with respect to the Warsaw basis. We adopt the notation of Ref.~\cite{Gherardi:2020det}.}\label{tab:red-six}
 \end{center}
\end{table}
%
\begin{table}[t]
 \begin{center}
  \resizebox{\textwidth}{!}{\begin{tabular}{cclcl}
   \toprule\\[-0.3cm]
   & \textbf{Operator} & \textbf{Notation} & \textbf{Operator} & \textbf{Notation}\\[0.5cm]
   \rotatebox[origin=c]{90}{\boldmath{$\phi^2 D^6$}} & $D^2\phi^\dagger D_\mu D_\nu D^\mu D^\nu\phi$ & $\mathcal{O}_{\phi^2}$ & \\[0.4cm]
   \hline\\[-0.3cm]
   \multirow{6}{*}{\rotatebox[origin=c]{90}{\boldmath{$\phi^4 D^4$}}} & $D_\mu\phi^\dagger D^\mu\phi(\phi^\dagger D^2\phi + \text{h.c.})$ & $\mathcal{O}_{\phi^4}^{(4)}$ & $D_\mu\phi^\dagger D^\mu\phi(\phi^\dagger \ii D^2\phi + \text{h.c.})$ & $\mathcal{O}_{\phi^4}^{(5)}$ \\
   & $(D_\mu\phi^\dagger \phi) (D^2\phi^\dagger D_\mu\phi) + \text{h.c.}$ & $\mathcal{O}_{\phi^4}^{(6)}$  & $(D_\mu\phi^\dagger \phi) (D^2\phi^\dagger \ii D_\mu\phi) + \text{h.c.}$ & $\mathcal{O}_{\phi^4}^{(7)}$ \\
& $(D^2\phi^\dagger\phi) (D^2\phi^\dagger\phi)+\text{h.c.}$ & $\mathcal{O}_{\phi^4}^{(8)}$ & $(D^2\phi^\dagger\phi) (\ii D^2\phi^\dagger\phi)+\text{h.c.}$ & $\mathcal{O}_{\phi^4}^{(9)}$ \\
& $(D^2\phi^\dagger D^2\phi) (\phi^\dagger\phi)$ & $\mathcal{O}_{\phi^4}^{(10)}$ & $(\phi^\dagger D^2\phi) (D^2\phi^\dagger\phi)$ & $\mathcal{O}_{\phi^4}^{(11)}$ \\
& $(D_\mu\phi^\dagger \phi)(D^\mu\phi^\dagger D^2\phi) + \text{h.c.}$ & $\mathcal{O}_{\phi^4}^{(12)}$ & $(D_\mu\phi^\dagger \phi)(D^\mu\phi^\dagger \ii D^2\phi) + \text{h.c.}$ & $\mathcal{O}_{\phi^4}^{(13)}$  \\[0.2cm]
   \hline\\[-0.3cm]
   \rotatebox[origin=c]{90}{\boldmath{$\phi^6 D^2$}}  & $(\phi^\dagger\phi)^2 (\phi^\dagger D^2\phi + \text{h.c.})$ & $\mathcal{O}_{\phi^6}^{(3)}$ & $(\phi^\dagger\phi)^2 D_\mu(\phi^\dagger\ii \overleftrightarrow{D}^\mu\phi)$ & $\mathcal{O}_{\phi^6}^{(4)}$ \\[0.4cm]
   \hline\\[-0.3cm]
   \multirow{2}{*}{\rotatebox[origin=c]{90}{\boldmath{$X \phi^4 D^2$}} } & $(\phi^\dag \phi) D_\nu W^{I \mu\nu}(D_\mu\phi^\dagger \sigma^I \phi + \text{h.c.}) $ & $\mathcal{O}_{W\phi^4 D^2}^{(5)}$  &
 $(\phi^\dag \phi) D_\nu W^{I \mu\nu}(D_\mu\phi^\dagger \text{i} \sigma^I \phi + \text{h.c.})$ & $\mathcal{O}_{W\phi^4 D^2}^{(6)}$  \\[0.2cm]
& $\epsilon^{IJK} (D_\mu \phi^\dag \sigma^I \phi) (\phi^\dag \sigma^J D_\nu \phi) W^{K \mu\nu} $ & $\mathcal{O}_{W\phi^4 D^2}^{(7)}$  & $ (\phi^{\dag} \phi) D_{\nu} B^{\mu\nu} (D_\mu \phi^\dagger \text{i} \phi + \text{h.c.})$ & $\mathcal{O}_{B\phi^4D^2}^{(3)}$\\[0.3cm]
   \bottomrule
  \end{tabular}}
  \caption{\it Independent dimension-eight bosonic operators involving the Higgs which are redundant with respect to the basis of Ref.~\cite{Murphy:2020rsh}. Redundant operators in the class $X^2\phi^2 D^2$ are not shown. The addition of h.c. when needed implies that all operators are hermitian.}\label{tab:red-eight}
 \end{center}
\end{table}

We are interested in the unknown $\mathcal{O}(E^4/\Lambda^4)$ piece of the renormalisation of bosonic operators. As such, we only provide this new contribution to the aforementioned divergences. The only exception are the Higgs kinetic term and the dimension-six redundant operators, for which we also compute $E^2/\Lambda^2$ corrections, as these generate $E^4/\Lambda^4$ terms when moving to the physical basis by means of field redefinitions.
Note also that, since we are dealing with only bosonic operators, we 
omit flavour indices. 
Flavourful couplings
are written in matrix
form (keeping the correct order in the matrix multiplication) and a
trace over
indices
is implicit. We also use the shorthand notation for
matrices $A^2 \equiv \mathrm{Tr}\,A^\dagger A$, where $A$ is an
arbitrary flavour matrix.

Thus, the divergences of the couplings of dimension $d\leq 4$ read:
\begin{align}
 \tilde{K}_\phi &\supset \frac{1}{2} (c_{\phi D}+2c_{\phi\Box})\frac{\mu^2}{\Lambda^2}\,,\\
 %
 \tilde{\lambda} &\supset -\frac{3}{2}(c_{\phi D}^2-4c_{\phi D}c_{\phi\Box}+8c_{\phi\Box}^2)\frac{\mu^4}{\Lambda^4}\,.
\end{align}
We use the symbol $\supset$ to make explicit that corrections irrelevant for the computation of the $E^4/\Lambda^4$ terms are disregarded.

The $\tilde{c}^{(6)}$ couplings in Eq.~(\ref{eq:divergences}) read:
\begin{align}
\tilde{c}_{\phi D} &\supset \jg{-}c_{\phi D} (5 c_{\phi D}-8c_{\phi\Box})\frac{\mu^2}{\Lambda^2}\,,\\
\tilde{c}_{\phi\Box} &\supset \jg{-} \frac{1}{4}(c_{\phi D}^2+24 c_{\phi D} c_{\phi\Box}-48 c_{\phi\Box}^2)\frac{\mu^2}{\Lambda^2}\,,\\
\tilde{c}_{\phi} &\supset \jg{-}6 c_\phi (3 c_{\phi D}-10 c_{\phi\Box})\frac{\mu^2}{\Lambda^2}\jg{-}12\lambda (c_{\phi D}^2-6 c_{\phi D} c_{\phi\Box}+12 c_{\phi\Box}^2)\frac{\mu^2}{\Lambda^2}\,,\\
\tilde{c}_{\phi D}^\prime &= \frac{1}{2}\bigg[(3g_2^2-3g_1^2-4\lambda)c_{\phi D} + (8\lambda-6g_2^2)c_{\phi\square} \bigg] \nonumber\\
&  - \bigg [3(  c_{u\phi} y^{u\dagger}+y^u c_{u\phi}^\dagger + y^d c_{d\phi}^\dagger+c_{d\phi} y^{d\dagger}) + y^e c_{e\phi}^\dagger+c_{e\phi} y^{e\dagger}
\bigg]\nonumber\\
& + 12 c_{\phi q}^{(3)} (y^u y^{u\dagger} + y^d y^{d\dagger}) - 6 (c_{\phi u d } y^{d\dagger} y^u +c_{\phi u d }^\dagger y^{u\dagger} y^d)  + 4 c_{\phi l}^{(3)} y^e y^{e\dagger}\\
&\jg{+}(c_{\phi D}^2+4 c_{\phi D} c_{\phi\Box}-8 c_{\phi\Box}^2) \frac{\mu^2}{\Lambda^2}
\,,\\ 
\tilde{c}_{\phi D}^{\prime\prime} &= -\frac{i}{2} \bigg [3(  c_{u\phi} y^{u\dagger}- y^u c_{u\phi}^\dagger + y^d c_{d\phi}^\dagger-c_{d\phi} y^{d\dagger}) + (y^e c_{e\phi}^\dagger-c_{e\phi} y^{e\dagger}) 
\bigg]
\,,\\
\tilde{c}_{BD\phi} &= -\frac{g_1}{6} \bigg[c_{\phi D} + c_{\phi\square}-4c_{\phi e}-4c_{\phi l}^{(1)}+8c_{\phi u}-4c_{\phi d}+4 c_{\phi q}^{(1)}\bigg]\,,\\
\tilde{c}_{WD\phi} &= -\frac{g_2}{6} \bigg[c_{\phi\square}+4c_{\phi l}^{(3)}+12c_{\phi q}^{(3)}\bigg]\,;
\end{align}
while for the $\tilde{c}^{(8)}$ we get:
\begin{align}
\tilde{c}_{\phi^8} &= \frac{3}{16}\bigg[336 c_\phi^2 + (g_1^4+g_2^4+2 g_1^2 g_2^2 + 160\lambda^2)c_{\phi D}^2 + 2304\lambda^2 c_{\phi\square}^2\nonumber\\
  &+ 512\lambda  \mr{c_{\phi D}}c_\phi -1920\lambda \mr{c_{\phi \square}}c_\phi -1152\lambda^2 c_{\phi\square}\mr{c_{\phi D}} \bigg]
\nonumber\\
&- \bigg[c_{e\phi} y^{e\dagger} c_{e\phi} y^{e\dagger}
    +c_{e\phi}^\dagger y^e c_{e\phi}^\dagger y^e
    +2 c_{e\phi} c_{e\phi}^\dagger y^e y^{e\dagger}
    +2 c_{e\phi}^\dagger c_{e\phi} y^{e\dagger} y^e\bigg]
\nonumber\\
&- 3\bigg[c_{u\phi} y^{u\dagger} c_{u\phi} y^{u\dagger}
    +c_{u\phi}^\dagger y^u c_{u\phi}^\dagger y^{u}
    +2 c_{u\phi} c_{u\phi}^\dagger y^u y^{u\dagger}
    +2 c_{u\phi}^\dagger c_{u\phi} y^{u\dagger} y^u\bigg]
\nonumber\\
&- 3\bigg[c_{d\phi} y^{d\dagger} c_{d\phi} y^{d\dagger}
    +c_{d\phi}^\dagger y^d c_{d\phi}^\dagger y^d
    +2 c_{d\phi} c_{d\phi}^\dagger y^d y^{d\dagger}
    +2 c_{d\phi}^\dagger c_{d\phi} y^{d\dagger} y^d\bigg]
\,,\\
\tilde{c}_{\phi^4}^{(1)} &=  \dfrac{1}{6} \bigg[ 11 c_{\phi D}^2 - 32 c_{\phi D}	c_{\phi \Box} +16 c_{\phi \Box}^2\nonumber+ 24 c_{\phi ud}^2 - 24 c_{\phi d}^2 - 8 c_{\phi e}^2 - 16 (c_{\phi l}^{(1)})^2\\
  &+ 16 (c_{\phi l}^{(3)})^2 - 48 (c_{\phi q}^{(1)})^2 + 
  48 (c_{\phi q}^{(3)})^2 - 24 c_{\phi u}^2\bigg]\,,\label{eq:cphi41}\\
\tilde{c}_{\phi^4}^{(2)} &= \dfrac{1}{6} \bigg[ 5 c_{\phi D}^2 + 16 c_{\phi D}	c_{\phi \Box} + 16 c_{\phi \Box}^2 + 24 c_{\phi d}^2 + 8 c_{\phi e}^2 + 16 (c_{\phi l}^{(1)})^2 + 
 16 (c_{\phi l}^{(3)})^2 + 48 (c_{\phi q}^{(1)})^2\nonumber\\
 &+ 48 (c_{\phi q}^{(3)})^2 + 24 c_{\phi u}^2\bigg]  \,,\label{eq:cphi42}\\
\tilde{c}_{\phi^4}^{(3)} &= \dfrac{1}{6} \left[ -7 c_{\phi D}^2 + 16 c_{\phi D} c_{\phi \Box}  + 40  c_{\phi \Box}^2 -32 (c_{\phi l}^{(3)})^2 - 96 (c_{\phi q}^{(3)})^2 - 24 c_{\phi ud}^2\right]  \,,\label{eq:cphi43}\\
\tilde{c}_{\phi^4}^{(4)} &= \dfrac{1}{2} \left[ - c_{\phi D}^2  -2  c_{\phi D} c_{\phi \Box}  +24  c_{\phi \Box}^2	\right] \,,\\
\tilde{c}_{\phi^4}^{(6)} &= c_{\phi D} \left(c_{\phi D} - c_{\phi \Box}\right) \,,\\
\tilde{c}_{\phi^4}^{(8)} &= \dfrac{1}{8} \left[ c_{\phi D}^2 -8  c_{\phi D} c_{\phi \Box}  + 32  c_{\phi \Box}^2	\right] \,,\\
\tilde{c}_{\phi^4}^{(10)} &= \dfrac{1}{2} c_{\phi D}^2 \,,\\
\tilde{c}_{\phi^4}^{(11)} &= \dfrac{1}{4} \left[c_{\phi D}^2 - 8 c_{\phi D} c_{\phi \Box}  +32   c_{\phi \Box}^2	\right] \,,\\
\tilde{c}_{\phi^4}^{(12)} &= \dfrac{1}{2} c_{\phi D} \left(c_{\phi D} +2  c_{\phi \Box}\right) \,,\\
\tilde{c}_{\phi^6}^{(1)} & = \frac{1}{4} \bigg[ 24 c_\phi (c_{\phi D} + 8 c_{\phi \Box}) - 2 c_{\phi D} c_{\phi \Box} (9 g_1^2 + 3 g_2^2 - 32 \lambda)\\
&+ c_{\phi D}^2 (-3 g_1^2 -9 g_2^2 + 34 \lambda) - 4 c_{\phi \Box}^2 (9 g_1^2 + 15 g_2^2 + 112 \lambda)\bigg] \nonumber \\
&  -3 (c_{d\phi}^2 + c_{u\phi}^2) - c_{e\phi}^2 - (3 c_{\phi l}^{(1)} + 5 c_{\phi l}^{(3)}) \bigg[y^e c_{e\phi}^\dagger + c_{e\phi} y^{e\dagger}\bigg] \nonumber  \\
& + 9 c_{\phi q }^{(1)} \bigg[ y^u c_{u\phi}^\dagger	+ c_{u \phi} y^{u\dagger} - y^d c_{d\phi}^\dagger	-  c_{d\phi} y^{d \dagger}\bigg] - 15 c_{\phi q }^{(3)} \bigg[ y^u c_{u\phi}^\dagger	+ c_{u \phi} y^{u\dagger}+ y^d c_{d\phi}^\dagger+ c_{d\phi} y^{d \dagger}\bigg]\nonumber \\
& - 3\bigg[ 3 (c_{\phi u } c_{u\phi}^\dagger y^u  	+ c_{u \phi} c_{\phi u} y^{u\dagger}  - c_{d\phi}c_{\phi d} y^{d\dagger}	- c_{\phi d} c_{d\phi}^\dagger y^d)   - c_{e\phi}c_{\phi e} y^{e\dagger}	- c_{\phi e} c_{e\phi}^\dagger y^e\bigg] \nonumber \\
& - 9 \bigg[ 2 ( c_{\phi q}^{(1)} - c_{\phi q}^{(3)} ) y^u c_{\phi u } y^{u\dagger} + 2 ( c_{\phi q}^{(1)} +  c_{\phi q}^{(3)} ) y^d c_{\phi d } y^{d\dagger} \bigg]+ 9 (c_{\phi q}^{(1)} c_{\phi q}^{(3)} + c_{\phi q}^{(3)} c_{\phi q}^{(1)}) \bigg[ - y^u y^{u\dagger} + y^d y^{d\dagger}\bigg]  \nonumber \\
& + 15c_{\phi q}^{(3)} \bigg[ y^u y^{u\dagger} + y^d y^{d\dagger}\bigg] c_{\phi q}^{(3)} + 9 c_{\phi q }^{(1)}  \bigg[ y^u y^{u\dagger} + y^d y^{d\dagger}\bigg] c_{\phi q}^{(1)} \nonumber \\
& -3\bigg[2 ( c_{\phi l}^{(1)} +  c_{\phi l}^{(3)} ) y^e c_{\phi e } y^{e\dagger} -  (c_{\phi l}^{(1)} c_{\phi l}^{(3)} +c_{\phi l}^{(3)} c_{\phi l}^{(1)}) y^e y^{e\dagger} \bigg] + 5 c_{\phi l}^{(3)} y^e y^{e\dagger} c_{\phi l}^{(3)} + 3 c_{\phi l }^{(1)} y^e y^{e\dagger} c_{\phi l}^{(1)} \nonumber \\
& +\frac{3}{2}\bigg[ c_{\phi u d}c_{\phi u d}^\dagger y^{u\dagger} y^u + c_{\phi u d}^\dagger c_{\phi u d} y^{d\dagger} y^d \bigg]  + 3 \bigg[3( c_{\phi u} y^{u\dagger} y^u c_{\phi u} + c_{\phi d} y^{d\dagger}y^d c_{\phi d}) + c_{\phi e} y^{e\dagger} y^e c_{\phi e} \bigg] \nonumber \\
& - 6 c_{\phi q}^{(3)} \bigg[ y^u c_{\phi u d} y^{d\dagger} + y^d c_{\phi u d}^\dagger y^{u\dagger} \bigg] + 3\bigg[ c_{\phi ud} c_{d\phi}^\dagger y^u  + c_{d\phi} c_{\phi u d}^\dagger y^{u\dagger} + c_{u \phi} c_{\phi u d} y^{d\dagger} + y^d c_{\phi ud}^\dagger c_{u\phi}^\dagger		\bigg] 
\,, \nonumber \\
\tilde{c}_{\phi^6}^{(2)} & = \frac{1}{8} \bigg[ 72 c_\phi c_{\phi D} - 24 c_{\phi\Box} (c_{\phi D} + 2 c_{\phi \Box}) g_1^2 - 3 c_{\phi D} (c_{\phi D} + 16 c_{\phi \Box}) g_2^2\\
& + 32 c_{\phi D} (3 c_{\phi D} - 8 c_{\phi \Box})  \lambda \bigg] - 3 \bigg[ c_{\phi u d } c_{d \phi}^\dagger y^u + c_{d \phi} c_{\phi u d}^{\dagger}y^{u\dagger}  + y^d c_{\phi u d}^\dagger c_{u\phi}^\dagger + c_{u\phi} c_{\phi u d} y^{d\dagger} \bigg] \nonumber \\
& - 2 \bigg[c_{\phi l}^{(1)} y^e c_{e\phi}^\dagger+ c_{e\phi} y^{e\dagger } c_{\phi l}^{(1)} -3 (c_{\phi q }^{(1)} y^u c_{u\phi}^\dagger + c_{u \phi} y^{u\dagger} c_{\phi q}^{(1)} - c_{\phi q }^{(1)} y^d c_{d\phi}^\dagger-c_{d\phi} y^{d\dagger } c_{\phi q}^{(1)})  \bigg] \nonumber \\
& + 2\bigg[ c_{\phi e} c_{e\phi}^\dagger y^e + c_{e\phi} c_{\phi e} y^{e\dagger}- 3 ( c_{\phi u} c_{u\phi}^\dagger y^u  + c_{u\phi} c_{\phi u} y^{u\dagger} - c_{\phi d } c_{d \phi}^\dagger y^d - c_{d\phi} c_{\phi d} y^{d\dagger})\bigg]\nonumber  \\
&- 4 \bigg[ c_{\phi l}^{(1)} + c_{\phi l}^{(3)}\bigg] y^e c_{\phi e} y^{e\dagger}  \nonumber - 12  \bigg[(c_{\phi q}^{(1)} - c_{\phi q}^{(3)}) y^u c_{\phi u} y^{u\dagger} + (c_{\phi q}^{(1)} + c_{\phi q}^{(3)}) y^d c_{\phi d} y^{d\dagger}\bigg]\\
& + 2\bigg[ c_{\phi l}^{(1)} c_{\phi l}^{(3)} +  c_{\phi l}^{(3)} c_{\phi l}^{(1)} + c_{\phi l}^{(1)}c_{\phi l}^{(1)} \bigg] y^e y^{e\dagger}+ 2  c_{\phi e } c_{\phi e }y^{e\dagger}y^e - 6\bigg[ c_{\phi q}^{(1)} c_{\phi q}^{(3)} + c_{\phi q}^{(3)} c_{\phi q}^{(1)}  - c_{\phi q}^{(1)}c_{\phi q}^{(1)}\bigg] y^u y^{u\dagger} \nonumber \\
&+6 c_{\phi u}c_{\phi u} y^{u\dagger}y^u
+ 6\bigg[ c_{\phi q}^{(1)} c_{\phi q}^{(3)} + c_{\phi q}^{(3)} c_{\phi q}^{(1)}  + c_{\phi q}^{(1)}c_{\phi q}^{(1)}\bigg] y^d y^{d\dagger} + 6c_{\phi d}c_{\phi d} y^{d\dagger}y^d \nonumber\\  
&- \frac{3}{2} \bigg[c_{\phi u d}c_{\phi u d}^\dagger y^{u\dagger}y^u  + c_{\phi u d}^\dagger c_{\phi u d} y^{d\dagger} y^d\bigg]
 + 6 c_{\phi q}^{(3)} \bigg[ y^d c_{\phi u d }^\dagger y^{u\dagger} + y^u c_{\phi u d} y^{d \dagger}\bigg] \nonumber
\,,\\
\tilde{c}_{\phi^6}^{(3)} &= \frac{1}{16}\bigg[432 c_{\phi} c_{\phi\square} + (448\lambda - 12g_1^2-12g_2^2) c_{\phi D} c_{\phi\square} -48c_{\phi}c_{\phi D} \\
&- 8(3 g_1^2+3 g_2^2 + 128 \lambda) c_{\phi\square}^2 - (3g_2^2+12\lambda)c_{\phi D}^2 \bigg] \nonumber \\
& - c_{e\phi}^2 - 3(c_{d\phi}^2 + c_{u\phi}^2) -\frac{1}{2} \bigg[ (c_{\phi l}^{(1)} + c_{\phi l}^{(3)})(y^e c_{e\phi}^\dagger + c_{e\phi} y^{e\dagger} ) - c_{\phi e} c_{e\phi}^\dagger y^e - c_{e\phi} c_{\phi e} y^{e\dagger}\bigg] \nonumber\\
& - \bigg[c_{\phi l}^{(1)} + c_{\phi l}^{(3)}\bigg]y^e c_{\phi e} y^{e\dagger}  + \frac{1}{2} \bigg[ c_{\phi l}^{(1)} c_{\phi l}^{(3)} 
  + c_{\phi l}^{(1)}c_{\phi l}^{(1)}+c_{\phi l}^{(3)} c_{\phi l}^{(1)} +c_{\phi l}^{(3)}c_{\phi l}^{(3)} \bigg]  y^e y^{e\dagger} + \frac{1}{2} c_{\phi e}c_{\phi e} y^{e\dagger}y^e\nonumber \\
& -\frac{3}{2} \left[ c_{\phi q}^{(1)} +c_{\phi q}^{(3)}  \right] (y^d c_{d\phi}^\dagger + c_{d\phi} y^{d\dagger})+ \frac{3}{2} \left[c_{\phi d} c_{d \phi}^\dagger y^d +c_{d\phi} c_{\phi d} y^{d\dagger}  \right] - 3 \left[ c_{\phi q}^{(1)} + c_{\phi q}^{(3)}\right] y^d c_{\phi d} y^{d\dagger} \nonumber \\
&+\frac{3}{2}\bigg[c_{\phi q}^{(1)} c_{\phi q}^{(3)}  + 
c_{\phi q}^{(1)}c_{\phi q}^{(1)} + c_{\phi q}^{(3)} c_{\phi q}^{(1)}+c_{\phi q}^{(3)}c_{\phi q}^{(3)} \bigg] y^dy^{d\dagger}+ \frac{3}{2} c_{\phi d}c_{\phi d} y^{d\dagger}y^d \nonumber \\
& - \frac{3}{2} \bigg[(-c_{\phi q}^{(1)} +c_{\phi q}^{(3)}  )(y^u c_{u\phi}^\dagger + c_{u\phi} y^{u\dagger} ) +c_{\phi u} c_{u \phi}^\dagger y^u +c_{u\phi} c_{\phi u} y^{u\dagger} \bigg] -3\left[ c_{\phi q}^{(1)} - c_{\phi q}^{(3)}\right] y^u c_{\phi u} y^{u\dagger}\nonumber \\
& + \frac{3}{2}\bigg[- c_{\phi q}^{(1)} c_{\phi q}^{(3)}  + c_{\phi q}^{(1)}c_{\phi q}^{(1)} - c_{\phi q}^{(3)} c_{\phi q}^{(1)} + c_{\phi q}^{(3)}c_{\phi q}^{(3)}\bigg] y^u y^{u\dagger}+\frac{3}{2} c_{\phi u}c_{\phi u} y^{u\dagger} y^u \nonumber \,,\\
\tilde{c}_{\phi^6}^{(4)} &= -\frac{3\text{i}}{2}\bigg[ c_{d \phi}^\dagger \left( c_{\phi q}^{(1)} + c_{\phi q}^{(3)}\right) y^d - c_{\phi d}c_{d\phi}^\dagger y^d +  c_{d\phi} c_{\phi d} y^{d\dagger} -  \left( c_{\phi q}^{(1)} + c_{\phi q}^{(3)}\right)c_{d\phi} y^{d\dagger}\, \\
&+ c_{u \phi}^\dagger \left( c_{\phi q}^{(1)} - c_{\phi q}^{(3)}\right) y^u - c_{\phi u}c_{u\phi}^\dagger y^u +  c_{u\phi} c_{\phi u} y^{u\dagger} -  \left( c_{\phi q}^{(1)} - c_{\phi q}^{(3)}\right)c_{u\phi} y^{u\dagger} \bigg] \; \nonumber \\
&-\frac{\text{i}}{2} \bigg[ c_{e \phi}^\dagger \left( c_{\phi l}^{(1)} + c_{\phi l}^{(3)}\right) y^e - c_{\phi e}c_{e\phi}^\dagger y^e +  c_{e\phi} c_{\phi e} y^{e\dagger} -  \left( c_{\phi l}^{(1)} + c_{\phi l}^{(3)}\right)c_{e\phi} y^{e\dagger} \bigg]\,,\nonumber\\
\tilde{c}_{W^2 \phi^4}^{(1)} &= \dfrac{g_2^2}{12}  \left[ c_{\phi D}^2 - 3  c_{\phi D} c_{\phi \Box} -12 (c_{\phi l}^{(3)})^2 - 36 (c_{\phi q}^{(3)})^2 + 9 c_{\phi ud}^2\right]	\,,\\
\tilde{c}_{W^2 \phi^4}^{(3)} &= \dfrac{g_2^2}{48} \left[ c_{\phi D}^2- 12 c_{\phi D} c_{\phi \Box} -24 c_{\phi d}^2 -8 c_{\phi e}^2 - 16 (c_{\phi l}^{(1)})^2 - 48 (c_{\phi q}^{(1)})^2 - 24 c_{\phi u}^2 + 12 c_{\phi ud}^2 \right]\,,\\
\tilde{c}_{W B \phi^4}^{(1)} &= \dfrac{g_1 g_2}{24} \left[ c_{\phi D}^2 - 12 c_{\phi D} c_{\phi \Box} -8 c_{\phi e}^2 - 16 (c_{\phi l}^{(1)})^2 - 48 (c_{\phi q}^{(1)})^2 - 24 c_{\phi u}^2 + 12 c_{\phi ud} - 24 c_{\phi d }^2 \right]    \,,\\
\tilde{c}_{B^2 \phi^4}^{(1)} &= -\dfrac{g_1^2}{48} \bigg[ 3c_{\phi D}^2 +24 c_{\phi d}^2 + 8 c_{\phi e}^2 + 16 (c_{\phi l}^{(1)})^2 - 48 (c_{\phi l}^{(3)})^2 + 48 (c_{\phi q}^{(1)})^2\nonumber\\
&- 144 (c_{\phi q}^{(3)})^2 + 24 c_{\phi u}^2 + 24 c_{\phi u d}^2	\bigg] \,,\\
\tilde{c}_{W \phi^4 D^2}^{(1)} &= \dfrac{g_2}{6}  \bigg[ 5 c_{\phi D}^2 - 24  c_{\phi D}c_{\phi \Box} -8 c_{\phi e}^2 -16 (c_{\phi l}^{(1)})^2 - 48 (c_{\phi l}^{(3)})^2 - 48 (c_{\phi q}^{(1)})^2\nonumber\\
&- 144 (c_{\phi q}^{(3)})^2 - 24 c_{\phi u}^2 + 48 c_{\phi ud}^2 - 24 c_{\phi d}^2		\bigg]\,,\\
\tilde{c}_{B\phi^4 D^2}^{(1)}&=-\dfrac{ g_1}{6} \bigg[24 c_{\phi d}^2+3 c_{\phi D}^2+8 (c_{\phi e}^2+2 (c_{\phi l}^{(1)})^2-6 (c_{\phi l}^{(3)})^2+6 (c_{\phi q}^{(1)})^2-18 (c_{\phi q}^{(3)})^2+3 c_{\phi u}^2)+24 c_{\phi ud}^2\bigg]\,,\\
\tilde{c}_{W\phi^4 D^2}^{(6)}&=\dfrac{g_2}{24} \bigg[5 c_{\phi D}^2-20 c_{\phi D} c_{\phi \Box}+16 (c_{\phi \Box}^2-2 (c_{\phi l}^{(3)})^2-6 (c_{\phi q}^{(3)})^2)+48 c_{\phi ud}^2\bigg]\,,\\
\tilde{c}_{W\phi^4 D^2}^{(7)}&=\dfrac{ g_2}{6} \bigg[24 c_{\phi d}^2-c_{\phi D} (c_{\phi D}-12 c_{\phi \Box})+8 (c_{\phi e}^2+2 (c_{\phi l}^{(1)})^2+6 (c_{\phi q}^{(1)})^2+3 c_{\phi u}^2)-12 c_{\phi ud}^2\bigg]\,,\\
\tilde{c}_{B\phi^4 D^2}^{(3)}&=\dfrac{g_1}{24} \bigg[-3 c_{\phi D}^2+4 c_{\phi D} c_{\phi \Box}+16 (c_{\phi \Box}^2+4 (c_{\phi l}^{(3)})^2+12 (c_{\phi q}^{(3)})^2)-24 c_{\phi ud}^2\bigg]\,.
\end{align}
All other relevant counterterms ($\tilde{\mu}^2$, $\tilde{c}_{D \phi}$, $\tilde{c}_{\phi^4}^{(5,7,9,13)}$, etc.) vanish at the order of $E/\Lambda$ we are interested in.

\section{The structure of the renormalisation group equations}
\label{sec:rgeform}
The explicit form of the divergences in the Green basis, shown in the
equations above, is of utmost importance, or else the computation of
RGEs at higher orders or involving other light degrees of
freedom could not be built on our
results~\cite{Criado:2018sdb}. Subsequently, though, one can reduce the redundant
operators to the physical basis. We do that following  the results in
Appendix~\ref{sec:redundancies}. 

In the physical basis, the RGEs of the different dimension-eight couplings read:
\begin{equation}
 16\pi^2 \mu \frac{d c_i^{(8)}}{d\mu} = -c_i^{(8)}\sum_j x_j n_j\frac{\partial}{\partial x_j}\frac{\tilde{c}_i^{(8)}}{c_i^{(8)}}\,,
\end{equation}
with $x$ running over all couplings, renormalisable or not, and with
$n$ representing the corresponding tree-level anomalous dimension, defined as the value required to keep the couplings dimensionless in $4-2\epsilon$ dimensions. The
minus sign results from requiring that the counterterms cancel the
divergences. The complete set of RGEs can be found in
  Appendix~\ref{sec:fullrges}, including those of renormalisable and dimension-six terms. In this section, we limit ourselves
to discussing the structure of the $\gamma'$ matrices defined in
Eq.~(\ref{eq:gprime}). 

Since the contribution comes from pairs of
dimension-six operators, we provide a (symmetric) matrix for each
$c_i^{(8)}$ in which we represent with a $\times$ a non-vanishing entry,
with $0$ a trivial zero, for which all contributions in the Green basis
vanish (for example in some cases there are no diagrams contributing
to the corresponding amplitude), and with $\emptyset$ a non-trivial zero, for
which several non-vanishing contributions cancel in the physical
basis.
We find:\\  
\begin{equation}\nonumber
 \footnotesize 
 \def\arraystretch{1}
 \begin{array}{c|c@{\hspace{0.2em}}c@{\hspace{0.2em}}c@{\hspace{0.2em}}c@{\hspace{0.2em}}c@{\hspace{0.2em}}c@{\hspace{0.2em}}c@{\hspace{0.2em}}c}
\mathbf{\gamma'_{c_\phi^8}\,\,}&c_{\phi} & c_{\phi D} & c_{\phi\Box} & c_{\phi\psi_L}^{(1)} & c_{\phi\psi_L}^{(3)} & c_{\phi\psi_R} & c_{\phi ud} & c_{\psi_R\phi}\\\hline
 c_{\phi} & \times & \times & \times & 0 & \times & 0 & \times & \times\\
 c_{\phi D}& & \times & \times & \times & \times & \times & \times & \times\\
 c_{\phi\Box} & & & \times & 0 & \times & 0 & \times & \times\\
 c_{\phi\psi_L}^{(1)} & & & & \times & \times & \times & 0 & \times\\
 c_{\phi\psi_L}^{(3)} & & & & & \times & \times & 0 & \times\\
 c_{\phi\psi_R} & & & & & & \times & 0 & \times\\
 c_{\phi ud} & & & & & & & \times & 0\\
 c_{\psi_R \phi} & & & & & & & & \times
 \end{array}
 \hspace{1cm}
 \begin{array}{c|c@{\hspace{0.2em}}c@{\hspace{0.2em}}c@{\hspace{0.2em}}c@{\hspace{0.2em}}c@{\hspace{0.2em}}c@{\hspace{0.2em}}c@{\hspace{0.2em}}c}
\mathbf{\gamma'_{c_{\phi^4}^{(1)}} }&c_{\phi} & c_{\phi D} & c_{\phi\Box} & c_{\phi\psi_L}^{(1)} & c_{\phi\psi_L}^{(3)} & c_{\phi\psi_R} & c_{\phi ud} & c_{\psi_R\phi}\\\hline
 c_{\phi} & 0 & 0 & 0 & 0 & 0 & 0 & 0 & 0\\
 c_{\phi D}& & \times & \times & 0 & 0 & 0 & 0 & 0\\
 c_{\phi\Box} & & & \times & 0 & 0 & 0 & 0 & 0\\
 c_{\phi\psi_L}^{(1)} & & & & \times& 0 & 0 & 0 & 0\\
 c_{\phi\psi_L}^{(3)} & & & & & \times & 0 & 0 & 0\\
 c_{\phi\psi_R} & & & & & & \times & 0 & 0\\
 c_{\phi ud} & & & & & & & \times & 0\\
 c_{\psi_R \phi} & & & & & & & & 0
 \end{array}\\[0.2cm]
 \end{equation}
\begin{equation}\nonumber
 \footnotesize 
 \def\arraystretch{1}
 \begin{array}{c|c@{\hspace{0.2em}}c@{\hspace{0.2em}}c@{\hspace{0.2em}}c@{\hspace{0.2em}}c@{\hspace{0.2em}}c@{\hspace{0.2em}}c@{\hspace{0.2em}}c}
\mathbf{\gamma'_{c_{\phi^4}^{(2)}} }&c_{\phi} & c_{\phi D} & c_{\phi\Box} & c_{\phi\psi_L}^{(1)} & c_{\phi\psi_L}^{(3)} & c_{\phi\psi_R} & c_{\phi ud} & c_{\psi_R\phi}\\\hline
 c_{\phi} & 0 & 0 & 0 & 0 & 0 & 0 & 0 & 0\\
 c_{\phi D}& & \times & \times & 0 & 0 & 0 & 0 & 0\\
 c_{\phi\Box} & & & \times & 0 & 0 & 0 & 0 & 0\\
 c_{\phi\psi_L}^{(1)} & & & & \times& 0 & 0 & 0 & 0\\
 c_{\phi\psi_L}^{(3)} & & & & & \times & 0 & 0 & 0\\
 c_{\phi\psi_R} & & & & & & \times & 0 & 0\\
 c_{\phi ud} & & & & & & & 0 & 0\\
 c_{\psi_R \phi} & & & & & & & & 0
 \end{array}\\[0.cm]
 \hspace{1cm}
 \begin{array}{c|c@{\hspace{0.2em}}c@{\hspace{0.2em}}c@{\hspace{0.2em}}c@{\hspace{0.2em}}c@{\hspace{0.2em}}c@{\hspace{0.2em}}c@{\hspace{0.2em}}c}
\mathbf{\gamma'_{c_{\phi^4}^{(3)}} }&c_{\phi} & c_{\phi D} & c_{\phi\Box} & c_{\phi\psi_L}^{(1)} & c_{\phi\psi_L}^{(3)} & c_{\phi\psi_R} & c_{\phi ud} & c_{\psi_R\phi}\\\hline
 c_{\phi} & 0 & 0 & 0 & 0 & 0 & 0 & 0 & 0\\
 c_{\phi D}& & \times & \times & 0 & 0 & 0 & 0 & 0\\
 c_{\phi\Box} & & & \times & 0 & 0 & 0 & 0 & 0\\
 c_{\phi\psi_L}^{(1)} & & & & 0& 0 & 0 & 0 & 0\\
 c_{\phi\psi_L}^{(3)} & & & & & \times & 0 & 0 & 0\\
 c_{\phi\psi_R} & & & & & & 0 & 0 & 0\\
 c_{\phi ud} & & & & & & & \times & 0\\
 c_{\psi_R \phi} & & & & & & & & 0
 \end{array}\\[0.2cm]
 \end{equation}
\begin{equation}\nonumber
 \footnotesize 
 \def\arraystretch{1}
\hspace{0.cm} \begin{array}{c|c@{\hspace{0.2em}}c@{\hspace{0.2em}}c@{\hspace{0.2em}}c@{\hspace{0.2em}}c@{\hspace{0.2em}}c@{\hspace{0.2em}}c@{\hspace{0.2em}}c}
\mathbf{\gamma'_{c_{\phi^6}^{(1)}}}&c_{\phi} & c_{\phi D} & c_{\phi\Box} & c_{\phi\psi_L}^{(1)} & c_{\phi\psi_L}^{(3)} & c_{\phi\psi_R} & c_{\phi ud} & c_{\psi_R\phi}\\\hline
 c_{\phi} & 0 & \times & \times & 0 &0 & 0 & 0 & 0\\
 c_{\phi D}& & \times & \times & \times & \times & \times & \times & \times\\
 c_{\phi\Box} & & & \times & 0 & \times & 0 & \times & \times\\
 c_{\phi\psi_L}^{(1)} & & & & \times & \times & \times & 0 & \times\\
 c_{\phi\psi_L}^{(3)} & & & & & \times & \times & \times & \times\\
 c_{\phi\psi_R} & & & & & & \times & 0 & \times\\
 c_{\phi ud} & & & & & & & \times & \times\\
 c_{\psi_R \phi} & & & & & & & & \times
 \end{array}
 \hspace{1cm}
 \begin{array}{c|c@{\hspace{0.2em}}c@{\hspace{0.2em}}c@{\hspace{0.2em}}c@{\hspace{0.2em}}c@{\hspace{0.2em}}c@{\hspace{0.2em}}c@{\hspace{0.2em}}c}
\mathbf{\gamma'_{c_{\phi^6}^{(2)}}}&c_{\phi} & c_{\phi D} & c_{\phi\Box} & c_{\phi\psi_L}^{(1)} & c_{\phi\psi_L}^{(3)} & c_{\phi\psi_R} & c_{\phi ud} & c_{\psi_R\phi}\\\hline
 c_{\phi} & 0 & \times & 0 & 0 &0 & 0 & 0 & 0\\
 c_{\phi D}& & \times & \times & \times & \times & \times & \times & \times\\
 c_{\phi\Box} & & & \times & 0 & 0 & 0 & 0 & 0\\
 c_{\phi\psi_L}^{(1)} & & & & \times & \times & \times & 0 & \times\\
 c_{\phi\psi_L}^{(3)} & & & & & \times & \times & \times & 0\\
 c_{\phi\psi_R} & & & & & & \times & 0 & \times\\
 c_{\phi ud} & & & & & & & \times & \times\\
 c_{\psi_R \phi} & & & & & & & & \times
 \end{array}\\[0.5cm]
 \end{equation}
\begin{equation}\nonumber
 \footnotesize 
 \def\arraystretch{1}
 \begin{array}{c|c@{\hspace{0.2em}}c@{\hspace{0.2em}}c@{\hspace{0.2em}}c@{\hspace{0.2em}}c@{\hspace{0.2em}}c@{\hspace{0.2em}}c@{\hspace{0.2em}}c}
\mathbf{\gamma'_{c_{W^2\phi^4}^{(1)}}}&c_{\phi} & c_{\phi D} & c_{\phi\Box} & c_{\phi\psi_L}^{(1)} & c_{\phi\psi_L}^{(3)} & c_{\phi\psi_R} & c_{\phi ud} & c_{\psi_R\phi}\\\hline
 c_{\phi} & 0 & 0 & 0 & 0 &0 & 0 & 0 & 0\\
 c_{\phi D}& & \times & \emptyset & 0 &0 & 0 & 0 & 0\\
 c_{\phi\Box} & & & 0 & 0 &0 & 0 & 0 & 0\\
 c_{\phi\psi_L}^{(1)} & & & & \times &0 & 0 & 0 & 0\\
 c_{\phi\psi_L}^{(3)} & & & & & \times & 0 & 0 & 0\\
 c_{\phi\psi_R} & & & & & & \times & 0 & 0\\
 c_{\phi ud} & & & & & & & \times & 0\\
 c_{\psi_R \phi} & & & & & & & & 0
 \end{array}
 \hspace{1cm}
 \begin{array}{c|c@{\hspace{0.2em}}c@{\hspace{0.2em}}c@{\hspace{0.2em}}c@{\hspace{0.2em}}c@{\hspace{0.2em}}c@{\hspace{0.2em}}c@{\hspace{0.2em}}c}
\mathbf{\gamma'_{c_{W^2\phi^4}^{(3)}}}&c_{\phi} & c_{\phi D} & c_{\phi\Box} & c_{\phi\psi_L}^{(1)} & c_{\phi\psi_L}^{(3)} & c_{\phi\psi_R} & c_{\phi ud} & c_{\psi_R\phi}\\\hline
 c_{\phi} & 0 & 0 & 0 & 0 &0 & 0 & 0 & 0\\
 c_{\phi D}& & \emptyset &  \emptyset & 0 &0 & 0 & 0 & 0\\
 c_{\phi\Box} & & & 0 & 0 &0 & 0 & 0 & 0\\
 c_{\phi\psi_L}^{(1)} & & & &  \emptyset &0 & 0 & 0 & 0\\
 c_{\phi\psi_L}^{(3)} & & & & & 0 & 0 & 0 & 0\\
 c_{\phi\psi_R} & & & & & &  \emptyset & 0 & 0\\
 c_{\phi ud} & & & & & & &  \emptyset & 0\\
 c_{\psi_R \phi} & & & & & & & & 0
 \end{array}\\[1cm]
 \end{equation}
\begin{equation}\nonumber
 \footnotesize 
 \def\arraystretch{1}
 \begin{array}{c|c@{\hspace{0.2em}}c@{\hspace{0.2em}}c@{\hspace{0.2em}}c@{\hspace{0.2em}}c@{\hspace{0.2em}}c@{\hspace{0.2em}}c@{\hspace{0.2em}}c}
\mathbf{\gamma'_{c_{WB\phi^4}^{(1)}}}&c_{\phi} & c_{\phi D} & c_{\phi\Box} & c_{\phi\psi_L}^{(1)} & c_{\phi\psi_L}^{(3)} & c_{\phi\psi_R} & c_{\phi ud} & c_{\psi_R\phi}\\\hline
 c_{\phi} & 0 & 0 & 0 & 0 &0 & 0 & 0 & 0\\
 c_{\phi D}& & \emptyset &  \emptyset & 0 &0 & 0 & 0 & 0\\
 c_{\phi\Box} & & & 0 & 0 &0 & 0 & 0 & 0\\
 c_{\phi\psi_L}^{(1)} & & & &  \emptyset &0 & 0 & 0 & 0\\
 c_{\phi\psi_L}^{(3)} & & & & & 0 & 0 & 0 & 0\\
 c_{\phi\psi_R} & & & & & &  \emptyset & 0 & 0\\
 c_{\phi ud} & & & & & & &  \emptyset & 0\\
 c_{\psi_R \phi} & & & & & & & & 0
 \end{array}
 \hspace{1cm}
 \begin{array}{c|c@{\hspace{0.2em}}c@{\hspace{0.2em}}c@{\hspace{0.2em}}c@{\hspace{0.2em}}c@{\hspace{0.2em}}c@{\hspace{0.2em}}c@{\hspace{0.2em}}c}
\mathbf{\gamma'_{c_{B^2\phi^4}^{(1)}}}&c_{\phi} & c_{\phi D} & c_{\phi\Box} & c_{\phi\psi_L}^{(1)} & c_{\phi\psi_L}^{(3)} & c_{\phi\psi_R} & c_{\phi ud} & c_{\psi_R\phi}\\\hline
 c_{\phi} & 0 & 0 & 0 & 0 &0 & 0 & 0 & 0\\
 c_{\phi D}& & \times & 0 & 0 &0 & 0 & 0 & 0\\
 c_{\phi\Box} & & & 0 & 0 &0 & 0 & 0 & 0\\
 c_{\phi\psi_L}^{(1)} & & & &  \times &0 & 0 & 0 & 0\\
 c_{\phi\psi_L}^{(3)} & & & & & \times & 0 & 0 & 0\\
 c_{\phi\psi_R} & & & & & &  \times & 0 & 0\\
 c_{\phi ud} & & & & & & &  \times & 0\\
 c_{\psi_R \phi} & & & & & & & & 0
 \end{array}\\[1cm]
 \end{equation}
\begin{equation}\nonumber
 \footnotesize 
 \def\arraystretch{1}
 \hspace{-0.1cm}\begin{array}{c|c@{\hspace{0.2em}}c@{\hspace{0.2em}}c@{\hspace{0.2em}}c@{\hspace{0.2em}}c@{\hspace{0.2em}}c@{\hspace{0.2em}}c@{\hspace{0.2em}}c}
\mathbf{\gamma'_{c_{W\phi^4 D^2}^{(1)}}}&c_{\phi} & c_{\phi D} & c_{\phi\Box} & c_{\phi\psi_L}^{(1)} & c_{\phi\psi_L}^{(3)} & c_{\phi\psi_R} & c_{\phi ud} & c_{\psi_R\phi}\\\hline
 c_{\phi} & 0 & 0 & 0 & 0 &0 & 0 & 0 & 0\\
 c_{\phi D}& & \times & \emptyset & 0 &0 & 0 & 0 & 0\\
 c_{\phi\Box} & & & 0 & 0 &0 & 0 & 0 & 0\\
 c_{\phi\psi_L}^{(1)} & & & &  \times &0 & 0 & 0 & 0\\
 c_{\phi\psi_L}^{(3)} & & & & & \times & 0 & 0 & 0\\
 c_{\phi\psi_R} & & & & & &  \times & 0 & 0\\
 c_{\phi ud} & & & & & & &  \times & 0\\
 c_{\psi_R \phi} & & & & & & & & 0
 \end{array}
 \hspace{0.9cm}
 \begin{array}{c|c@{\hspace{0.2em}}c@{\hspace{0.2em}}c@{\hspace{0.2em}}c@{\hspace{0.2em}}c@{\hspace{0.2em}}c@{\hspace{0.2em}}c@{\hspace{0.2em}}c}
\mathbf{\gamma'_{c_{B \phi^4 D^2}^{(1)}}}&c_{\phi} & c_{\phi D} & c_{\phi\Box} & c_{\phi\psi_L}^{(1)} & c_{\phi\psi_L}^{(3)} & c_{\phi\psi_R} & c_{\phi ud} & c_{\psi_R\phi}\\\hline
 c_{\phi} & 0 & 0 & 0 & 0 &0 & 0 & 0 & 0\\
 c_{\phi D}& & \times & 0 & 0 &0 & 0 & 0 & 0\\
 c_{\phi\Box} & & & 0 & 0 &0 & 0 & 0 & 0\\
 c_{\phi\psi_L}^{(1)} & & & &  \times &0 & 0 & 0 & 0\\
 c_{\phi\psi_L}^{(3)} & & & & & \times & 0 & 0 & 0\\
 c_{\phi\psi_R} & & & & & &  \times & 0 & 0\\
 c_{\phi ud} & & & & & & &  \times & 0\\
 c_{\psi_R \phi} & & & & & & & & 0
 \end{array}\\[1.5cm]
 \end{equation}
All other $\gamma'$ matrices vanish identically, with all their zeros being trivial.

Finally, in Table~\ref{tab:rgeform} we provide a different view on the
global structure of the anomalous dimensions, by showing, for each
pair of dimension-six interactions, the dimension-eight operators that
get renormalised by them. Despite being not explicitly shown, contributions proportional to two fermionic operators involve only leptons or quarks, but not both.
\begin{landscape}
\begin{table}[t]
\resizebox{1.35\textwidth}{!}
{\begin{tabular}{ccccccccc}
& \large{$\bm{c_{\phi}}$} & \large{$\bm{c_{\phi D}}$} & \large{$\bm{c_{\phi\Box}}$} & \large{$\bm{c_{\phi \psi_L}^{(1)}}$}  & \large{$\bm{c_{\phi \psi_L}^{(3)}}$} & \large{$\bm{c_{\phi \psi_R}}$}  & \large{$\bm{c_{\phi ud}}$} & \large{$\bm{c_{\psi_R \phi}}$} \\\\\cline{1-9}
\large{$\bm{c_{\phi}}$} & \parbox{1cm}{\begin{align*}
    & \phi^8 \end{align*}} & \parbox{1cm}{\begin{align*}
    & \phi^8,\,{\phi^6}^{(1)},\,{\phi^6}^{(2)} \end{align*}} & \parbox{1cm}{\begin{align*}
    & \phi^8,\,{\phi^6}^{(1)} \end{align*}} & & \parbox{1cm}{\begin{align*}
    & \phi^8 \end{align*}}& & \parbox{1cm}{\begin{align*}
    & \phi^8 \end{align*}} & \parbox{1cm}{\begin{align*}
    & \phi^8 \end{align*}}\\ \cline{2-9}
\large{$\bm{c_{\phi D}}$} & & \parbox{1cm}{\begin{align*}
    & \phi^8,\,{\phi^6}^{(1)},\,{\phi^6}^{(2)} \\
    & {\phi^4}^{(1)}\,,{\phi^4}^{(2)}\,,{\phi^4}^{(3)}  \\
    & {V^2 \phi^4}^{(1)}\,,{V \phi^4 D^2}^{(1)}\end{align*}}&\parbox{1cm}{\begin{align*}
    & \phi^8,\,{\phi^6}^{(1)},\,{\phi^6}^{(2)} \\
    & {\phi^4}^{(1)}\,,{\phi^4}^{(2)}\,,{\phi^4}^{(3)} \end{align*}} & \parbox{1cm}{\begin{align*}
    & \phi^8,\,{\phi^6}^{(1)},\,{\phi^6}^{(2)} \end{align*}}& \parbox{1cm}{\begin{align*}
    & \phi^8,\,{\phi^6}^{(1)},\,{\phi^6}^{(2)}  \end{align*}}& \parbox{1cm}{\begin{align*}
    & \phi^8,\,{\phi^6}^{(1)},\,{\phi^6}^{(2)}   \end{align*}}& \parbox{1cm}{\begin{align*}
    & \phi^8,\,{\phi^6}^{(1)},\,{\phi^6}^{(2)}  \end{align*}}&\parbox{1cm}{\begin{align*}
    & \phi^8,\,{\phi^6}^{(1)},\,{\phi^6}^{(2)}  \end{align*}} \\ \cline{3-9}
\large{$\bm{c_{\phi 	\Box}}$} & & &\parbox{1cm}{\begin{align*}
    & \phi^8,\,{\phi^6}^{(1)},\,{\phi^6}^{(2)} \\
    & {\phi^4}^{(1)}\,,{\phi^4}^{(2)}\,,{\phi^4}^{(3)}  \end{align*}} & & \parbox{1cm}{\begin{align*}
    & \phi^8,\,{\phi^6}^{(1)}  \end{align*}}& &\parbox{1cm}{\begin{align*}
    & \phi^8,\,{\phi^6}^{(1)}   \end{align*}} &\parbox{1cm}{\begin{align*}
    & \phi^8,\,{\phi^6}^{(1)} \end{align*}} \\ \cline{4-9}
\large{$\bm{c_{\phi \psi_L}^{(1)}}$} & & & &\parbox{1cm}{\begin{align*}
    & \phi^8,\,{\phi^6}^{(1)},\,{\phi^6}^{(2)} \\
    & {\phi^4}^{(1)}\,,{\phi^4}^{(2)}  \\
    &  {V^2 \phi^4}^{(1)}\,,{V \phi^4 D^2}^{(1)}\end{align*}} & \parbox{1cm}{\begin{align*}
    & \phi^8,\,{\phi^6}^{(1)},\,{\phi^6}^{(2)}  \end{align*}}& \parbox{1cm}{\begin{align*}
    & \phi^8,\,{\phi^6}^{(1)},\,{\phi^6}^{(2)}   \end{align*}}& &\parbox{1cm}{\begin{align*}
    & \phi^8,\,{\phi^6}^{(1)},\,{\phi^6}^{(2)}  \end{align*}} \\ \cline{5-9}
\large{$\bm{c_{\phi \psi_L}^{(3)}}$} & & & & & \parbox{1cm}{\begin{align*}
    & \phi^8,\,{\phi^6}^{(1)},\,{\phi^6}^{(2)} \\
    & {\phi^4}^{(1)}\,,{\phi^4}^{(2)}\,,{\phi^4}^{(3)}   \\
    &  {V^2 \phi^4}^{(1)}\,,{V \phi^4 D^2}^{(1)}\end{align*}}& \parbox{1cm}{\begin{align*}
    & \phi^8,\,{\phi^6}^{(1)},\,{\phi^6}^{(2)}  \end{align*}}&\parbox{1cm}{\begin{align*}
    & {\phi^6}^{(1)},\,{\phi^6}^{(2)}   \end{align*}} &\parbox{1cm}{\begin{align*}
    & \phi^8,\,{\phi^6}^{(1)}   \end{align*}} \\ \cline{6-9}
\large{$\bm{c_{\phi \psi_R}}$} & & & & & & \parbox{1cm}{\begin{align*}
    & \phi^8,\,{\phi^6}^{(1)},\,{\phi^6}^{(2)} \\
    & {\phi^4}^{(1)}\,,{\phi^4}^{(2)}  \\
    &  {V^2 \phi^4}^{(1)}\,,{V \phi^4 D^2}^{(1)}\end{align*}}& &\parbox{1cm}{\begin{align*}
    & \phi^8,\,{\phi^6}^{(1)},\,{\phi^6}^{(2)}  \end{align*}} \\ \cline{7-9}
\large{$\bm{c_{\phi ud}}$} & & & & & & &\parbox{1cm}{\begin{align*}
    & \phi^8,\,{\phi^6}^{(1)},\,{\phi^6}^{(2)} \\
    & {\phi^4}^{(1)}\,,{\phi^4}^{(3)}  \\
    &  {V^2 \phi`^4}^{(1)}\,,{V \phi^4 D^2}^{(1)}\end{align*}} &\parbox{1cm}{\begin{align*}
    & {\phi^6}^{(1)},\,{\phi^6}^{(2)}   \end{align*}} \\ \cline{8-9}
\large{$\bm{c_{\psi_R \phi}}$} & & & & & & & &\parbox{1cm}{\begin{align*}
    & \phi^8,\,{\phi^6}^{(1)},\,{\phi^6}^{(2)} \end{align*}}  \\ \cline{9-9}
 \end{tabular}
}
\caption{\it Dimension-eight operators that are renormalised by the pairs of dimension-six interactions in the corresponding column and row. $V$ stands for either $W$ or $B$.}\label{tab:rgeform}
\end{table}
\end{landscape}

\section{Discussion and outlook}
\label{sec:discussion}

We conclude this article highlighting several observations that can be
made on the basis of the RGE matrices above: 

\textbf{1.} 
All the dimension-eight operators that are renormalised can arise at
tree level in UV completions of the SM~\cite{Craig:2019wmo}. The
reason is simply that those operators that arise only at loop level
involve two Higgs fields (see Table~\ref{tab:dim8ops}), unlike
any one-loop diagram containing two insertions of the dimension-six
terms. 
The same holds for dimension-six operators.
Thus, we conclude that within the bosonic sector of the SMEFT,
dimension-six tree-level operators do not mix into loop-level
operators to order $E^4/\Lambda^4$. This extends previous findings at
order $E^2/\Lambda^2$~\cite{Cheung:2015aba}.\\ 

\textbf{2.} Several of the $\gamma'$ matrices above exhibit a number of zeros
(denoted by $0$) for which all contributions in the Green basis are
vanishing (they can result simply from the absence of Feynman
diagrams or from CP conservation reasons). For example, the first
row in $\gamma'_{c_{\phi^4}^{(1)}}$ 
reflects that there are no one-particle-irreducible diagrams with four Higgses involving
the insertion of one six-Higgs operator and one four-Higgs
interaction. Instead,
those denoted by $\emptyset$  ensue from non-trivial cancellations between
different counterterms in the Green basis which, on-shell, add to
zero. For example, the (23) entry of $\gamma'_{c_{W^2\phi^4}^{(1)}}$
vanishes because the terms proportional to $c_{\phi D}c_{\phi\square}$
in $\tilde{c}_{W^2\phi^4}^{(1)}$ and $\tilde{c}_{W\phi^4 D^2}^{(7)}$
cancel in Eq.~(\ref{eq:W2H41}). Zeros as this one might be understood
on the basis of the helicity-amplitude formalism~\cite{EliasMiro:2020tdv,Baratella:2020lzz}.\\ 

\textbf{3.} Related to the previous point, we find the very surprising
result that the Peskin-Takeuchi parameters~\cite{Peskin:1990zt} 
$S$ and $U$ are not renormalised by tree-level dimension-six operators
to order $v^4/\Lambda^4$. Indeed, these observables 
read~\cite{Hays:2020scx,Murphy:2020rsh}:
\begin{align}
\frac{1}{16\pi} S = \frac{v^2}{\Lambda^2}\left[c_{\phi WB} + c_{WB\phi^4}^{(1)}\frac{v^4}{\Lambda^4}\right]\,,\quad
\frac{1}{16\pi} U = \frac{v^4}{\Lambda^4} c_{W^2\phi^4}^{(3)}\,,
\end{align}
with $\mathcal{O}_{\phi WB} = (\phi^\dagger\sigma^I\phi) W_{\mu\nu}^I B^{\mu\nu}$. (Note that $U$ arises only at dimension eight.) What we find is that $c_{WB\phi^4}^{(1)}$
and $c_{W^2\phi^4}^{(3)}$ do not renormalise because the direct contribution cancels that from the redundant
operator $\mathcal{O}_{W\phi^4 D^2}^{(7)}$. This fact, together with the non-renormalisation of $c_{\phi WB}$ found in
Refs.~\cite{Alonso:2013hga}, shows that both $S$ and $U$ are not triggered by dimension-six tree-level interactions at one loop.\\

\textbf{4.} The Wilson coefficients $c_{\phi^4}^{(1)}$, $c_{\phi^4}^{(2)}$ and $c_{\phi^4}^{(3)}$ are subject to positivity constraints~\cite{Remmen:2019cyz}. In particular, $c_{\phi^4}^{(2)}\geq 0$, $c_{\phi^4}^{(1)}+c_{\phi^4}^{(2)}\geq 0$ and $c_{\phi^4}^{(1)}+c_{\phi^4}^{(2)}+c_{\phi^4}^{(3)}\geq 0$. These inequalities should reflect in the corresponding matrices.

To see how, let us first note that there exist well-behaved UV completions of the SM that induce, at tree level, the operators $c_\phi$, $c_{\phi D}$ and $c_{\phi\square}$ with \textit{arbitrary values}; see Appendix~\ref{sec:UVmodel} for a particular example. The values of the dimension-eight Wilson coefficients $c_{\phi^4}^{(1)}$, $c_{\phi^4}^{(2)}$ and $c_{\phi^4}^{(3)}$ at any energy $\mu<M$ triggered by double insertions of the dimension-six operators scale differently with the model couplings than the tree-level contribution (which in general can not be avoided). In particular, within the model of Appendix~\ref{sec:UVmodel}, we have $\sim \kappa^4/M^4$ versus $\sim \kappa^2/M^2$. This suggests that both contributions must satisfy the positivity bounds separately. Indeed, in the limit of scale-invariant dimension-six Wilson coefficients, we can check that:
\begin{align}
 16\pi^2 c_{\phi^4}^{(2)} &= \frac{1}{3}(5c_{\phi D}^2+16c_{\phi D}c_{\phi\square} + 16 c^2_{\phi\square})\log{\frac{M}{\mu}} > 0\,,\\[0.2cm]
 16\pi^2 \left[c_{\phi^4}^{(1)}+c_{\phi^4}^{(2)}\right] &= \frac{16}{3}(c_{\phi D}^2-c_{\phi D}c_{\phi\square}+2 c_{\phi\square}^2)\log{\frac{M}{\mu}} > 0\,,\\[0.2cm]
 16\pi^2 \left[c_{\phi^4}^{(1)}+c_{\phi^4}^{(2)}+c_{\phi^4}^{(3)}\right] &= 3(c_{\phi D}^2+8 c_{\phi\square}^2)\log{\frac{M}{\mu}}>0\,;
\end{align}
for arbitrary values of $c_{\phi D}$ and $c_{\phi\square}$. (Fermionic Wilson coefficients do not modify these relations because they contribute as sums of modulus squared and therefore positively, as a result also of very fine cancellations between positive and negative terms in Eqs.~\ref{eq:cphi41}--\ref{eq:cphi43}.) Note that these inequalities hold non-trivially; for example $c_{\phi^4}^{(1)}$ is negative in a neighbourhood of its minimum.
It should be possible to extend this kind of analysis to other operators (which do not renormalise within the assumptions we make in this work), thus providing interesting cross-checks of the anomalous dimensions (or new bounds on Wilson coefficients).\\

\textbf{5.} Among the non-vanishing entries in the different $\gamma'$s, we find values that depart significantly from the naive estimate of $\mathcal{O}(1)$. The most notable of these, not suppressed by gauge or $\lambda$ couplings, are the $126$ in $\gamma'_{\phi^8}$ and the $96$ in $\gamma'_{c_{\phi^6}^{(1)}}$; see Appendix~\ref{sec:fullrges}. As we discuss below, these large numbers can have important low-energy implications.\\

\textbf{6.}  Although we do not aim to exhaust all possible phenomenological implications of the running of the dimension-eight operators,
we would like to stress that the $T$ parameter, defined by~\cite{Murphy:2020rsh}
\begin{equation}
\alpha T = -\frac{1}{2}\frac{v^2}{\Lambda^2}\left[c_{\phi D} + c_{\phi^6}^{(2)}\frac{v^2}{\Lambda^2}\right]\,,
\end{equation}
with $\alpha\sim 1/137$ being the fine-structure constant, receives contributions from the operator $\mathcal{O}_{\phi ud}$ 
only at order $v^4/\Lambda^4$ (because $c_{\phi D}$ is not renormalised by one insertion of $\mathcal{O}_{\phi ud}$; see Ref.~\cite{Jenkins:2013wua}).
Using bounds on $T$ from Ref.~\cite{deBlas:2016nqo}, and assuming that
only $c_{\phi t b}$ is non-vanishing, we obtain $c_{\phi tb} \leq
5.9$ for $\Lambda = 1$ TeV. This constraint is competitive with the
value $c_{\phi tb} \leq 5.3$ reported in Ref.~\cite{Maltoni:2019aot}.
 \\

\textbf{7.} We would also like to emphasize the importance of one-loop $v^4/\Lambda^4$ effects for the EW phase transition (EWPT) ensuing from modifications of the Higgs potential~\cite{Zhang:1992fs,Grojean:2004xa,Bodeker:2004ws,Delaunay:2007wb,deVries:2017ncy,Chala:2018ari}. To this aim, let us assume that $c_\phi$ is the only non-vanishing Wilson coefficient in the UV, and let us neglect gauge and Yukawa couplings. The Higgs potential in the infrared is then provided by running $\mathcal{L}_\text{UV}$ down to the EW scale. In the leading-logarithm approximation, this reads:
\begin{align}
 V \sim \jg{-}\mu^2|\phi|^2 + \lambda|\phi|^4 + \frac{c_\phi}{\Lambda^2}\left(1-\frac{108}{16\pi^2}\lambda\log{\frac{\Lambda}{v}}\right) |\phi|^6 + \frac{126}{16\pi^2\Lambda^4}\log{\frac{\Lambda}{v}} c^2_\phi \,|\phi|^8\,,
\end{align}
where $c_\phi$ is evaluated in the UV, and the renormalisable couplings are assumed scale-invariant for simplicity. (The first logarithm can be read from Ref.~\cite{Jenkins:2013zja}.) In both $|\phi|^6$ and $|\phi|^8$ we have included only the dominant corrections.

Following Refs.~\cite{Chala:2018ari,Caprini:2019egz}, we know that the EWPT is first order and strong as required by EW baryogenesis~\cite{Kuzmin:1985mm} provided that $500\,\,\text{GeV}\lesssim\Lambda/\sqrt{c_\text{eff}}\lesssim 750$ GeV, where we have defined $c_\text{eff} = c_\phi + 3/2\, v^2/\Lambda^2 c_{\phi^8}$. Fixing, as a matter of example, $\Lambda=1$ TeV, it can be easily checked that this occurs for:
\begin{align}
 1.7\, \text{TeV}^{-2}\lesssim c_\phi\lesssim 3.7\,\text{TeV}^{-2}\,
\end{align}
if the running of $c_{\phi^8}$ is neglected, whereas if we account for it we obtain:
\begin{align}
 1.5\, \text{TeV}^{-2}\lesssim c_\phi\lesssim 2.6\,\text{TeV}^{-2}\,.
\end{align}%
The $30$\,\% difference in the upper limit evidences the potential importance of both dimension-eight operators and their running. \\

To conclude, let us remark that our results comprise one step further towards the one-loop renormalisation of the SMEFT to order $v^4/\Lambda^4$. This endeavor was initiated in Refs.~\cite{Grojean:2013kd,Elias-Miro:2013gya,Elias-Miro:2013mua,Jenkins:2013zja,Jenkins:2013wua,Alonso:2013hga} (see also Refs.~\cite{Chankowski:1993tx,Babu:1993qv,Antusch:2001ck} for the renormalisation of the Weinbeg operator) and continued in Ref.~\cite{Alonso:2014zka} (for baryon-number violating interactions), Ref.~\cite{Davidson:2018zuo} (which includes the renormalisation of dimension-six operators by pairs of Weinberg interactions), Refs.~\cite{Liao:2016hru,Liao:2019tep} (which involves the renormalisation of dimension-seven operators triggered by relevant couplings) and Ref.~\cite{Chala:2021juk} (in which neutrino masses are renormalised to order $v^3/\Lambda^3$, including arbitrary combinations of dimension-five, -six and -seven operators). Some partial results of renormalisation within the dimension-eight sector of the SMEFT can be found in Refs.~\cite{Panico:2018hal,Jiang:2020mhe}. See Table~\ref{tab:summary} for a summary of the state of the art.

In forecoming works, we plan to extend the results of this paper with the inclusion, in the UV, of the operators  of dimension eight that can be generated at tree level. We will also consider the renormalisation of non-bosonic operators. The latter can be induced, in particular, by field redefinitions aimed at removing the operators $\mathcal{O}_{BD\phi}$, $\mathcal{O}_{\phi D}'$, $\mathcal{O}_{WD\phi}$ and $\mathcal{O}_{W\phi^4 D^2}^{(6)}$; see Appendix~\ref{sec:redundancies}. Consequently, our current findings  lay the basis for future work in this direction.

\begin{table}[t]
 \begin{center}
 \resizebox{1.\textwidth}{!}{\begin{tabular}{l|ccccccccccc}
   & $d_5$ & $d_5^2$ & $d_6$ & $d_5^3$ & $d_5\times d_6$ & $d_7$ & $d_5^4$ & $d_5^2\times d_6$ & $d_6^2$ & $d_5\times d_7$ & $d_8$\\
   \toprule
   $d_{\leq 4}$ (bosonic) &  &  & $\gtick$~\cite{Jenkins:2013zja} &  &   &  &  &  & \colorbox{gray}{\textcolor{white}{This\, work}} &  & $\rtick$ \\
   $d_{\leq 4}$ (fermionic) &  &  & $\gtick$~\cite{Jenkins:2013zja} &  &   &  &  &  & $\rtick$ &  & $\rtick$ \\
   $d_5$ & $\gtick$~\cite{Chankowski:1993tx,Babu:1993qv,Antusch:2001ck} &  &  &  & $\gtick$~\cite{Chala:2021juk} & $\otick$~\cite{Chala:2021juk} &  &  &  & & \\
   $d_6$ (bosonic) &  & $\gtick$~\cite{Davidson:2018zuo} & $\gtick$~\cite{Jenkins:2013zja,Jenkins:2013wua,Alonso:2013hga} & & & &  & $\rtick$ & \colorbox{gray}{\textcolor{white}{This\, work}} & $\rtick$ & $\rtick$ \\
   $d_6$ (fermionic) &  & $\gtick$~\cite{Davidson:2018zuo} & $\gtick$~\cite{Jenkins:2013zja,Jenkins:2013wua,Alonso:2013hga,Alonso:2014zka} & & & & & $\rtick$ & $\rtick$ & $\rtick$ & $\rtick$\\
   $d_7$ &  &  & & $\otick$~\cite{Chala:2021juk} & $\otick$~\cite{Chala:2021juk} & $\gtick$~\cite{Liao:2016hru,Liao:2019tep}\\
   $d_8$ (bosonic) &  &  & & & & & $\rtick$ & $\rtick$ &  \colorbox{gray}{\textcolor{white}{This\, work}} & $\rtick$ & $\rtick$\\
   $d_8$ (fermionic) &  &  & & & & & $\rtick$ & $\rtick$ & $\rtick$ & $\rtick$ & $\rtick$
   \\\bottomrule
 \end{tabular}}
 \caption{\it State of the art of SMEFT renormalisation. The rows represent the operators (defined by their dimension $d$) being renormalised, while the columns show the operators entering the loops. Note that there are no bosonic interactions at odd dimension. Blank entries vanish. A tick $\gtick$ represents that the complete contribution is known. The $\otick$ indicates that only (but substantial) partial results have been already obtained. The $\rtick$ indicates that nothing, or very little, is known. The contribution made in this paper is marked by $\colorbox{gray}{\,\,\,}$.}\label{tab:summary}
 \end{center}
\end{table}

\section*{Acknowledgments}
We would like to thank Jorge de Blas for useful discussions.~\mr{We would also like to thank A. Helset, E. E. Jenkins and A. V. Manohar for pointing out typos in previous versions of the manuscript.}
MC and JS are supported by the Ministry of Science and Innovation under grant number FPA2016-78220-C3-1/3-P (FEDER), SRA under grant 
PID2019-106087GB-C21/C22 (10.13039/501100011033), and 
by the Junta de Andaluc\'ia grants FQM 101, A-FQM-211-UGR18 and P18-FR-4314 (FEDER).
MC is also supported by the Spanish MINECO under the Ram\'on y Cajal programme. 
GG and MR acknowledge support by LIP 
(FCT, COMPETE2020-Portugal2020, FEDER, POCI-01-0145-FEDER-007334) as well as by FCT under project CERN/FIS-PAR/0024/2019.
GG is also supported by FCT under the grant SFRH/BD/144244/2019.
MR is also supported by Funda\c{c}\~ao para a Ci\^encia e Tecnologia (FCT) under the grant PD/BD/142773/2018.

\appendix

\section{Removing redundant operators}
\label{sec:redundancies}
The redundant operators generated in the process of renormalisation can be removed upon performing suitable perturbative field redefinitions, for example $\phi\to\phi + \frac{1}{\Lambda^2}\mathcal{O}$, where $\mathcal{O}$ is called the \textit{perturbation}. We are interested in the effect of these field redefinitions to linear order in the perturbation (because $\mathcal{O}$ is loop suppressed and therefore quadratic powers of this term are formally two-loop corrections),
which can be implemeted through the equations of motion of the SMEFT to order $v^2/\Lambda^2$~\cite{Criado:2018sdb}. These read~\cite{Barzinji:2018xvu}:
\begin{align}
D^2 \phi^i &= \mu^2 \phi^i - 2\lambda (\phi^\dagger \phi)\phi^i + \frac{1}{\Lambda^2} \bigg\{ 3c_\phi (\phi^\dagger \phi)^2 \phi^i + 2 c_{\phi\Box}\phi^i \Box(\phi^\dagger \phi) \nonumber \\
&- c_{\phi D} \left[ \left(D^\mu \phi\right)^i \left( \phi^\dagger \overleftrightarrow{D}_\mu \phi\right) + \phi^i \partial^\mu \left(\phi^\dagger  D_\mu \phi\right)\right] \bigg\}+\cdots\,, \\[0.2cm]
\partial^\nu B_{\mu\nu} &= \frac{g_1}{2} \phi^\dagger i \overleftrightarrow{D}_\mu \phi
+ \frac{c_{\phi D}}{\Lambda^2}\frac{g_1}{2} \left(\phi^\dagger\phi\right) \left( \phi^\dagger i \overleftrightarrow{D}_\mu \phi \right) +\cdots \, , \\[0.2cm]
D^\nu W^I_{\mu\nu} &= \frac{g_2}{2}  \phi^\dagger i \overleftrightarrow{D}^I_\mu \phi
+ \frac{c_{\phi D}}{\Lambda^2} \frac{g_2}{\mr{2}}\left(\phi^\dagger \sigma^I \phi\right) \left(\phi^\dagger i\overleftrightarrow{D}_\mu \phi \right) +\cdots\,,
\end{align}
where the ellipses represent fermionc operators, on which we are not interested.
The following relations hold on-shell:
\begin{align}
\mathcal{O}_{BD\phi}& =  \frac{g_1}{2} \bigg[ \mathcal{O}_{\phi \Box} + 4  \mathcal{O}_{\phi D} \jg{+} \frac{\mu^2}{\Lambda^2} c_{\phi D} \mathcal{O}_{\phi }\bigg]+\frac{g_1}{2}  \frac{c_{\phi D}}{\Lambda^2} \bigg[ - 2\lambda\mathcal{O}_{\phi^{8}}  + 3 \mathcal{O}_{\phi^{6}}^{(1)}  + 2 \mathcal{O}_{\phi^{6}}^{(2)} \bigg]+\cdots\,, \\
\label{eq:HDp}
\mathcal{O}_{\phi D}' & = -\frac{1}{2} \bigg[-\mathcal{O}_{\phi \Box} \jg{+} 2 \mu^2 |\phi|^4 - 4 \lambda \mathcal{O}_{\phi} \jg{-} \frac{\mu^2}{\Lambda^2}\left(c_{\phi D} - 8 c_{\phi \Box} \right)\mathcal{O}_\phi\bigg]  \\
& - \frac{1}{2\Lambda^2} \bigg[ \left(6c_\phi - 16\lambda c_{\phi\square} + 2 \lambda c_{\phi D}\right) \mathcal{O}_{\phi^{8}} +\left(8c_{\phi\square} + c_{\phi D} \right) \mathcal{O}_ {\phi^{6}}^{(1)} + 2 c_{\phi D} \mathcal{O}_ {\phi^{6}}^{(2)}\bigg]+\cdots\,, \nonumber\\
\label{eq:DW}
 \mathcal{O}_{WD\phi} & = -g_2 \bigg[-\frac{3}{2}\mathcal{O}_{\phi \Box} \jg{+} 2 \mu^2 |\phi|^4 - 4 \lambda \mathcal{O}_{\phi} \jg{-} \frac{\mu^2}{\Lambda^2}\left( \mr{\frac{3}{2}}c_{\phi D} - 8 c_{\phi \Box} \right)\mathcal{O}_\phi\bigg]   \\
& - \frac{g_2}{\Lambda^2}  \bigg[ \left(6c_\phi - 16\lambda c_{\phi\square} + \mr{3} \lambda c_{\phi D}\right) \mathcal{O}_{\phi^{8}}
+\left(8c_{\phi\square} -\mr{\frac{1}{2}} c_{\phi D} \right) \mathcal{O}_ {\phi^{6}}^{(1)} %
\mr{+ c_{\phi D}  \mathcal{O}_ {\phi^{6}}^{(2)}}\bigg]+\cdots\,;\nonumber
\end{align}
where the ellipses encode again fermionic interactions.
The operator $\mathcal{O}_{\phi D}''$ gives no contributions to the bosonic sector.

To arrive at these results, we used the following identities:
\begin{align}
|\phi|^2 \left(\phi^\dagger D_\mu \phi \right) \left( D^\mu \phi^\dagger \phi\right) & = \frac{1}{2} \bigg[ \mathcal{O}_{\phi^{6}}^{(1)} + \mathcal{O}_{\phi^{6}}^{(2)} \bigg]\,, \\[0.2cm]
|\phi|^2 \left(\phi^\dagger \text{i} \overleftrightarrow{D} \phi\right)^2 & = 3 \mathcal{O}_{\phi^{6}}^{(1)}  + 2 \mathcal{O}_{\phi^{6}}^{(2)} + \frac{1}{2} \mathcal{O}_{\phi^{6}}^{(3)}\,,\\[0.2cm]
|\phi|^4 \Box |\phi|^2 & = 2 \mathcal{O}_{\phi^{6}}^{(1)} +\mathcal{O}_{\phi^{6}}^{(3)}\,.
\end{align}
In turn, the redundant dimension-eight operators become, on-shell:
\begin{align}
\mathcal{O}_{\phi^{6}}^{(3)}  &= 2\mu^2 \mathcal{O}_{\phi}-4 \lambda \mathcal{O}_{\phi^{8}}\,,\\
\mathcal{O}_{\phi^{4}}^{(4)} &= \jg{-}\mu^2\bigg[-\mr{\mathcal{O}_{\phi \Box}} \jg{+}2 \mu^2 |\phi|^4 - 4 \lambda \mathcal{O}_{\phi} \bigg]  - 4 \lambda \mathcal{O}_ {\phi^{6}}^{(1)}\,, \\
\mathcal{O}_{\phi^{4}}^{(6)}  & = 2\mu^2 \mathcal{O}_{\phi D} -2 \lambda \left[ \mathcal{O}_{\phi^{6}}^{(1)} + \mathcal{O}_{\phi^{6}}^{(2)} \right] \,, \\
\mathcal{O}_{\phi^{4}}^{(8)} &= \jg{-}2 \mu^2 \bigg[\jg{-} \mu^2 |\phi|^4 + 4 \lambda \mathcal{O}_\phi\bigg]+8 \lambda^2 \mathcal{O}_{\phi^{8}} \,, \\
\mathcal{O}_{\phi^{4}}^{(10)} &=  \jg{-}\mu^2 \bigg[\jg{-} \mu^2 |\phi|^4 + 4 \lambda \mathcal{O}_\phi\bigg]+4 \lambda^2\mathcal{O}_{\phi^{8}}\,, \\
\mathcal{O}_{\phi^{4}}^{(11)} &= \jg{-}\mu^2 \bigg[ \jg{-}\mu^2 |\phi|^4 + 4 \lambda \mathcal{O}_\phi\bigg]+4 \lambda^2\mathcal{O}_{\phi^{8}} \,, \\
\mathcal{O}_{\phi^{4}}^{(12)} &= \jg{-}\mu^2 \bigg[2 \mathcal{O}_{\phi D} + \mathcal{O}_{\phi \Box} - 2 \lambda \mathcal{O}_\phi\bigg]+2 \left[ 2\lambda \mathcal{O}_ {\phi^{6}}^{(1)} + \lambda \mathcal{O}_ {\phi^{6}}^{(2)} - 2 \lambda^2 \mathcal{O}_{\phi^{8}} \right] \,, \\
\mathcal{O}_{W\phi^4 D^2}^{(6)} & =  -\frac{g_2}{2}  \left[\mu^2 \mathcal{O}_\phi+5%
 \mathcal{O}_{\phi^6}^{(1)} - 2\lambda\mathcal{O}_{\phi^8} + \cdots 
\right] \,, \\
\mathcal{O}_{W\phi^4 D^2}^{(7)}  &= \frac{1}{4} \bigg\lbrace\frac{g_2}{2} \left[-2\mu^2 \mathcal{O}_\phi+\mathcal{O}_{W^2 \phi^4}^{(1)} +  \mathcal{O}_{W^2 \phi^4}^{(3)} \right]+  g_1 \mathcal{O}_{WB \phi^4}^{(1)} + 8 \mathcal{O}_{W\phi^4 D^2}^{(1)} \nonumber \\
&- \mr{4} g_2 \mathcal{O}_{\phi^6}^{(1)}  \mr{-} g_2 \mathcal{O}_{\phi^6}^{(2)} + 2 g_2 \lambda \mathcal{O}_{\phi^8}^{(1)} \bigg\rbrace\,,\\
\mathcal{O}_{B\phi^4D^2}^{(3)} &= -\frac{g_1}{2} \left[\mu^2 \mathcal{O}_\phi+3\mathcal{O}_{\phi^6}^{(1)} + 2\mathcal{O}_{\phi^6}^{(2)} - 2 \lambda \mathcal{O}_{\phi^8} \right]\,;
\end{align}
where again the ellipses represent terms on which we are not interested in this work. The operator $\mathcal{O}_{\phi^6}^{(4)}$ contributes only to fermionic interactions.

Altogether, these equations lead to:
\begin{align}
 \lambda &\to \lambda \jg{+} \frac{\mu^2}{\Lambda^2} \left(c'_{\phi D} \mr{+ 2 g_2 c_{WD\phi}}\right)-  \frac{\mu^4}{\Lambda^4} \bigg[ - 2\left( c_{\phi^4}^{(4)} - c_{\phi^4}^{(8)}\right) + c_{\phi^4}^{(10)} + c_{\phi^4}^{(11)} \bigg]\,,\\
 \label{eq:cphi}
 c_{\phi} &\to  c_\phi +  2 \lambda \left(c'_{\phi D} \mr{+ 2 g_2 c_{WD\phi}} \right)\jg{-} \frac{\mu^2}{2 \Lambda^4} \bigg[ 
 -\left(c_{\phi D} - 8 c_{\phi \Box}\right)(c'_{\phi D} + 2 g_2 c_{WD\phi}) \mr{- } g_2 c_{\phi D} c_{WD\phi}\nonumber \\
 &  - g_1 c_{\phi D} c_{BD\phi} - 4 c_{\phi^6}^{(3)} + 4 \lambda \left\{ -2 c_{\phi^4}^{(4)} + 4c_{\phi^4}^{(8)} + 2 c_{\phi^4}^{(10)} + 2 c_{\phi^4}^{(11)}  - c_{\phi^4}^{(12)} \right\} + g_ 2 c_{W\phi^4 D^2}^{(6)} \nonumber \\
 & + \frac{g_2}{2}c_{W\phi^4 D^2}^{(7)} + g_1 c_{B\phi^4 D^2}^{(3)}
  \bigg] \mr{- 3 K_{\phi} c_{\phi}}
 \,,\\
  \label{eq:cphiD}
 c_{\phi D} &\to c_{\phi D}  \mr{+ 2 g_1 c_{BD\phi}}    \jg{-}   2 \frac{\mu^2}{ \Lambda^4} \bigg[
 - c_{\phi^4}^{(6)} +  c_{\phi^4}^{(12)} \bigg]\mr{- 2 K_{\phi} c_{\phi D}}
 \,,\\
  \label{eq:cphiB}
 c_{\phi\Box} &\to c_{\phi\Box} + \frac{1}{2} \left(c'_{\phi D} \mr{+ g_1 c_{BD\phi} + 3 g_2 c_{WD\phi}}\right)	\jg{-}\frac{\mu^2}{\Lambda^4}\bigg[ 
 - c_{\phi^4}^{(4)} + c_{\phi^4}^{(12)} \bigg]\mr{- 2 K_{\phi} c_{\phi \square}}\,,
\end{align}
where we have already normalised canonically the Higgs kinetic term;
as well as
\begin{align}
 c_{\phi^8} & \to c_{\phi^8} -g_1\lambda  c_{BD\phi} c_{\phi D} - \left(c'_{\phi D} + 2 g_2 c_{WD\phi} \right)  \left(3c_\phi - 8\lambda c_{\phi\square} + \lambda c_{\phi D}\right) \\
 &\mr{- } g_2 \lambda  c_{WD\phi}c_{\phi D}  - 4 \lambda c_{\phi^6}^{(3)} + 4\lambda^2 \bigg[ 2 c_{\phi^4}^{(8)} +c_{\phi^4}^{(10)} + c_{\phi^4}^{(11)} - c_{\phi^4}^{(12)} \bigg] \nonumber \\
 & +  g_2 \lambda  c_{W\phi^4 D^2}^{(6)} + \frac{g_2}{2} \lambda c_{W\phi^4 D^2}^{(7)} + g_1 \lambda c_{B\phi^4 D^2}^{(3)}\nonumber\,,\\[0.2cm]
c_{\phi^6}^{(1)} & \to c_{\phi^6}^{(1)} + \frac{3}{2} g_1 c_{BD\phi} c_{\phi D} - \frac{ \left(c'_{\phi D} + 2 g_2 c_{WD\phi} \right) }{2} \left( 8 c_{\phi \square} + c_{\phi D}\right) + \mr{\frac{3}{2}} g_2 c_{WD\phi} c_{\phi D} \\
& - 2 \lambda \bigg[ 2 c_{\phi^4}^{(4)} + c_{\phi^4}^{(6)} - 2 c_{\phi^4}^{(12)} \bigg] - \frac{5}{2} g_2 c_{W\phi^4 D^2}^{(6)} \mr{-}  g_2 c_{W\phi^4 D^2}^{(7)} - \frac{3}{2} g_1 c_{B\phi^4 D^2}^{(3)}\,,\nonumber\\[0.2cm]
c_{\phi^6}^{(2)} & \to c_{\phi^6}^{(2)} +  c_{\phi D} \left(g_1 c_{BD\phi}  \mr{-g_2 c_{WD\phi}}-  c'_{\phi D}\right) - 2 \lambda \bigg[ c_{\phi^4}^{(6)} - c_{\phi^4}^{(12)} \bigg] \mr{-} \frac{g_2}{4} c_{W\phi^4 D^2}^{(7)} - g_1 c_{B\phi^4 D^2}^{(3)}\,,\nonumber\\[0.2cm]
c_{W^2 \phi^4}^{(1)} &\to c_{W^2 \phi^4}^{(1)} + \frac{g_2}{8 } c_{W\phi^4 D^2}^{(7)}\,,\label{eq:W2H41}\\[0.2cm]
c_{W^2 \phi^4}^{(3)} &\to c_{W^2 \phi^4}^{(3)} + \frac{g_2}{8} c_{W\phi^4 D^2}^{(7)}\,,\\[0.2cm]
c_{W B \phi^4}^{(1)} &\to c_{W B \phi^4}^{(1)} + \frac{g_1}{4} c_{W\phi^4 D^2}^{(7)}\,\\[0.2cm]
c_{W\phi^4 D^2}^{(1)} &\to c_{W\phi^4 D^2}^{(1)} + 2 c_{W\phi^4 D^2}^{(7)}\,.
\end{align}
\section{Renormalisation group equations}
\label{sec:fullrges}
For a given coupling $c$, we define:
\begin{equation}
\dot{c} \equiv 16\pi^2 \mu \frac{\text{d}c}{d\mu}.
\end{equation}
Thus, we have:
\begin{align}
 \dot{\lambda} &\supset \left(5 c_{\phi D}^2 - 24 c_{\phi D} c_{\phi \Box} + 24 c_{\phi \Box}^2 \right)
 \frac{\mu^4}{\Lambda^4}\,,
\end{align}
\begin{align}
    \dot{c}_{\phi D} &\supset  \left(10 c_{\phi D}^2\jg{-} 4c_{\phi D}c_{\phi\Box} \right) \frac{\mu^2}{\Lambda^2} \,,
\end{align}
\begin{align}
     \dot{c}_{\phi\Box} &\supset  \left(\frac{3}{2}c_{\phi D}^2\jg{+}14c_{\phi D} c_{\phi\Box} \jg{-} 36c_{\phi\Box}^2\right)\frac{\mu^2}{\Lambda^2}\,,
\end{align}
\begin{align}
\dot{c}_{\phi} &\supset\frac{\mu ^2}{24 \Lambda ^2} \left\{\left(37 g_1^2-15 g_2^2+840 \lambda \right) c_{{\phi D}}^2-8 \left(2 \left(\left(\text{Tr}\left[c_{{\phi d}}\right]+\text{Tr}\left[c_{{\phi e}}\right]+\text{Tr}\left[c^{(1)}_{{\phi l}}\right]-\text{Tr}\left[c^{(1)}_{{\phi q}}\right] \right.\right.\right.\right. \nonumber \\ 
&\left. \left. \left. \left.-2 \text{Tr}\left[c_{{\phi u}}\right]\right) g_1^2+c_{{\phi\square}} \left(13 g_1^2-28 g_2^2+480 \lambda \right)-3 g_2^2 \left(\text{Tr}\left[c^{(3)}_{{\phi l}}\right]+3 \text{Tr}\left[c^{(3)}_{{\phi q}}\right]\right)\right) \right.\right.\nonumber\\
&\left.\left. -3 \left(3 \text{Tr}\left[c_{{d\phi }}y^{{d\dagger }}\right]+\text{Tr}\left[c_{{e\phi }}y^{{l\dagger }}\right]+3 \text{Tr}\left[c_{{u\phi }}y^{{u\dagger }}\right]+3 \text{Tr}\left[\left(c_{{d\phi }}\right){}^{\dagger }y^d\right]+\text{Tr}\left[\left(c_{{e\phi }}\right){}^{\dagger }y^l\right]\right.\right.\right. \nonumber \\
&\left.\left.\left.+3 \text{Tr}\left[\left(c_{{u\phi }}\right){}^{\dagger }y^u\right]-2 \left(2 \text{Tr}\left[c^{(3)}_{{\phi l}}y^ly^{{l\dagger }}\right]+6 \text{Tr}\left[c^{(3)}_{{\phi q}}y^dy^{{d\dagger }}\right]+6 \text{Tr}\left[c^{(3)}_{{\phi q}}y^uy^{{u\dagger }}\right] \right.\right.\right.\right. \nonumber \\
&\left.\left.\left.\left. -3 \left(\text{Tr}\left[c_{{\phi ud}}y^{{d\dagger }}y^u\right]+\text{Tr}\left[\left(c_{{\phi ud}}\right){}^{\dagger }y^{{u\dagger }}y^d\right]\right)\right)\right)\right) c_{{\phi D}}+72 c_{\phi } \left(17 c_{{\phi D}}-74 c_{{\phi\square}}\right)\right.\nonumber \\
&\left.+8 \left(20 \left(g_1^2-3 g_2^2+96 \lambda \right) c_{{\phi\square}}^2-8 \left(9 \text{Tr}\left[c_{{d\phi }}y^{{d\dagger }}\right]+3 \text{Tr}\left[c_{{e\phi }}y^{{l\dagger }}\right]+9 \text{Tr}\left[c_{{u\phi }}y^{{u\dagger }}\right] \right.\right.\right. \nonumber \\
& \left.\left.\left.+9 \text{Tr}\left[\left(c_{{d\phi }}\right){}^{\dagger }y^d\right]+3 \text{Tr}\left[\left(c_{{e\phi }}\right){}^{\dagger }y^l\right]+9 \text{Tr}\left[\left(c_{{u\phi }}\right){}^{\dagger }y^u\right]+2 \left(2 \left(\text{Tr}\left[c^{(3)}_{{\phi l}}\right]+3 \text{Tr}\left[c^{(3)}_{{\phi q}}\right]\right) g_2^2\right.\right.\right.\right. \nonumber \\
&\left.\left.\left.\left.-6 \text{Tr}\left[c^{(3)}_{{\phi l}}y^ly^{{l\dagger }}\right]-18 \text{Tr}\left[c^{(3)}_{{\phi q}}y^dy^{{d\dagger }}\right]-18 \text{Tr}\left[c^{(3)}_{{\phi q}}y^uy^{{u\dagger }}\right]+9 \text{Tr}\left[c_{{\phi ud}}y^{{d\dagger }}y^u\right]\right.\right.\right.\right. \nonumber \\
&\left.\left.\left.\left.+9 \text{Tr}\left[\left(c_{{\phi ud}}\right){}^{\dagger }y^{{u\dagger }}y^d\right]\right)\right) c_{{\phi\square}}+8 g_1^2 \left(\text{Tr}\left[c^{(3)}_{{\phi l}}\left(c^{(3)}_{{\phi l}}\right){}^{\dagger }\right]+3 \text{Tr}\left[c^{(3)}_{{\phi q}}\left(c^{(3)}_{{\phi q}}\right){}^{\dagger }\right]\right)\right.\right. \nonumber \\
&\left.\left. +2 g_2^2 \left(3 \text{Tr}\left[c_{{\phi d}}\left(c_{{\phi d}}\right){}^{\dagger }\right]+\text{Tr}\left[c_{{\phi e}}\left(c_{{\phi e}}\right){}^{\dagger }\right]+2 \text{Tr}\left[c^{(1)}_{{\phi l}}\left(c^{(1)}_{{\phi l}}\right){}^{\dagger }\right]-2 \text{Tr}\left[c^{(3)}_{{\phi l}}\left(c^{(3)}_{{\phi l}}\right){}^{\dagger }\right]\right.\right.\right. \nonumber \\
&\left.\left.\left.+6 \text{Tr}\left[c^{(1)}_{{\phi q}}\left(c^{(1)}_{{\phi q}}\right){}^{\dagger }\right]-6 \text{Tr}\left[c^{(3)}_{{\phi q}}\left(c^{(3)}_{{\phi q}}\right){}^{\dagger }\right]+3 \text{Tr}\left[c_{{\phi u}}\left(c_{{\phi u}}\right){}^{\dagger }\right]\right)\right.\right. \nonumber \\
&\left.\left.+3 \left(g_2^2-g_1^2\right) \text{Tr}\left[c_{{\phi ud}}\left(c_{{\phi ud}}\right){}^{\dagger }\right]+6 \left(6 \text{Tr}\left[c_{{d\phi }}\left(c_{{d\phi }}\right){}^{\dagger }\right]+2 \text{Tr}\left[c_{{e\phi }}\left(c_{{e\phi }}\right){}^{\dagger }\right]\right.\right. \right.\nonumber \\
&\left.\left.\left.+6 \text{Tr}\left[c_{{u\phi }}\left(c_{{u\phi }}\right){}^{\dagger }\right]+3 \text{Tr}\left[c_{{d\phi }}y^{{d\dagger }}c^{(1)}_{{\phi q}}\right]+3 \text{Tr}\left[c_{{d\phi }}y^{{d\dagger }}c^{(3)}_{{\phi q}}\right]-3 \text{Tr}\left[c_{{d\phi }}c_{{\phi d}}y^{{d\dagger }}\right]\right.\right. \right.\nonumber \\
&\left.\left.\left.+\text{Tr}\left[c_{{e\phi }}y^{{l\dagger }}c^{(1)}_{{\phi l}}\right]+\text{Tr}\left[c_{{e\phi }}y^{{l\dagger }}c^{(3)}_{{\phi l}}\right]-\text{Tr}\left[c_{{e\phi }}c_{{\phi e}}y^{{l\dagger }}\right]-3 \text{Tr}\left[c^{(1)}_{{\phi q}}y^u\left(c_{{u\phi }}\right){}^{\dagger }\right]\right.\right. \right.\nonumber \\
&\left.\left.\left.-3 \text{Tr}\left[c^{(1)}_{{\phi q}}c_{{u\phi }}y^{{u\dagger }}\right]+3 \text{Tr}\left[c^{(3)}_{{\phi q}}y^u\left(c_{{u\phi }}\right){}^{\dagger }\right]+3 \text{Tr}\left[c^{(3)}_{{\phi q}}c_{{u\phi }}y^{{u\dagger }}\right]+3 \text{Tr}\left[c_{{\phi u}}y^{{u\dagger }}c_{{u\phi }}\right]\right.\right. \right.\nonumber \\
&\left.\left.\left.+3 \text{Tr}\left[c_{{\phi u}}\left(c_{{u\phi }}\right){}^{\dagger }y^u\right]-3 \text{Tr}\left[\left(c_{{d\phi }}\right){}^{\dagger }y^dc_{{\phi d}}\right]+3 \text{Tr}\left[\left(c_{{d\phi }}\right){}^{\dagger }c^{(1)}_{{\phi q}}y^d\right]+3 \text{Tr}\left[\left(c_{{d\phi }}\right){}^{\dagger }c^{(3)}_{{\phi q}}y^d\right]\right.\right. \right.\nonumber \\
&\left.\left.\left.-\text{Tr}\left[\left(c_{{e\phi }}\right){}^{\dagger }y^lc_{{\phi e}}\right]+\text{Tr}\left[\left(c_{{e\phi }}\right){}^{\dagger }c^{(1)}_{{\phi l}}y^l\right]+\text{Tr}\left[\left(c_{{e\phi }}\right){}^{\dagger }c^{(3)}_{{\phi l}}y^l\right]-3 \text{Tr}\left[c_{{\phi d}}y^{{d\dagger }}y^dc_{{\phi d}}\right]\right.\right. \right.\nonumber \\
&\left.\left.\left.+6 \text{Tr}\left[c_{{\phi d}}y^{{d\dagger }}c^{(1)}_{{\phi q}}y^d\right]+6 \text{Tr}\left[c_{{\phi d}}y^{{d\dagger }}c^{(3)}_{{\phi q}}y^d\right]-\text{Tr}\left[c_{{\phi e}}y^{{l\dagger }}y^lc_{{\phi e}}\right]+2 \text{Tr}\left[c_{{\phi e}}y^{{l\dagger }}c^{(1)}_{{\phi l}}y^l\right]\right.\right. \right.\nonumber \\
&\left.\left.\left.+2 \text{Tr}\left[c_{{\phi e}}y^{{l\dagger }}c^{(3)}_{{\phi l}}y^l\right]-\text{Tr}\left[c^{(1)}_{{\phi l}}y^ly^{{l\dagger }}c^{(1)}_{{\phi l}}\right]-\text{Tr}\left[c^{(1)}_{{\phi l}}y^ly^{{l\dagger }}c^{(3)}_{{\phi l}}\right]-\text{Tr}\left[c^{(1)}_{{\phi l}}c^{(3)}_{{\phi l}}y^ly^{{l\dagger }}\right]\right.\right. \right.\nonumber \\
&\left.\left.\left.-\text{Tr}\left[c^{(3)}_{{\phi l}}y^ly^{{l\dagger }}c^{(3)}_{{\phi l}}\right]-3 \left(\text{Tr}\left[c^{(1)}_{{\phi q}}y^dy^{{d\dagger }}c^{(1)}_{{\phi q}}\right]+\text{Tr}\left[c^{(1)}_{{\phi q}}y^dy^{{d\dagger }}c^{(3)}_{{\phi q}}\right]+\text{Tr}\left[c^{(1)}_{{\phi q}}y^uy^{{u\dagger }}c^{(1)}_{{\phi q}}\right]\right.\right. \right.\right.\nonumber \\
&\left.\left.\left.\left.-\text{Tr}\left[c^{(1)}_{{\phi q}}y^uy^{{u\dagger }}c^{(3)}_{{\phi q}}\right]-2 \text{Tr}\left[c^{(1)}_{{\phi q}}y^uc_{{\phi u}}y^{{u\dagger }}\right]+\text{Tr}\left[c^{(1)}_{{\phi q}}c^{(3)}_{{\phi q}}y^dy^{{d\dagger }}\right]-\text{Tr}\left[c^{(1)}_{{\phi q}}c^{(3)}_{{\phi q}}y^uy^{{u\dagger }}\right]\right.\right. \right.\right.\nonumber \\
&\left.\left.\left.\left.+\text{Tr}\left[c^{(3)}_{{\phi q}}y^dy^{{d\dagger }}c^{(3)}_{{\phi q}}\right]+\text{Tr}\left[c^{(3)}_{{\phi q}}y^uy^{{u\dagger }}c^{(3)}_{{\phi q}}\right]+2 \text{Tr}\left[c^{(3)}_{{\phi q}}y^uc_{{\phi u}}y^{{u\dagger }}\right]\right.\right. \right.\right.\nonumber \\
&\left.\left.\left.\left.+\text{Tr}\left[c_{{\phi u}}y^{{u\dagger }}y^uc_{{\phi u}}\right]\right)\right)\right)\right\}\,,
\end{align}
\begin{align}
\dot{c}_{\phi^{8}}& = -60 c_{{\phi D}}^2 \lambda ^2-864 c_{{\phi \square}}^2 \lambda ^2+432 c_{{\phi D}} c_{{\phi \square}} \lambda ^2-4 \left(c_{{\phi D}}^2-10 c_{{\phi \square}} c_{{\phi D}}+32 c_{{\phi \square}}^2\right) \lambda ^2\nonumber \\
&+48 c_{\phi } \left(15 c_{{\phi \square}}-4 c_{{\phi D}}\right) \lambda +\frac{1}{12} g_1^2 \left(3 c_{{\phi D}}^2-4 c_{{\phi \square}} c_{{\phi D}}-8 \left(2 c_{{\phi \square}}^2+8 \text{Tr}\left[\left(c^{(3)}_{{\phi l}}\right){}^{\dagger }c^{(3)}_{{\phi l}}\right]\right.\right. \nonumber \\
&\left.\left.+24 \text{Tr}\left[\left(c^{(3)}_{{\phi q}}\right){}^{\dagger }c^{(3)}_{{\phi q}}\right]-3 \text{Tr}\left[\left(c_{{\phi ud}}\right){}^{\dagger }c_{{\phi ud}}\right]\right)\right) \lambda +\frac{1}{6} g_2^2 \left(c_{{\phi D}}^2-12 c_{{\phi \square}} c_{{\phi D}}\right. \nonumber \\
&\left.-8 \left(3 \text{Tr}\left[\left(c_{{\phi d}}\right){}^{\dagger }c_{{\phi d}}\right]+\text{Tr}\left[\left(c_{{\phi e}}\right){}^{\dagger }c_{{\phi e}}\right]+2 \text{Tr}\left[\left(c^{(1)}_{{\phi l}}\right){}^{\dagger }c^{(1)}_{{\phi l}}\right]+6 \text{Tr}\left[\left(c^{(1)}_{{\phi q}}\right){}^{\dagger }c^{(1)}_{{\phi q}}\right]\right.\right. \nonumber \\
&\left.\left.+3 \text{Tr}\left[\left(c_{{\phi u}}\right){}^{\dagger }c_{{\phi u}}\right]\right)+12 \text{Tr}\left[\left(c_{{\phi ud}}\right){}^{\dagger }c_{{\phi ud}}\right]\right) \lambda -\frac{1}{12} g_2^2 \left(5 c_{{\phi D}}^2-20 c_{{\phi \square}} c_{{\phi D}}\right.\nonumber \\
&\left.+16 \left(c_{{\phi \square}}^2-2 \text{Tr}\left[\left(c^{(3)}_{{\phi l}}\right){}^{\dagger }c^{(3)}_{{\phi l}}\right]-6 \text{Tr}\left[\left(c^{(3)}_{{\phi q}}\right){}^{\dagger }c^{(3)}_{{\phi q}}\right]+3 \text{Tr}\left[\left(c_{{\phi ud}}\right){}^{\dagger }c_{{\phi ud}}\right]\right)\right) \lambda \nonumber\\
&+\frac{1}{2} \left(-3 \left(g_2^2+4 \lambda \right) c_{{\phi D}}^2-4 c_{{\phi \square}} \left(3 \left(g_1^2+g_2^2\right)-112 \lambda \right) c_{{\phi D}}-48 c_{\phi } \left(c_{{\phi D}}-9 c_{{\phi \square}}\right)\right.\nonumber\\
&\left.-8 c_{{\phi \square}}^2 \left(128 \lambda +3 \left(g_1^2+g_2^2\right)\right)+8 \left(-6 \text{Tr}\left[\left(c_{{d\phi }}\right){}^{\dagger }c_{{d\phi }}\right]-2 \text{Tr}\left[\left(c_{{e\phi }}\right){}^{\dagger }c_{{e\phi }}\right]-6 \text{Tr}\left[\left(c_{{u\phi }}\right){}^{\dagger }c_{{u\phi }}\right]\right.\right.\nonumber \\ 
&\left.\left.-3 \text{Tr}\left[y^d\left(c_{{d\phi }}\right){}^{\dagger }c^{(1)}_{{\phi q}}\right]-3 \text{Tr}\left[y^d\left(c_{{d\phi }}\right){}^{\dagger }c^{(3)}_{{\phi q}}\right]+3 \text{Tr}\left[y^{{d\dagger }}c_{{d\phi }}c_{{\phi d}}\right]-\text{Tr}\left[y^l\left(c_{{e\phi }}\right){}^{\dagger }c^{(1)}_{{\phi l}}\right]\right.\right.\nonumber \\ 
&\left.\left.-\text{Tr}\left[y^l\left(c_{{e\phi }}\right){}^{\dagger }c^{(3)}_{{\phi l}}\right]+\text{Tr}\left[y^{{l\dagger }}c_{{e\phi }}c_{{\phi e}}\right]-3 \text{Tr}\left[y^uc_{{\phi u}}\left(c_{{u\phi }}\right){}^{\dagger }\right]+3 \text{Tr}\left[y^{{u\dagger }}c^{(1)}_{{\phi q}}c_{{u\phi }}\right]\right.\right.\nonumber \\ 
&\left.\left.-3 \text{Tr}\left[y^{{u\dagger }}c^{(3)}_{{\phi q}}c_{{u\phi }}\right]-3 \text{Tr}\left[c_{{u\phi }}c_{{\phi u}}y^{{u\dagger }}\right]+3 \text{Tr}\left[c_{{\phi d}}\left(c_{{d\phi }}\right){}^{\dagger }y^d\right]+\text{Tr}\left[c_{{\phi e}}\left(c_{{e\phi }}\right){}^{\dagger }y^l\right]\right.\right.\nonumber \\ 
&\left.\left.-\text{Tr}\left[c^{(1)}_{{\phi l}}c_{{e\phi }}y^{{l\dagger }}\right]-\text{Tr}\left[c^{(3)}_{{\phi l}}c_{{e\phi }}y^{{l\dagger }}\right]-3 \text{Tr}\left[c^{(1)}_{{\phi q}}c_{{d\phi }}y^{{d\dagger }}\right]-3 \text{Tr}\left[c^{(3)}_{{\phi q}}c_{{d\phi }}y^{{d\dagger }}\right]\right.\right.\nonumber \\ 
&\left.\left.+3 \text{Tr}\left[\left(c_{{u\phi }}\right){}^{\dagger }c^{(1)}_{{\phi q}}y^u\right]-3 \text{Tr}\left[\left(c_{{u\phi }}\right){}^{\dagger }c^{(3)}_{{\phi q}}y^u\right]-6 \text{Tr}\left[y^dc_{{\phi d}}y^{{d\dagger }}c^{(1)}_{{\phi q}}\right]-6 \text{Tr}\left[y^dc_{{\phi d}}y^{{d\dagger }}c^{(3)}_{{\phi q}}\right]\right.\right.\nonumber \\ 
&\left.\left.+3 \text{Tr}\left[y^dc_{{\phi d}}c_{{\phi d}}y^{{d\dagger }}\right]-2 \text{Tr}\left[y^lc_{{\phi e}}y^{{l\dagger }}c^{(1)}_{{\phi l}}\right]-2 \text{Tr}\left[y^lc_{{\phi e}}y^{{l\dagger }}c^{(3)}_{{\phi l}}\right]+\text{Tr}\left[y^lc_{{\phi e}}c_{{\phi e}}y^{{l\dagger }}\right]\right.\right.\nonumber \\ 
&\left.\left.+\text{Tr}\left[y^{{l\dagger }}c^{(1)}_{{\phi l}}c^{(1)}_{{\phi l}}y^l\right]+\text{Tr}\left[y^{{l\dagger }}c^{(1)}_{{\phi l}}c^{(3)}_{{\phi l}}y^l\right]+\text{Tr}\left[y^{{l\dagger }}c^{(3)}_{{\phi l}}c^{(1)}_{{\phi l}}y^l\right]+\text{Tr}\left[y^{{l\dagger }}c^{(3)}_{{\phi l}}c^{(3)}_{{\phi l}}y^l\right]\right.\right.\nonumber \\ 
&\left.\left.+3 \left(\text{Tr}\left[y^{{d\dagger }}c^{(1)}_{{\phi q}}c^{(1)}_{{\phi q}}y^d\right]+\text{Tr}\left[y^{{d\dagger }}c^{(1)}_{{\phi q}}c^{(3)}_{{\phi q}}y^d\right]+\text{Tr}\left[y^{{d\dagger }}c^{(3)}_{{\phi q}}c^{(1)}_{{\phi q}}y^d\right]+\text{Tr}\left[y^{{d\dagger }}c^{(3)}_{{\phi q}}c^{(3)}_{{\phi q}}y^d\right]\right.\right.\right.\nonumber \\ 
&\left.\left.\left.+\text{Tr}\left[y^uc_{{\phi u}}c_{{\phi u}}y^{{u\dagger }}\right]-2 \text{Tr}\left[y^{{u\dagger }}c^{(1)}_{{\phi q}}y^uc_{{\phi u}}\right]+\text{Tr}\left[y^{{u\dagger }}c^{(1)}_{{\phi q}}c^{(1)}_{{\phi q}}y^u\right]-\text{Tr}\left[y^{{u\dagger }}c^{(1)}_{{\phi q}}c^{(3)}_{{\phi q}}y^u\right]\right.\right.\right.\nonumber \\ 
&\left.\left.\left.+2 \text{Tr}\left[y^{{u\dagger }}c^{(3)}_{{\phi q}}y^uc_{{\phi u}}\right]-\text{Tr}\left[y^{{u\dagger }}c^{(3)}_{{\phi q}}c^{(1)}_{{\phi q}}y^u\right]+\text{Tr}\left[y^{{u\dagger }}c^{(3)}_{{\phi q}}c^{(3)}_{{\phi q}}y^u\right]\right)\right)\right) \lambda \nonumber \\
& -\frac{1}{3} c_{{\phi D}} g_2^2 \left(c_{{\phi \square}}+4 \left(\text{Tr}\left[c^{(3)}_{{\phi l}}\right]+3 \text{Tr}\left[c^{(3)}_{{\phi q}}\right]\right)\right) \lambda -\frac{1}{3} c_{{\phi D}} g_1^2 \left(c_{{\phi D}}+c_{{\phi \square}}-4 \left(\text{Tr}\left[c_{{\phi d}}\right]\right.\right.\nonumber \\ 
&\left.\left.+\text{Tr}\left[c_{{\phi e}}\right]+\text{Tr}\left[c^{(1)}_{{\phi l}}\right]-\text{Tr}\left[c^{(1)}_{{\phi q}}\right]-2 \text{Tr}\left[c_{{\phi u}}\right]\right)\right) \lambda -126 c_{\phi }^2-\frac{3}{8} c_{{\phi D}}^2 \left(g_1^2+g_2^2\right){}^2\nonumber \\ 
&+6 \left(\text{Tr}\left[y^d\left(c_{{d\phi }}\right){}^{\dagger }y^d\left(c_{{d\phi }}\right){}^{\dagger }\right]+2 \text{Tr}\left[y^d\left(c_{{d\phi }}\right){}^{\dagger }c_{{d\phi }}y^{{d\dagger }}\right]+\text{Tr}\left[y^{{d\dagger }}c_{{d\phi }}y^{{d\dagger }}c_{{d\phi }}\right]\right.\nonumber \\
&\left.+2 \text{Tr}\left[y^{{d\dagger }}c_{{d\phi }}\left(c_{{d\phi }}\right){}^{\dagger }y^d\right]\right)+2 \text{Tr}\left[y^l\left(c_{{e\phi }}\right){}^{\dagger }y^l\left(c_{{e\phi }}\right){}^{\dagger }\right]+4 \text{Tr}\left[y^l\left(c_{{e\phi }}\right){}^{\dagger }c_{{e\phi }}y^{{l\dagger }}\right]\nonumber \\
&+2 \text{Tr}\left[y^{{l\dagger }}c_{{e\phi }}y^{{l\dagger }}c_{{e\phi }}\right]+4 \text{Tr}\left[y^{{l\dagger }}c_{{e\phi }}\left(c_{{e\phi }}\right){}^{\dagger }y^l\right]+6 \left(\text{Tr}\left[y^u\left(c_{{u\phi }}\right){}^{\dagger }y^u\left(c_{{u\phi }}\right){}^{\dagger }\right]\right.\nonumber \\
&\left.+2 \text{Tr}\left[y^u\left(c_{{u\phi }}\right){}^{\dagger }c_{{u\phi }}y^{{u\dagger }}\right]+\text{Tr}\left[y^{{u\dagger }}c_{{u\phi }}y^{{u\dagger }}c_{{u\phi }}\right]+2 \text{Tr}\left[y^{{u\dagger }}c_{{u\phi }}\left(c_{{u\phi }}\right){}^{\dagger }y^u\right]\right)\nonumber \\
&-\frac{1}{3} \left(3 c_{\phi }+\lambda \left(c_{{\phi D}}-8 c_{{\phi \square}}\right)\right) \left(8 \left(\text{Tr}\left[c^{(3)}_{{\phi l}}\right]+3 \text{Tr}\left[c^{(3)}_{{\phi q}}\right]\right) g_2^2+3 c_{{\phi D}} \left(3 g_1^2-3 g_2^2+4 \lambda \right)\right.\nonumber \\
&\left.+4 c_{{\phi \square}} \left(5 g_2^2-6 \lambda \right)+6 \left(3 \text{Tr}\left[y^d\left(c_{{d\phi }}\right){}^{\dagger }\right]+3 \text{Tr}\left[y^{{d\dagger }}c_{{d\phi }}\right]+\text{Tr}\left[y^l\left(c_{{e\phi }}\right){}^{\dagger }\right]+\text{Tr}\left[y^{{l\dagger }}c_{{e\phi }}\right]\right.\right.\nonumber \\
&\left.\left.+3 \text{Tr}\left[y^u\left(c_{{u\phi }}\right){}^{\dagger }\right]+3 \text{Tr}\left[y^{{u\dagger }}c_{{u\phi }}\right]+6 \text{Tr}\left[y^d\left(c_{{\phi ud}}\right){}^{\dagger }y^{{u\dagger }}\right]-12 \text{Tr}\left[y^{{d\dagger }}c^{(3)}_{{\phi q}}y^d\right]\right.\right.\nonumber \\
&\left.\left.-4 \text{Tr}\left[y^{{l\dagger }}c^{(3)}_{{\phi l}}y^l\right]+6 \text{Tr}\left[y^uc_{{\phi ud}}y^{{d\dagger }}\right]-12 \text{Tr}\left[y^{{u\dagger }}c^{(3)}_{{\phi q}}y^u\right]\right)\right)\,,
\end{align}
\begin{align}
\dot{c}_{\phi^{6}}^{(1)}& =\left(\frac{g_1^2}{8}+\frac{161 g_2^2}{24}-23 \lambda \right) c_{{\phi D}}^2-\frac{1}{3} \left(6 \left(\text{Tr}\left[c_{{\phi d}}\right]+\text{Tr}\left[c_{{\phi e}}\right]+\text{Tr}\left[c^{(1)}_{{\phi l}}\right]-\text{Tr}\left[c^{(1)}_{{\phi q}}\right]\right.\right. \nonumber \\
&\left.\left.-2 \text{Tr}\left[c_{{\phi u}}\right]\right) g_1^2+6 c_{{\phi \square}} \left(g_1^2-6 g_2^2+32 \lambda \right)+3 \left(3 \text{Tr}\left[y^d\left(c_{{d\phi }}\right){}^{\dagger }\right]+3 \text{Tr}\left[y^{{d\dagger }}c_{{d\phi }}\right]\right.\right. \nonumber \\
&\left.\left.+\text{Tr}\left[y^l\left(c_{{e\phi }}\right){}^{\dagger }\right]+\text{Tr}\left[y^{{l\dagger }}c_{{e\phi }}\right]+3 \text{Tr}\left[y^u\left(c_{{u\phi }}\right){}^{\dagger }\right]+3 \text{Tr}\left[y^{{u\dagger }}c_{{u\phi }}\right]+6 \text{Tr}\left[y^d\left(c_{{\phi ud}}\right){}^{\dagger }y^{{u\dagger }}\right]\right.\right. \nonumber \\
&\left.\left.-12 \text{Tr}\left[y^{{d\dagger }}c^{(3)}_{{\phi q}}y^d\right]-4 \text{Tr}\left[y^{{l\dagger }}c^{(3)}_{{\phi l}}y^l\right]+6 \text{Tr}\left[y^uc_{{\phi ud}}y^{{d\dagger }}\right]-12 \text{Tr}\left[y^{{u\dagger }}c^{(3)}_{{\phi q}}y^u\right]\right)\right. \nonumber \\
&\left.-2 g_2^2 \left(\text{Tr}\left[c^{(3)}_{{\phi l}}\right]+3 \text{Tr}\left[c^{(3)}_{{\phi q}}\right]\right)\right) c_{{\phi D}}+352 \lambda c_{{\phi \square}}^2+20 c_{{\phi \square}}^2 g_1^2+\frac{20}{3} c_{{\phi \square}}^2 g_2^2-12 c_{\phi } \left(c_{{\phi D}}+8 c_{{\phi \square}}\right) \nonumber \\
&-24 c_{{\phi \square}} \text{Tr}\left[y^d\left(c_{{d\phi }}\right){}^{\dagger }\right]-24 c_{{\phi \square}} \text{Tr}\left[y^{{d\dagger }}c_{{d\phi }}\right]-8 c_{{\phi \square}} \text{Tr}\left[y^l\left(c_{{e\phi }}\right){}^{\dagger }\right]-8 c_{{\phi \square}} \text{Tr}\left[y^{{l\dagger }}c_{{e\phi }}\right] \nonumber \\
&-24 c_{{\phi \square}} \text{Tr}\left[y^u\left(c_{{u\phi }}\right){}^{\dagger }\right]-24 c_{{\phi \square}} \text{Tr}\left[y^{{u\dagger }}c_{{u\phi }}\right]+6 \text{Tr}\left[\left(c_{{d\phi }}\right){}^{\dagger }c_{{d\phi }}\right]+2 \text{Tr}\left[\left(c_{{e\phi }}\right){}^{\dagger }c_{{e\phi }}\right] \nonumber \\
&+6 \text{Tr}\left[\left(c_{{u\phi }}\right){}^{\dagger }c_{{u\phi }}\right]+8 g_2^2 \text{Tr}\left[\left(c_{{\phi d}}\right){}^{\dagger }c_{{\phi d}}\right]+\frac{8}{3} g_2^2 \text{Tr}\left[\left(c_{{\phi e}}\right){}^{\dagger }c_{{\phi e}}\right]+\frac{16}{3} g_2^2 \text{Tr}\left[\left(c^{(1)}_{{\phi l}}\right){}^{\dagger }c^{(1)}_{{\phi l}}\right] \nonumber \\
&+8 g_1^2 \text{Tr}\left[\left(c^{(3)}_{{\phi l}}\right){}^{\dagger }c^{(3)}_{{\phi l}}\right]-\frac{20}{3} g_2^2 \text{Tr}\left[\left(c^{(3)}_{{\phi l}}\right){}^{\dagger }c^{(3)}_{{\phi l}}\right]+16 g_2^2 \text{Tr}\left[\left(c^{(1)}_{{\phi q}}\right){}^{\dagger }c^{(1)}_{{\phi q}}\right] \nonumber \\
&+24 g_1^2 \text{Tr}\left[\left(c^{(3)}_{{\phi q}}\right){}^{\dagger }c^{(3)}_{{\phi q}}\right]-20 g_2^2 \text{Tr}\left[\left(c^{(3)}_{{\phi q}}\right){}^{\dagger }c^{(3)}_{{\phi q}}\right]+8 g_2^2 \text{Tr}\left[\left(c_{{\phi u}}\right){}^{\dagger }c_{{\phi u}}\right] \nonumber \\
&-3 g_1^2 \text{Tr}\left[\left(c_{{\phi ud}}\right){}^{\dagger }c_{{\phi ud}}\right]+6 g_2^2 \text{Tr}\left[\left(c_{{\phi ud}}\right){}^{\dagger }c_{{\phi ud}}\right]+18 \text{Tr}\left[y^d\left(c_{{d\phi }}\right){}^{\dagger }c^{(1)}_{{\phi q}}\right] \nonumber \\
&+30 \text{Tr}\left[y^d\left(c_{{d\phi }}\right){}^{\dagger }c^{(3)}_{{\phi q}}\right]-48 c_{{\phi \square}} \text{Tr}\left[y^d\left(c_{{\phi ud}}\right){}^{\dagger }y^{{u\dagger }}\right]-6 \text{Tr}\left[y^d\left(c_{{\phi ud}}\right){}^{\dagger }\left(c_{{u\phi }}\right){}^{\dagger }\right] \nonumber \\
&-18 \text{Tr}\left[y^{{d\dagger }}c_{{d\phi }}c_{{\phi d}}\right]+96 c_{{\phi \square}} \text{Tr}\left[y^{{d\dagger }}c^{(3)}_{{\phi q}}y^d\right]+6 \text{Tr}\left[y^l\left(c_{{e\phi }}\right){}^{\dagger }c^{(1)}_{{\phi l}}\right]+10 \text{Tr}\left[y^l\left(c_{{e\phi }}\right){}^{\dagger }c^{(3)}_{{\phi l}}\right] \nonumber \\
&-6 \text{Tr}\left[y^{{l\dagger }}c_{{e\phi }}c_{{\phi e}}\right]+32 c_{{\phi \square}} \text{Tr}\left[y^{{l\dagger }}c^{(3)}_{{\phi l}}y^l\right]+18 \text{Tr}\left[y^uc_{{\phi u}}\left(c_{{u\phi }}\right){}^{\dagger }\right]-48 c_{{\phi \square}} \text{Tr}\left[y^uc_{{\phi ud}}y^{{d\dagger }}\right] \nonumber \\
&-6 \text{Tr}\left[y^{{u\dagger }}c_{{d\phi }}\left(c_{{\phi ud}}\right){}^{\dagger }\right]-18 \text{Tr}\left[y^{{u\dagger }}c^{(1)}_{{\phi q}}c_{{u\phi }}\right]+96 c_{{\phi \square}} \text{Tr}\left[y^{{u\dagger }}c^{(3)}_{{\phi q}}y^u\right] \nonumber \\
&+30 \text{Tr}\left[y^{{u\dagger }}c^{(3)}_{{\phi q}}c_{{u\phi }}\right]+18 \text{Tr}\left[c_{{u\phi }}c_{{\phi u}}y^{{u\dagger }}\right]-6 \text{Tr}\left[c_{{u\phi }}c_{{\phi ud}}y^{{d\dagger }}\right]-18 \text{Tr}\left[c_{{\phi d}}\left(c_{{d\phi }}\right){}^{\dagger }y^d\right] \nonumber \\
&-6 \text{Tr}\left[c_{{\phi e}}\left(c_{{e\phi }}\right){}^{\dagger }y^l\right]+6 \text{Tr}\left[c^{(1)}_{{\phi l}}c_{{e\phi }}y^{{l\dagger }}\right]+10 \text{Tr}\left[c^{(3)}_{{\phi l}}c_{{e\phi }}y^{{l\dagger }}\right]+18 \text{Tr}\left[c^{(1)}_{{\phi q}}c_{{d\phi }}y^{{d\dagger }}\right] \nonumber \\
&+30 \text{Tr}\left[c^{(3)}_{{\phi q}}c_{{d\phi }}y^{{d\dagger }}\right]-6 \text{Tr}\left[c_{{\phi ud}}\left(c_{{d\phi }}\right){}^{\dagger }y^u\right]-18 \text{Tr}\left[\left(c_{{u\phi }}\right){}^{\dagger }c^{(1)}_{{\phi q}}y^u\right] \nonumber \\
&+30 \text{Tr}\left[\left(c_{{u\phi }}\right){}^{\dagger }c^{(3)}_{{\phi q}}y^u\right]+36 \text{Tr}\left[y^dc_{{\phi d}}y^{{d\dagger }}c^{(1)}_{{\phi q}}\right]+36 \text{Tr}\left[y^dc_{{\phi d}}y^{{d\dagger }}c^{(3)}_{{\phi q}}\right] \nonumber \\
&-18 \text{Tr}\left[y^dc_{{\phi d}}c_{{\phi d}}y^{{d\dagger }}\right]-3 \text{Tr}\left[y^d\left(c_{{\phi ud}}\right){}^{\dagger }c_{{\phi ud}}y^{{d\dagger }}\right]-18 \text{Tr}\left[y^{{d\dagger }}c^{(1)}_{{\phi q}}c^{(1)}_{{\phi q}}y^d\right] \nonumber \\
&-18 \text{Tr}\left[y^{{d\dagger }}c^{(1)}_{{\phi q}}c^{(3)}_{{\phi q}}y^d\right]+12 \text{Tr}\left[y^{{d\dagger }}c^{(3)}_{{\phi q}}y^uc_{{\phi ud}}\right]-18 \text{Tr}\left[y^{{d\dagger }}c^{(3)}_{{\phi q}}c^{(1)}_{{\phi q}}y^d\right] \nonumber \\
&-30 \text{Tr}\left[y^{{d\dagger }}c^{(3)}_{{\phi q}}c^{(3)}_{{\phi q}}y^d\right]+12 \text{Tr}\left[y^lc_{{\phi e}}y^{{l\dagger }}c^{(1)}_{{\phi l}}\right]+12 \text{Tr}\left[y^lc_{{\phi e}}y^{{l\dagger }}c^{(3)}_{{\phi l}}\right] \nonumber \\
&-6 \text{Tr}\left[y^lc_{{\phi e}}c_{{\phi e}}y^{{l\dagger }}\right]-6 \text{Tr}\left[y^{{l\dagger }}c^{(1)}_{{\phi l}}c^{(1)}_{{\phi l}}y^l\right]-6 \text{Tr}\left[y^{{l\dagger }}c^{(1)}_{{\phi l}}c^{(3)}_{{\phi l}}y^l\right]-6 \text{Tr}\left[y^{{l\dagger }}c^{(3)}_{{\phi l}}c^{(1)}_{{\phi l}}y^l\right] \nonumber \\
&-10 \text{Tr}\left[y^{{l\dagger }}c^{(3)}_{{\phi l}}c^{(3)}_{{\phi l}}y^l\right]-3 \left(6 \text{Tr}\left[y^uc_{{\phi u}}c_{{\phi u}}y^{{u\dagger }}\right]+\text{Tr}\left[y^uc_{{\phi ud}}\left(c_{{\phi ud}}\right){}^{\dagger }y^{{u\dagger }}\right] \right.\nonumber \\
&\left.-12 \text{Tr}\left[y^{{u\dagger }}c^{(1)}_{{\phi q}}y^uc_{{\phi u}}\right]+6 \text{Tr}\left[y^{{u\dagger }}c^{(1)}_{{\phi q}}c^{(1)}_{{\phi q}}y^u\right]-6 \text{Tr}\left[y^{{u\dagger }}c^{(1)}_{{\phi q}}c^{(3)}_{{\phi q}}y^u\right]\right.\nonumber \\
&\left.-4 \text{Tr}\left[y^{{u\dagger }}c^{(3)}_{{\phi q}}y^d\left(c_{{\phi ud}}\right){}^{\dagger }\right]+12 \text{Tr}\left[y^{{u\dagger }}c^{(3)}_{{\phi q}}y^uc_{{\phi u}}\right]-6 \text{Tr}\left[y^{{u\dagger }}c^{(3)}_{{\phi q}}c^{(1)}_{{\phi q}}y^u\right]\right.\nonumber \\
&\left.+10 \text{Tr}\left[y^{{u\dagger }}c^{(3)}_{{\phi q}}c^{(3)}_{{\phi q}}y^u\right]\right)-\frac{32}{3} c_{{\phi \square}} g_2^2 \text{Tr}\left[c^{(3)}_{{\phi l}}\right]-32 c_{{\phi \square}} g_2^2 \text{Tr}\left[c^{(3)}_{{\phi q}}\right]\,,
\end{align}
\begin{align}
\dot{c}_{\phi^{6}}^{(2)}& =-18 c_{\phi } c_{{\phi D}}+12 \text{Tr}\left[y^d\left(c_{{d\phi }}\right){}^{\dagger }c^{(1)}_{{\phi q}}\right]+6 \text{Tr}\left[y^d\left(c_{{\phi ud}}\right){}^{\dagger }\left(c_{{u\phi }}\right){}^{\dagger }\right]-12 \text{Tr}\left[y^{{d\dagger }}c_{{d\phi }}c_{{\phi d}}\right]\nonumber \\
&+4 \text{Tr}\left[y^l\left(c_{{e\phi }}\right){}^{\dagger }c^{(1)}_{{\phi l}}\right]-4 \text{Tr}\left[y^{{l\dagger }}c_{{e\phi }}c_{{\phi e}}\right]+12 \text{Tr}\left[y^uc_{{\phi u}}\left(c_{{u\phi }}\right){}^{\dagger }\right]+6 \text{Tr}\left[y^{{u\dagger }}c_{{d\phi }}\left(c_{{\phi ud}}\right){}^{\dagger }\right]\nonumber \\
&-12 \text{Tr}\left[y^{{u\dagger }}c^{(1)}_{{\phi q}}c_{{u\phi }}\right]+12 \text{Tr}\left[c_{{u\phi }}c_{{\phi u}}y^{{u\dagger }}\right]+6 \text{Tr}\left[c_{{u\phi }}c_{{\phi ud}}y^{{d\dagger }}\right]-12 \text{Tr}\left[c_{{\phi d}}\left(c_{{d\phi }}\right){}^{\dagger }y^d\right]\nonumber \\
&-4 \text{Tr}\left[c_{{\phi e}}\left(c_{{e\phi }}\right){}^{\dagger }y^l\right]+4 \text{Tr}\left[c^{(1)}_{{\phi l}}c_{{e\phi }}y^{{l\dagger }}\right]+12 \text{Tr}\left[c^{(1)}_{{\phi q}}c_{{d\phi }}y^{{d\dagger }}\right]+6 \text{Tr}\left[c_{{\phi ud}}\left(c_{{d\phi }}\right){}^{\dagger }y^u\right]\nonumber \\
&-12 \text{Tr}\left[\left(c_{{u\phi }}\right){}^{\dagger }c^{(1)}_{{\phi q}}y^u\right]+24 \text{Tr}\left[y^dc_{{\phi d}}y^{{d\dagger }}c^{(1)}_{{\phi q}}\right]+24 \text{Tr}\left[y^dc_{{\phi d}}y^{{d\dagger }}c^{(3)}_{{\phi q}}\right]\nonumber \\
&-12 \text{Tr}\left[y^dc_{{\phi d}}c_{{\phi d}}y^{{d\dagger }}\right]+3 \text{Tr}\left[y^d\left(c_{{\phi ud}}\right){}^{\dagger }c_{{\phi ud}}y^{{d\dagger }}\right]-12 \text{Tr}\left[y^{{d\dagger }}c^{(1)}_{{\phi q}}c^{(1)}_{{\phi q}}y^d\right]\nonumber \\
&-12 \text{Tr}\left[y^{{d\dagger }}c^{(1)}_{{\phi q}}c^{(3)}_{{\phi q}}y^d\right]-12 \text{Tr}\left[y^{{d\dagger }}c^{(3)}_{{\phi q}}y^uc_{{\phi ud}}\right]-12 \text{Tr}\left[y^{{d\dagger }}c^{(3)}_{{\phi q}}c^{(1)}_{{\phi q}}y^d\right]\nonumber \\
&+8 \text{Tr}\left[y^lc_{{\phi e}}y^{{l\dagger }}c^{(1)}_{{\phi l}}\right]+8 \text{Tr}\left[y^lc_{{\phi e}}y^{{l\dagger }}c^{(3)}_{{\phi l}}\right]-4 \text{Tr}\left[y^lc_{{\phi e}}c_{{\phi e}}y^{{l\dagger }}\right]-4 \text{Tr}\left[y^{{l\dagger }}c^{(1)}_{{\phi l}}c^{(1)}_{{\phi l}}y^l\right]\nonumber \\
&-4 \text{Tr}\left[y^{{l\dagger }}c^{(1)}_{{\phi l}}c^{(3)}_{{\phi l}}y^l\right]-4 \text{Tr}\left[y^{{l\dagger }}c^{(3)}_{{\phi l}}c^{(1)}_{{\phi l}}y^l\right]+3 \left(-4 \text{Tr}\left[y^uc_{{\phi u}}c_{{\phi u}}y^{{u\dagger }}\right]\right.\nonumber \\
&\left.+\text{Tr}\left[y^uc_{{\phi ud}}\left(c_{{\phi ud}}\right){}^{\dagger }y^{{u\dagger }}\right]+4 \left(2 \text{Tr}\left[y^{{u\dagger }}c^{(1)}_{{\phi q}}y^uc_{{\phi u}}\right]-\text{Tr}\left[y^{{u\dagger }}c^{(1)}_{{\phi q}}c^{(1)}_{{\phi q}}y^u\right]\right.\right.\nonumber \\
&\left.\left.+\text{Tr}\left[y^{{u\dagger }}c^{(1)}_{{\phi q}}c^{(3)}_{{\phi q}}y^u\right]-\text{Tr}\left[y^{{u\dagger }}c^{(3)}_{{\phi q}}y^d\left(c_{{\phi ud}}\right){}^{\dagger }\right]-2 \text{Tr}\left[y^{{u\dagger }}c^{(3)}_{{\phi q}}y^uc_{{\phi u}}\right]\right.\right.\nonumber \\
&\left.\left.+\text{Tr}\left[y^{{u\dagger }}c^{(3)}_{{\phi q}}c^{(1)}_{{\phi q}}y^u\right]\right)\right)+\frac{1}{12} \left(\left(-35 g_1^2+44 g_2^2-312 \lambda \right) c_{{\phi D}}^2+8 \left(2 c_{{\phi \square}} \left(5 g_1^2+5 g_2^2+48 \lambda \right)\right.\right.\nonumber \\
&\left.\left.-3 \left(3 \text{Tr}\left[y^d\left(c_{{d\phi }}\right){}^{\dagger }\right]+3 \text{Tr}\left[y^{{d\dagger }}c_{{d\phi }}\right]+\text{Tr}\left[y^l\left(c_{{e\phi }}\right){}^{\dagger }\right]+\text{Tr}\left[y^{{l\dagger }}c_{{e\phi }}\right]+3 \text{Tr}\left[y^u\left(c_{{u\phi }}\right){}^{\dagger }\right]\right.\right.\right.\nonumber \\
&\left.\left.\left.+3 \text{Tr}\left[y^{{u\dagger }}c_{{u\phi }}\right]+6 \text{Tr}\left[y^d\left(c_{{\phi ud}}\right){}^{\dagger }y^{{u\dagger }}\right]-12 \text{Tr}\left[y^{{d\dagger }}c^{(3)}_{{\phi q}}y^d\right]-4 \text{Tr}\left[y^{{l\dagger }}c^{(3)}_{{\phi l}}y^l\right]\right.\right.\right.\nonumber \\
&\left.\left.\left.+6 \text{Tr}\left[y^uc_{{\phi ud}}y^{{d\dagger }}\right]-12 \text{Tr}\left[y^{{u\dagger }}c^{(3)}_{{\phi q}}y^u\right]\right)-2 \left(\left(\text{Tr}\left[c_{{\phi d}}\right]+\text{Tr}\left[c_{{\phi e}}\right]+\text{Tr}\left[c^{(1)}_{{\phi l}}\right]-\text{Tr}\left[c^{(1)}_{{\phi q}}\right]\right.\right.\right.\right.\nonumber \\
&\left.\left.\left.\left.-2 \text{Tr}\left[c_{{\phi u}}\right]\right) g_1^2+g_2^2 \left(\text{Tr}\left[c^{(3)}_{{\phi l}}\right]+3 \text{Tr}\left[c^{(3)}_{{\phi q}}\right]\right)\right)\right) c_{{\phi D}}+32 g_1^2 \left(5 c_{{\phi \square}}^2+2 \text{Tr}\left[\left(c^{(3)}_{{\phi l}}\right){}^{\dagger }c^{(3)}_{{\phi l}}\right]\right.\right.\nonumber \\
&\left.\left.+6 \text{Tr}\left[\left(c^{(3)}_{{\phi q}}\right){}^{\dagger }c^{(3)}_{{\phi q}}\right]\right)+8 g_2^2 \left(3 \text{Tr}\left[\left(c_{{\phi d}}\right){}^{\dagger }c_{{\phi d}}\right]+\text{Tr}\left[\left(c_{{\phi e}}\right){}^{\dagger }c_{{\phi e}}\right]+2 \text{Tr}\left[\left(c^{(1)}_{{\phi l}}\right){}^{\dagger }c^{(1)}_{{\phi l}}\right]\right.\right.\nonumber \\
&\left.\left.+6 \text{Tr}\left[\left(c^{(1)}_{{\phi q}}\right){}^{\dagger }c^{(1)}_{{\phi q}}\right]+3 \text{Tr}\left[\left(c_{{\phi u}}\right){}^{\dagger }c_{{\phi u}}\right]\right)-12 \left(2 g_1^2+g_2^2\right) \text{Tr}\left[\left(c_{{\phi ud}}\right){}^{\dagger }c_{{\phi ud}}\right]\right)\,,
\end{align}
\begin{align}%
\dot{c}_{\phi^{4}}^{(1)}& = \frac{1}{3} \bigg(24 \text{Tr}[(c_{{\phi d}}){}^{\dagger }c_{{\phi d}}]+8 \text{Tr}[(c_{{\phi e}}){}^{\dagger }c_{{\phi e}}]+16 \text{Tr}[(c_{\phi l}^{(1)}){}^{\dagger }c_{\phi l}^{(1)}]-16 \text{Tr}[(c_{\phi l}^{(3)}){}^{\dagger }c_{\phi l}^{(3)}] \, \\
&+48 \text{Tr}[(c_{\phi q}^{(1)}){}^{\dagger }c_{\phi q}^{(1)}]-48 \text{Tr}[(c_{\phi q}^{(3)}){}^{\dagger }c_{\phi q}^{(3)}]+24 \text{Tr}[(c_{{\phi u}}){}^{\dagger }c_{{\phi u}}]-24 \text{Tr}[(c_{{\phi ud}}){}^{\dagger }c_{{\phi ud}}]+32 c_{{\phi D}} c_{\phi \Box}\, \nonumber \\
&-11 c_{{\phi D}}^2-16 c_{\phi \Box}^2\bigg) \,, \nonumber \\
\dot{c}_{\phi^{4}}^{(2)}& =\frac{1}{3} \bigg(-8  (3 \text{Tr}[(c_{{\phi d}}){}^{\dagger }c_{{\phi d}}]+\text{Tr}[(c_{{\phi e}}){}^{\dagger }c_{{\phi e}}]+2 \text{Tr}[(c_{\phi l}^{(1)}){}^{\dagger }c_{\phi l}^{(1)}]+2\text{Tr}[(c_{\phi l}^{(3)}){}^{\dagger }c_{\phi l}^{(3)}]\, \\
&+ 6\text{Tr}[(c_{\phi q}^{(1)}){}^{\dagger }c_{\phi q}^{(1)}]+6\text{Tr}[(c_{\phi q}^{(3)}){}^{\dagger }c_{\phi q}^{(3)}]+3 \text{Tr}[(c_{{\phi u}}){}^{\dagger }c_{{\phi u}}])-16 c_{{\phi D}} c_{\phi \Box}-5 c_{{\phi D}}^2-16 c_{\phi \Box}^2 \bigg) \,, \nonumber \\
\dot{c}_{\phi^{4}}^{(3)}& =\frac{1}{3} \bigg(32 \text{Tr}[(c_{\phi l}^{(3)}){}^{\dagger }c_{\phi l}^{(3)}]+96 \text{Tr}[(c_{\phi q}^{(3)}){}^{\dagger }c_{\phi q}^{(3)}]+24 \text{Tr}[(c_{{\phi ud}}){}^{\dagger }c_{{\phi ud}}]-16 c_{{\phi D}} c_{{\phi \Box}}\, \\ 
& + 7 c_{{\phi D}}^2 -40 c_{{\phi \Box}}^2 \bigg) \,, \nonumber \\
\dot{c}_{W ^2 \phi^4}^{(1)} & = -\frac{1}{24} g_2^2 \bigg(3 c_{{\phi D}}^2 + 24 \text{Tr}[(c_{{\phi d}}){}^{\dagger }c_{{\phi d}}] - 48 \text{Tr}[(c_{\phi l}^{(3)}){}^{\dagger }c_{\phi l}^{(3)}]+48 \text{Tr}[(c_{\phi q}^{(1)}){}^{\dagger }c_{\phi q}^{(1)}] \, \\
&- 144 \text{Tr}[(c_{\phi q}^{(3)}){}^{\dagger }c_{\phi q}^{(3)}]+24 \text{Tr}[(c_{{\phi u}}){}^{\dagger }c_{{\phi u}}]+ 24 \text{Tr}[(c_{{\phi ud}}){}^{\dagger }c_{{\phi ud}}]+8 \text{Tr}[(c_{{\phi e}}){}^{\dagger }c_{{\phi e}}] \, \nonumber \\
&+16 \text{Tr}[(c_{\phi l}^{(1)}){}^{\dagger }c_{\phi l}^{(1)}]\bigg) \,, \nonumber \\
\dot{c}_{W\phi^4 D^2}^{(1)} & = -\frac{1}{3} g_2 \bigg(3 c_{{\phi D}}^2 + 24 \text{Tr}[(c_{{\phi d}}){}^{\dagger }c_{{\phi d}}]-48 \text{Tr}[(c_{\phi l}^{(3)}){}^{\dagger }c_{\phi l}^{(3)}]+48 \text{Tr}[(c_{\phi q}^{(1)}){}^{\dagger }c_{\phi q}^{(1)}] \, \\
&- 144 \text{Tr}[(c_{\phi q}^{(3)}){}^{\dagger }c_{\phi q}^{(3)}]+24 \text{Tr}[(c_{{\phi u}}){}^{\dagger }c_{{\phi u}}]+24 \text{Tr}[(c_{{\phi ud}}){}^{\dagger }c_{{\phi ud}}]+8 \text{Tr}[(c_{{\phi e}}){}^{\dagger }c_{{\phi e}}] \, \nonumber \\
&+16 \text{Tr}[(c_{\phi l}^{(1)}){}^{\dagger }c_{\phi l}^{(1)}] \bigg) \, , \nonumber \\
\dot{c}_{B ^2 \phi^4}^{(1)} & = \frac{1}{24} g_1^2 \bigg(3 c_{{\phi D}}^2+24 \text{Tr}[(c_{{\phi d}}){}^{\dagger }c_{{\phi d}}]-48 \text{Tr}[(c_{\phi l}^{(3)}){}^{\dagger }c_{\phi l}^{(3)}]+48 \text{Tr}[(c_{\phi q}^{(1)}){}^{\dagger }c_{\phi q}^{(1)}] \, \\
&-144 \text{Tr}[(c_{\phi q}^{(3)}){}^{\dagger }c_{\phi q}^{(3)}]+24 \text{Tr}[(c_{{\phi u}}){}^{\dagger }c_{{\phi u}}]+24 \text{Tr}[(c_{{\phi ud}}){}^{\dagger }c_{{\phi ud}}]+8 \text{Tr}[(c_{{\phi e}}){}^{\dagger }c_{{\phi e}}] \, \nonumber \\
&+16 \text{Tr}[(c_{\phi l}^{(1)}){}^{\dagger }c_{\phi l}^{(1)}] \bigg) \, , \nonumber \\
\dot{c}_{B\phi^4 D^2}^{(1)} & =\frac{1}{3} g_1 \bigg( 3 c_{{\phi D}}^2 + 24 \text{Tr}[(c_{{\phi d}}){}^{\dagger }c_{{\phi d}}] - 48 \text{Tr}[(c_{\phi l}^{(3)}){}^{\dagger }c_{\phi l}^{(3)}]+48 \text{Tr}[(c_{\phi q}^{(1)}){}^{\dagger }c_{\phi q}^{(1)}] \, \\
&- 144 \text{Tr}[(c_{\phi q}^{(3)}){}^{\dagger }c_{\phi q}^{(3)}]+ 24 \text{Tr}[(c_{{\phi u}}){}^{\dagger }c_{{\phi u}}]+24 \text{Tr}[(c_{{\phi ud}}){}^{\dagger }c_{{\phi ud}}]+8 \text{Tr}[(c_{{\phi e}}){}^{\dagger }c_{{\phi e}}] \, \nonumber \\
&+16 \text{Tr}[(c_{\phi l}^{(1)}){}^{\dagger }c_{\phi l}^{(1)}] \bigg) \, . \nonumber
\end{align}

\section{Ultraviolet completion of the Standard Model}
\label{sec:UVmodel}
The purpose of this appendix is proving that there exists at least one UV completion of the SM that induces arbitrary values of $c_\phi$, $c_{\phi D}$ and $c_{\phi\square}$.
To this aim, let us extend the SM (for $\mu^2=0$) with three colorless scalars: $\mathcal{S}\sim(1,1)_0$, $\Xi_0\sim(1,3)_0$ and $\Xi_1\sim(1,3)_1$. The numbers in parentheses and the subscript indicate the representations of $SU(3)_c$, $SU(2)_L$ and $U(1)_Y$, respectively.

Let us assume that they all have mass $M$ much larger than the EW scale, and that the new physics interaction Lagrangian is:
\begin{align}
 \mathcal{L}_\text{NP} = \kappa_\mathcal{S}\mathcal{S}\phi^\dagger\phi +\lambda_\mathcal{S}\mathcal{S}^2\phi^\dagger\phi + \kappa_{\Xi_0}\phi^\dagger\Xi_0^a\sigma_a\phi + \left(\kappa_{\Xi_1}\Xi_1^{a\dagger} \tilde{\phi}^\dagger\sigma_a\phi + \text{h.c.}\right)\,.
\end{align}
(Other triple and quartic terms are allowed, but we just ignored them for simplicity.) Then, by integrating out the heavy modes at tree level at the scale $M$, we obtain~\cite{deBlas:2014mba}:
\begin{align}
 \frac{c_\phi}{\Lambda^2} &= -\lambda_\mathcal{S}\frac{\kappa_\mathcal{S}^2}{M^4}\,, \nonumber\\[0.2cm]
 \frac{c_{\phi D}}{\Lambda^2} &= \frac{2}{M^4}(2 \kappa_{\Xi_1}^2-\kappa_{\Xi_0}^2)\,, \nonumber\\[0.2cm]
 \frac{c_{\phi\square}}{\Lambda^2} &= \frac{1}{2 M^4}(4\kappa_{\Xi_1}^2 + \kappa_{\Xi_0}^2 - \kappa_\mathcal{S}^2)\,.
\end{align}
Obviously, $c_\phi$ can have arbitrary sign by just tuning $\lambda_\mathcal{S}$. Likewise, $c_{\phi D}$ can be made arbitrarily negative provided $\kappa_{\Xi_1}/\kappa_{\Xi_0}\ll 1$, and positive otherwise. Notwithstanding this later choice, $c_{\phi\square}$ will be positive for small enough $\kappa_\mathcal{S}$ and negative for large values of this parameter. In summary, the signs of the three tree-level generated dimension-six operators are arbitrary and uncorrelated.

In the process of integrating out the fields of mass $M$, dimension-eight operators arise too. With the help of \texttt{MatchingTools}~\cite{Criado:2017khh}, we find that (see also Ref.~\cite{Remmen:2019cyz}):
\begin{align}
 \frac{c_{\phi^4}^{(1)}}{\Lambda^4} = 4\frac{\kappa_{\Xi_0}^2}{M^6}\,,\qquad
 \frac{c_{\phi^4}^{(2)}}{\Lambda^4} = 8\frac{\kappa_{\Xi_1}^2}{M^6}\,,\qquad
 \frac{c_{\phi^4}^{(3)}}{\Lambda^4} = \frac{2}{M^6}(\kappa_\mathcal{S}^2-\kappa_{\Xi_0}^2)\,.
\end{align}
Contrary to the dimension-six Wilson coefficients above, these couplings fullfill the positivity bounds $c_{\phi^4}^{(2)}\geq 0$, $c_{\phi^4}^{(1)}+c_{\phi^4}^{(2)}\geq 0$ and $c_{\phi^4}^{(1)}+c_{\phi^4}^{(2)}+c_{\phi^4}^{(3)}\geq 0$ obtained in Ref.~\cite{Remmen:2019cyz} for arbitrary values of the $\kappa$s.

\bibliographystyle{SciPost_bibstyle} 
\bibliography{refs} 

\begin{thebibliography}{10}
\providecommand{\url}[1]{\texttt{#1}}
\providecommand{\urlprefix}{URL }
\expandafter\ifx\csname urlstyle\endcsname\relax
  \providecommand{\doi}[1]{doi:\discretionary{}{}{}#1}\else
  \providecommand{\doi}{doi:\discretionary{}{}{}\begingroup
  \urlstyle{rm}\Url}\fi
\providecommand{\eprint}[2][]{\href{https://arxiv.org/abs/#2}{#2}}

\bibitem{Brivio:2017vri}
I.~Brivio and M.~Trott,
\newblock \emph{{The Standard Model as an Effective Field Theory}},
\newblock Phys. Rept. \textbf{793}, 1 (2019),
\newblock \doi{10.1016/j.physrep.2018.11.002},
\newblock \eprint{1706.08945}.

\bibitem{Yuan:2020fyf}
L.~Yuan,
\newblock \emph{{Exotics and BSM in ATLAS and CMS (Non dark matter searches)}},
\newblock PoS \textbf{CORFU2019}, 051 (2020),
\newblock \doi{10.22323/1.376.0051}.

\bibitem{Ellis:2020unq}
J.~Ellis, M.~Madigan, K.~Mimasu, V.~Sanz and T.~You,
\newblock \emph{{Top, Higgs, Diboson and Electroweak Fit to the Standard Model
  Effective Field Theory}},
\newblock JHEP \textbf{04}, 279 (2021),
\newblock \doi{10.1007/JHEP04(2021)279},
\newblock \eprint{2012.02779}.

\bibitem{Grojean:2013kd}
C.~Grojean, E.~E. Jenkins, A.~V. Manohar and M.~Trott,
\newblock \emph{{Renormalization Group Scaling of Higgs Operators and
  \textbackslash{}Gamma(h -\ensuremath{>} \textbackslash{}gamma
  \textbackslash{}gamma)}},
\newblock JHEP \textbf{04}, 016 (2013),
\newblock \doi{10.1007/JHEP04(2013)016},
\newblock \eprint{1301.2588}.

\bibitem{Elias-Miro:2013gya}
J.~Elias-Mir\'o, J.~R. Espinosa, E.~Masso and A.~Pomarol,
\newblock \emph{{Renormalization of dimension-six operators relevant for the
  Higgs decays $h\rightarrow \gamma\gamma,\gamma Z$}},
\newblock JHEP \textbf{08}, 033 (2013),
\newblock \doi{10.1007/JHEP08(2013)033},
\newblock \eprint{1302.5661}.

\bibitem{Elias-Miro:2013mua}
J.~Elias-Miro, J.~R. Espinosa, E.~Masso and A.~Pomarol,
\newblock \emph{{Higgs windows to new physics through d=6 operators:
  constraints and one-loop anomalous dimensions}},
\newblock JHEP \textbf{11}, 066 (2013),
\newblock \doi{10.1007/JHEP11(2013)066},
\newblock \eprint{1308.1879}.

\bibitem{Jenkins:2013zja}
E.~E. Jenkins, A.~V. Manohar and M.~Trott,
\newblock \emph{{Renormalization Group Evolution of the Standard Model
  Dimension Six Operators I: Formalism and lambda Dependence}},
\newblock JHEP \textbf{10}, 087 (2013),
\newblock \doi{10.1007/JHEP10(2013)087},
\newblock \eprint{1308.2627}.

\bibitem{Jenkins:2013wua}
E.~E. Jenkins, A.~V. Manohar and M.~Trott,
\newblock \emph{{Renormalization Group Evolution of the Standard Model
  Dimension Six Operators II: Yukawa Dependence}},
\newblock JHEP \textbf{01}, 035 (2014),
\newblock \doi{10.1007/JHEP01(2014)035},
\newblock \eprint{1310.4838}.

\bibitem{Alonso:2013hga}
R.~Alonso, E.~E. Jenkins, A.~V. Manohar and M.~Trott,
\newblock \emph{{Renormalization Group Evolution of the Standard Model
  Dimension Six Operators III: Gauge Coupling Dependence and Phenomenology}},
\newblock JHEP \textbf{04}, 159 (2014),
\newblock \doi{10.1007/JHEP04(2014)159},
\newblock \eprint{1312.2014}.

\bibitem{Buchalla:2019wsc}
G.~Buchalla, A.~Celis, C.~Krause and J.-N. Toelstede,
\newblock \emph{{Master Formula for One-Loop Renormalization of Bosonic SMEFT
  Operators}}  (2019),
\newblock \eprint{1904.07840}.

\bibitem{Descotes-Genon:2018foz}
S.~Descotes-Genon, A.~Falkowski, M.~Fedele, M.~Gonz\'alez-Alonso and J.~Virto,
\newblock \emph{{The CKM parameters in the SMEFT}},
\newblock JHEP \textbf{05}, 172 (2019),
\newblock \doi{10.1007/JHEP05(2019)172},
\newblock \eprint{1812.08163}.

\bibitem{Bissmann:2019gfc}
S.~Bi\ss{}mann, J.~Erdmann, C.~Grunwald, G.~Hiller and K.~Kr\"oninger,
\newblock \emph{{Constraining top-quark couplings combining top-quark and
  $\boldsymbol{B}$ decay observables}},
\newblock Eur. Phys. J. C \textbf{80}(2), 136 (2020),
\newblock \doi{10.1140/epjc/s10052-020-7680-9},
\newblock \eprint{1909.13632}.

\bibitem{Terol-Calvo:2019vck}
J.~Terol-Calvo, M.~T\'ortola and A.~Vicente,
\newblock \emph{{High-energy constraints from low-energy neutrino nonstandard
  interactions}},
\newblock Phys. Rev. D \textbf{101}(9), 095010 (2020),
\newblock \doi{10.1103/PhysRevD.101.095010},
\newblock \eprint{1912.09131}.

\bibitem{Degrande:2020evl}
C.~Degrande, G.~Durieux, F.~Maltoni, K.~Mimasu, E.~Vryonidou and C.~Zhang,
\newblock \emph{{Automated one-loop computations in the standard model
  effective field theory}},
\newblock Phys. Rev. D \textbf{103}(9), 096024 (2021),
\newblock \doi{10.1103/PhysRevD.103.096024},
\newblock \eprint{2008.11743}.

\bibitem{Azatov:2016sqh}
A.~Azatov, R.~Contino, C.~S. Machado and F.~Riva,
\newblock \emph{{Helicity selection rules and noninterference for BSM
  amplitudes}},
\newblock Phys. Rev. D \textbf{95}(6), 065014 (2017),
\newblock \doi{10.1103/PhysRevD.95.065014},
\newblock \eprint{1607.05236}.

\bibitem{Chala:2018ari}
M.~Chala, C.~Krause and G.~Nardini,
\newblock \emph{{Signals of the electroweak phase transition at colliders and
  gravitational wave observatories}},
\newblock JHEP \textbf{07}, 062 (2018),
\newblock \doi{10.1007/JHEP07(2018)062},
\newblock \eprint{1802.02168}.

\bibitem{Murphy:2020rsh}
C.~W. Murphy,
\newblock \emph{{Dimension-8 operators in the Standard Model Eective Field
  Theory}},
\newblock JHEP \textbf{10}, 174 (2020),
\newblock \doi{10.1007/JHEP10(2020)174},
\newblock \eprint{2005.00059}.

\bibitem{Corbett:2021eux}
T.~Corbett, A.~Helset, A.~Martin and M.~Trott,
\newblock \emph{{EWPD in the SMEFT to dimension eight}}  (2021),
\newblock \eprint{2102.02819}.

\bibitem{Panico:2018hal}
G.~Panico, A.~Pomarol and M.~Riembau,
\newblock \emph{{EFT approach to the electron Electric Dipole Moment at the
  two-loop level}},
\newblock JHEP \textbf{04}, 090 (2019),
\newblock \doi{10.1007/JHEP04(2019)090},
\newblock \eprint{1810.09413}.

\bibitem{Ardu:2021koz}
M.~Ardu and S.~Davidson,
\newblock \emph{{What is Leading Order for LFV in SMEFT?}}  (2021),
\newblock \eprint{2103.07212}.

\bibitem{Lehman:2014jma}
L.~Lehman,
\newblock \emph{{Extending the Standard Model Effective Field Theory with the
  Complete Set of Dimension-7 Operators}},
\newblock Phys. Rev. D \textbf{90}(12), 125023 (2014),
\newblock \doi{10.1103/PhysRevD.90.125023},
\newblock \eprint{1410.4193}.

\bibitem{Liao:2016hru}
Y.~Liao and X.-D. Ma,
\newblock \emph{{Renormalization Group Evolution of Dimension-seven Baryon- and
  Lepton-number-violating Operators}},
\newblock JHEP \textbf{11}, 043 (2016),
\newblock \doi{10.1007/JHEP11(2016)043},
\newblock \eprint{1607.07309}.

\bibitem{Hays:2018zze}
C.~Hays, A.~Martin, V.~Sanz and J.~Setford,
\newblock \emph{{On the impact of dimension-eight SMEFT operators on Higgs
  measurements}},
\newblock JHEP \textbf{02}, 123 (2019),
\newblock \doi{10.1007/JHEP02(2019)123},
\newblock \eprint{1808.00442}.

\bibitem{Ellis:2019zex}
J.~Ellis, S.-F. Ge, H.-J. He and R.-Q. Xiao,
\newblock \emph{{Probing the scale of new physics in the $ZZ\gamma$ coupling at
  $e^+e^-$ colliders}},
\newblock Chin. Phys. C \textbf{44}(6), 063106 (2020),
\newblock \doi{10.1088/1674-1137/44/6/063106},
\newblock \eprint{1902.06631}.

\bibitem{Alioli:2020kez}
S.~Alioli, R.~Boughezal, E.~Mereghetti and F.~Petriello,
\newblock \emph{{Novel angular dependence in Drell-Yan lepton production via
  dimension-8 operators}},
\newblock Phys. Lett. B \textbf{809}, 135703 (2020),
\newblock \doi{10.1016/j.physletb.2020.135703},
\newblock \eprint{2003.11615}.

\bibitem{Ellis:2020ljj}
J.~Ellis, H.-J. He and R.-Q. Xiao,
\newblock \emph{{Probing new physics in dimension-8 neutral gauge couplings at
  $e^{+}e^{-}$ colliders}},
\newblock Sci. China Phys. Mech. Astron. \textbf{64}(2), 221062 (2021),
\newblock \doi{10.1007/s11433-020-1617-3},
\newblock \eprint{2008.04298}.

\bibitem{Hays:2020scx}
C.~Hays, A.~Helset, A.~Martin and M.~Trott,
\newblock \emph{{Exact SMEFT formulation and expansion to
  $\mathcal{O}(v^4/\Lambda^4)$}},
\newblock JHEP \textbf{11}, 087 (2020),
\newblock \doi{10.1007/JHEP11(2020)087},
\newblock \eprint{2007.00565}.

\bibitem{Gu:2020ldn}
J.~Gu, L.-T. Wang and C.~Zhang,
\newblock \emph{{An unambiguous test of positivity at lepton colliders}}
  (2020),
\newblock \eprint{2011.03055}.

\bibitem{Craig:2019wmo}
N.~Craig, M.~Jiang, Y.-Y. Li and D.~Sutherland,
\newblock \emph{{Loops and Trees in Generic EFTs}},
\newblock JHEP \textbf{08}, 086 (2020),
\newblock \doi{10.1007/JHEP08(2020)086},
\newblock \eprint{2001.00017}.

\bibitem{Davidson:2018zuo}
S.~Davidson, M.~Gorbahn and M.~Leak,
\newblock \emph{{Majorana neutrino masses in the renormalization group
  equations for lepton flavor violation}},
\newblock Phys. Rev. D \textbf{98}(9), 095014 (2018),
\newblock \doi{10.1103/PhysRevD.98.095014},
\newblock \eprint{1807.04283}.

\bibitem{Zhang:2018shp}
C.~Zhang and S.-Y. Zhou,
\newblock \emph{{Positivity bounds on vector boson scattering at the LHC}},
\newblock Phys. Rev. D \textbf{100}(9), 095003 (2019),
\newblock \doi{10.1103/PhysRevD.100.095003},
\newblock \eprint{1808.00010}.

\bibitem{Bi:2019phv}
Q.~Bi, C.~Zhang and S.-Y. Zhou,
\newblock \emph{{Positivity constraints on aQGC: carving out the physical
  parameter space}},
\newblock JHEP \textbf{06}, 137 (2019),
\newblock \doi{10.1007/JHEP06(2019)137},
\newblock \eprint{1902.08977}.

\bibitem{Remmen:2019cyz}
G.~N. Remmen and N.~L. Rodd,
\newblock \emph{{Consistency of the Standard Model Effective Field Theory}},
\newblock JHEP \textbf{12}, 032 (2019),
\newblock \doi{10.1007/JHEP12(2019)032},
\newblock \eprint{1908.09845}.

\bibitem{Remmen:2020vts}
G.~N. Remmen and N.~L. Rodd,
\newblock \emph{{Flavor Constraints from Unitarity and Analyticity}},
\newblock Phys. Rev. Lett. \textbf{125}(8), 081601 (2020),
\newblock \doi{10.1103/PhysRevLett.125.081601},
\newblock \eprint{2004.02885}.

\bibitem{Remmen:2020uze}
G.~N. Remmen and N.~L. Rodd,
\newblock \emph{{Signs, Spin, SMEFT: Positivity at Dimension Six}}  (2020),
\newblock \eprint{2010.04723}.

\bibitem{Bonnefoy:2020yee}
Q.~Bonnefoy, E.~Gendy and C.~Grojean,
\newblock \emph{{Positivity bounds on Minimal Flavor Violation}},
\newblock JHEP \textbf{04}, 115 (2021),
\newblock \doi{10.1007/JHEP04(2021)115},
\newblock \eprint{2011.12855}.

\bibitem{Bellazzini:2020cot}
B.~Bellazzini, J.~Elias~Mir\'o, R.~Rattazzi, M.~Riembau and F.~Riva,
\newblock \emph{{Positive Moments for Scattering Amplitudes}}  (2020),
\newblock \eprint{2011.00037}.

\bibitem{Grzadkowski:2010es}
B.~Grzadkowski, M.~Iskrzynski, M.~Misiak and J.~Rosiek,
\newblock \emph{{Dimension-Six Terms in the Standard Model Lagrangian}},
\newblock JHEP \textbf{10}, 085 (2010),
\newblock \doi{10.1007/JHEP10(2010)085},
\newblock \eprint{1008.4884}.

\bibitem{Li:2020gnx}
H.-L. Li, Z.~Ren, J.~Shu, M.-L. Xiao, J.-H. Yu and Y.-H. Zheng,
\newblock \emph{{Complete Set of Dimension-8 Operators in the Standard Model
  Effective Field Theory}}  (2020),
\newblock \eprint{2005.00008}.

\bibitem{Elias-Miro:2014eia}
J.~Elias-Miro, J.~R. Espinosa and A.~Pomarol,
\newblock \emph{{One-loop non-renormalization results in EFTs}},
\newblock Phys. Lett. B \textbf{747}, 272 (2015),
\newblock \doi{10.1016/j.physletb.2015.05.056},
\newblock \eprint{1412.7151}.

\bibitem{Cheung:2015aba}
C.~Cheung and C.-H. Shen,
\newblock \emph{{Nonrenormalization Theorems without Supersymmetry}},
\newblock Phys. Rev. Lett. \textbf{115}(7), 071601 (2015),
\newblock \doi{10.1103/PhysRevLett.115.071601},
\newblock \eprint{1505.01844}.

\bibitem{Bern:2019wie}
Z.~Bern, J.~Parra-Martinez and E.~Sawyer,
\newblock \emph{{Nonrenormalization and Operator Mixing via On-Shell Methods}},
\newblock Phys. Rev. Lett. \textbf{124}(5), 051601 (2020),
\newblock \doi{10.1103/PhysRevLett.124.051601},
\newblock \eprint{1910.05831}.

\bibitem{delAguila:2000rc}
F.~del Aguila, M.~Perez-Victoria and J.~Santiago,
\newblock \emph{{Observable contributions of new exotic quarks to quark
  mixing}},
\newblock JHEP \textbf{09}, 011 (2000),
\newblock \doi{10.1088/1126-6708/2000/09/011},
\newblock \eprint{hep-ph/0007316}.

\bibitem{delAguila:2008pw}
F.~del Aguila, J.~de~Blas and M.~Perez-Victoria,
\newblock \emph{{Effects of new leptons in Electroweak Precision Data}},
\newblock Phys. Rev. D \textbf{78}, 013010 (2008),
\newblock \doi{10.1103/PhysRevD.78.013010},
\newblock \eprint{0803.4008}.

\bibitem{delAguila:2010mx}
F.~del Aguila, J.~de~Blas and M.~Perez-Victoria,
\newblock \emph{{Electroweak Limits on General New Vector Bosons}},
\newblock JHEP \textbf{09}, 033 (2010),
\newblock \doi{10.1007/JHEP09(2010)033},
\newblock \eprint{1005.3998}.

\bibitem{deBlas:2014mba}
J.~de~Blas, M.~Chala, M.~Perez-Victoria and J.~Santiago,
\newblock \emph{{Observable Effects of General New Scalar Particles}},
\newblock JHEP \textbf{04}, 078 (2015),
\newblock \doi{10.1007/JHEP04(2015)078},
\newblock \eprint{1412.8480}.

\bibitem{deBlas:2017xtg}
J.~de~Blas, J.~C. Criado, M.~Perez-Victoria and J.~Santiago,
\newblock \emph{{Effective description of general extensions of the Standard
  Model: the complete tree-level dictionary}},
\newblock JHEP \textbf{03}, 109 (2018),
\newblock \doi{10.1007/JHEP03(2018)109},
\newblock \eprint{1711.10391}.

\bibitem{Alloul:2013bka}
A.~Alloul, N.~D. Christensen, C.~Degrande, C.~Duhr and B.~Fuks,
\newblock \emph{{FeynRules 2.0 - A complete toolbox for tree-level
  phenomenology}},
\newblock Comput. Phys. Commun. \textbf{185}, 2250 (2014),
\newblock \doi{10.1016/j.cpc.2014.04.012},
\newblock \eprint{1310.1921}.

\bibitem{Hahn:2000kx}
T.~Hahn,
\newblock \emph{{Generating Feynman diagrams and amplitudes with FeynArts 3}},
\newblock Comput. Phys. Commun. \textbf{140}, 418 (2001),
\newblock \doi{10.1016/S0010-4655(01)00290-9},
\newblock \eprint{hep-ph/0012260}.

\bibitem{Hahn:1998yk}
T.~Hahn and M.~Perez-Victoria,
\newblock \emph{{Automatized one loop calculations in four-dimensions and
  D-dimensions}},
\newblock Comput. Phys. Commun. \textbf{118}, 153 (1999),
\newblock \doi{10.1016/S0010-4655(98)00173-8},
\newblock \eprint{hep-ph/9807565}.

\bibitem{matchmaker}
A.~Carmona, A.~Lazopoulos, P.~Olgoso and J.~Santiago,
\newblock \emph{{MatchMaker: automated one-loop matching}},
\newblock (to appear) .

\bibitem{Gherardi:2020det}
V.~Gherardi, D.~Marzocca and E.~Venturini,
\newblock \emph{{Matching scalar leptoquarks to the SMEFT at one loop}},
\newblock JHEP \textbf{07}, 225 (2020),
\newblock \doi{10.1007/JHEP07(2020)225},
\newblock [Erratum: JHEP 01, 006 (2021)],
\newblock \eprint{2003.12525}.

\bibitem{Criado:2018sdb}
J.~C. Criado and M.~P\'erez-Victoria,
\newblock \emph{{Field redefinitions in effective theories at higher orders}},
\newblock JHEP \textbf{03}, 038 (2019),
\newblock \doi{10.1007/JHEP03(2019)038},
\newblock \eprint{1811.09413}.

\bibitem{EliasMiro:2020tdv}
J.~Elias~Mir\'o, J.~Ingoldby and M.~Riembau,
\newblock \emph{{EFT anomalous dimensions from the S-matrix}},
\newblock JHEP \textbf{09}, 163 (2020),
\newblock \doi{10.1007/JHEP09(2020)163},
\newblock \eprint{2005.06983}.

\bibitem{Baratella:2020lzz}
P.~Baratella, C.~Fernandez and A.~Pomarol,
\newblock \emph{{Renormalization of Higher-Dimensional Operators from On-shell
  Amplitudes}},
\newblock Nucl. Phys. B \textbf{959}, 115155 (2020),
\newblock \doi{10.1016/j.nuclphysb.2020.115155},
\newblock \eprint{2005.07129}.

\bibitem{Peskin:1990zt}
M.~E. Peskin and T.~Takeuchi,
\newblock \emph{{A New constraint on a strongly interacting Higgs sector}},
\newblock Phys. Rev. Lett. \textbf{65}, 964 (1990),
\newblock \doi{10.1103/PhysRevLett.65.964}.

\bibitem{deBlas:2016nqo}
J.~de~Blas, M.~Ciuchini, E.~Franco, S.~Mishima, M.~Pierini, L.~Reina and
  L.~Silvestrini,
\newblock \emph{{Electroweak precision constraints at present and future
  colliders}},
\newblock PoS \textbf{ICHEP2016}, 690 (2017),
\newblock \doi{10.22323/1.282.0690},
\newblock \eprint{1611.05354}.

\bibitem{Maltoni:2019aot}
F.~Maltoni, L.~Mantani and K.~Mimasu,
\newblock \emph{{Top-quark electroweak interactions at high energy}},
\newblock JHEP \textbf{10}, 004 (2019),
\newblock \doi{10.1007/JHEP10(2019)004},
\newblock \eprint{1904.05637}.

\bibitem{Zhang:1992fs}
X.-m. Zhang,
\newblock \emph{{Operators analysis for Higgs potential and cosmological bound
  on Higgs mass}},
\newblock Phys. Rev. D \textbf{47}, 3065 (1993),
\newblock \doi{10.1103/PhysRevD.47.3065},
\newblock \eprint{hep-ph/9301277}.

\bibitem{Grojean:2004xa}
C.~Grojean, G.~Servant and J.~D. Wells,
\newblock \emph{{First-order electroweak phase transition in the standard model
  with a low cutoff}},
\newblock Phys. Rev. D \textbf{71}, 036001 (2005),
\newblock \doi{10.1103/PhysRevD.71.036001},
\newblock \eprint{hep-ph/0407019}.

\bibitem{Bodeker:2004ws}
D.~Bodeker, L.~Fromme, S.~J. Huber and M.~Seniuch,
\newblock \emph{{The Baryon asymmetry in the standard model with a low
  cut-off}},
\newblock JHEP \textbf{02}, 026 (2005),
\newblock \doi{10.1088/1126-6708/2005/02/026},
\newblock \eprint{hep-ph/0412366}.

\bibitem{Delaunay:2007wb}
C.~Delaunay, C.~Grojean and J.~D. Wells,
\newblock \emph{{Dynamics of Non-renormalizable Electroweak Symmetry
  Breaking}},
\newblock JHEP \textbf{04}, 029 (2008),
\newblock \doi{10.1088/1126-6708/2008/04/029},
\newblock \eprint{0711.2511}.

\bibitem{deVries:2017ncy}
J.~de~Vries, M.~Postma, J.~van~de Vis and G.~White,
\newblock \emph{{Electroweak Baryogenesis and the Standard Model Effective
  Field Theory}},
\newblock JHEP \textbf{01}, 089 (2018),
\newblock \doi{10.1007/JHEP01(2018)089},
\newblock \eprint{1710.04061}.

\bibitem{Caprini:2019egz}
C.~Caprini \emph{et~al.},
\newblock \emph{{Detecting gravitational waves from cosmological phase
  transitions with LISA: an update}},
\newblock JCAP \textbf{03}, 024 (2020),
\newblock \doi{10.1088/1475-7516/2020/03/024},
\newblock \eprint{1910.13125}.

\bibitem{Kuzmin:1985mm}
V.~A. Kuzmin, V.~A. Rubakov and M.~E. Shaposhnikov,
\newblock \emph{{On the Anomalous Electroweak Baryon Number Nonconservation in
  the Early Universe}},
\newblock Phys. Lett. B \textbf{155}, 36 (1985),
\newblock \doi{10.1016/0370-2693(85)91028-7}.

\bibitem{Chankowski:1993tx}
P.~H. Chankowski and Z.~Pluciennik,
\newblock \emph{{Renormalization group equations for seesaw neutrino masses}},
\newblock Phys. Lett. B \textbf{316}, 312 (1993),
\newblock \doi{10.1016/0370-2693(93)90330-K},
\newblock \eprint{hep-ph/9306333}.

\bibitem{Babu:1993qv}
K.~S. Babu, C.~N. Leung and J.~T. Pantaleone,
\newblock \emph{{Renormalization of the neutrino mass operator}},
\newblock Phys. Lett. B \textbf{319}, 191 (1993),
\newblock \doi{10.1016/0370-2693(93)90801-N},
\newblock \eprint{hep-ph/9309223}.

\bibitem{Antusch:2001ck}
S.~Antusch, M.~Drees, J.~Kersten, M.~Lindner and M.~Ratz,
\newblock \emph{{Neutrino mass operator renormalization revisited}},
\newblock Phys. Lett. B \textbf{519}, 238 (2001),
\newblock \doi{10.1016/S0370-2693(01)01127-3},
\newblock \eprint{hep-ph/0108005}.

\bibitem{Alonso:2014zka}
R.~Alonso, H.-M. Chang, E.~E. Jenkins, A.~V. Manohar and B.~Shotwell,
\newblock \emph{{Renormalization group evolution of dimension-six baryon number
  violating operators}},
\newblock Phys. Lett. B \textbf{734}, 302 (2014),
\newblock \doi{10.1016/j.physletb.2014.05.065},
\newblock \eprint{1405.0486}.

\bibitem{Liao:2019tep}
Y.~Liao and X.-D. Ma,
\newblock \emph{{Renormalization Group Evolution of Dimension-seven Operators
  in Standard Model Effective Field Theory and Relevant Phenomenology}},
\newblock JHEP \textbf{03}, 179 (2019),
\newblock \doi{10.1007/JHEP03(2019)179},
\newblock \eprint{1901.10302}.

\bibitem{Chala:2021juk}
M.~Chala and A.~Titov,
\newblock \emph{{Neutrino masses in the Standard Model effective field theory}}
   (2021),
\newblock \eprint{2104.08248}.

\bibitem{Jiang:2020mhe}
M.~Jiang, T.~Ma and J.~Shu,
\newblock \emph{{Renormalization Group Evolution from On-shell SMEFT}},
\newblock JHEP \textbf{01}, 101 (2021),
\newblock \doi{10.1007/JHEP01(2021)101},
\newblock \eprint{2005.10261}.

\bibitem{Barzinji:2018xvu}
A.~Barzinji, M.~Trott and A.~Vasudevan,
\newblock \emph{{Equations of Motion for the Standard Model Effective Field
  Theory: Theory and Applications}},
\newblock Phys. Rev. D \textbf{98}(11), 116005 (2018),
\newblock \doi{10.1103/PhysRevD.98.116005},
\newblock \eprint{1806.06354}.

\bibitem{Criado:2017khh}
J.~C. Criado,
\newblock \emph{{MatchingTools: a Python library for symbolic effective field
  theory calculations}},
\newblock Comput. Phys. Commun. \textbf{227}, 42 (2018),
\newblock \doi{10.1016/j.cpc.2018.02.016},
\newblock \eprint{1710.06445}.

\end{thebibliography}

\nolinenumbers

\end{document}